# Collapse and coexistence for a molecular braid with an attractive interaction component subject to mechanical forces

Dominic J. (O') Lee[1,a.)]

[1]Department of Chemistry, Imperial College London, SW7 2AZ, London, UK

## Abstract

Dual mechanical braiding experiments provide a useful tool with which to investigate the nature of interactions between rod-like molecules, for instance actin and DNA. In conditions close to molecular condensation, one would expect an appearance of a local minimum in the interaction potential between the two molecules. We investigate this situation, introducing an attractive component into the interaction potential, using a model developed for describing such experiments. We consider attraction that does not depend on molecular structure, as well that which depends on a DNA-like helix structure. In braiding experiments, an attractive term may lead to certain effects. A local minimum may cause molecules to collapse from a loosely braided configuration into a tight one, occurring at a critical value of the moment applied about the axis of the braid. For a fixed number of braid pitches, this may lead to coexistence between the two braiding states, tight and loose. Coexistence implies certain proportions of the braid are in each state, their relative size depending on the number of braid pitches. This manifests itself as linear dependence in numerically calculated quantities as functions of the number of braid pitches. Also, in the collapsed state, the braid radius stays roughly constant. Furthermore, if the attractive interaction is helix dependent, the left-right handed braid symmetry is broken. For a DNA like charge distribution, using the Kornyshev-Leikin interaction model, our results suggest that significant braid collapse and coexistence only occurs for left handed braids. Regardless of the interaction model, the study highlights the possible qualitative physics of braid collapse and coexistence; and the role helix specific forces might play, if important. The model could also be used to connect other microscopic theories of interaction with braiding experiments.

## 1. Introduction

The mechanical braiding of two long rod-like molecules [1,2,3] may provide a useful tool in probing the interaction forces between them, as well as their bending elasticity. In the case of DNA and other helical molecules, an interesting question to ask is: can such forces be chiral, as a dependence on helical structure would suggest [4,5,6,7]? Indeed, it would be difficult to imagine the formation of cholesteric phase [5,8], occurring for DNA, in molecular assemblies without some kind of chiral force, even if only effectively steric in origin. To reformulate the question another way: does the helical structure of molecules significantly influence forces between them under certain conditions? Perhaps, mechanical braiding experiments can give a definitive answer to this question, as well as provide other information about intermolecular forces for a molecular pair. For instance, how these forces depend on the relative inter-axial separation. Single

---

[a.)] Electronic Mail: domolee@hotmail.com

molecule braiding experiments, such as the ones performed in Refs [9,10,11,12,13,14,15,16], have indeed probed the separation dependence of intermolecular interactions. Also, in describing such experiments, there have been many theoretical studies [17,18,19,20,21,22,23,24,25,26,27,28,29]; in some, the structure-independent, repulsive electrostatic forces, between molecules, has been included. Nevertheless, dual molecule braiding [1,2,3] has certain advantages in investigating the chiral nature of interaction forces, which we discuss below.

In the braiding experiments of Refs. [1] and [3], two molecules were attached to a single magnetic bead and to a substrate; both the number of turns of the bead, as well as the pulling force, controlled by a magnetic field. Once the bead is rotated beyond half a turn, a braided section starts to form. By changing the number of turns and the pulling force, not only does length of the braid change; but the distance between centre lines and the (tilt) angle between them, within the braid, should change as well. Therefore, by constructing a reliable model that relates the intermolecular interaction potential to experimental observables, one indeed should be able to probe the potential, as a function of both inter-axial separation and tilt angle. One particular advantage of dual molecule braiding over single molecule braiding is that, if the molecules are nicked (or allowed some other way to relieve their torsional stress), there is no coupling between molecular bending and twisting. Consequently, there is no denaturisation when we rotate in one particular direction. In the particular case of twisting a single molecule of DNA, pulling forces of the order of $1 \mathrm{pN}$ denature the DNA, when we rotate the bead in the clockwise direction [25]. Because dual braiding does not suffer from this problem, when twisting and bending are decoupled, such experiments should also provide a useful tool to see what effect interactions might have on the left/handed braid symmetry; the degree of their chirality.

To construct a model that describes the braiding of two semi-flexible, rod-like molecules requires the inclusion of worm like chain fluctuations, limited by mechanical forces and the braid geometry. This should also account for the fact that braid geometry is, in turn, determined by the mechanical forces and fluctuations. In the pioneering studies Refs. [30] and [31] a statistical mechanical theory to describe mechanical dual molecular braiding for semi-flexible molecules was developed. Also, an elaborate study of the mechanics of braiding in the ground state, allowing for a general description of braid structures in response to different applied mechanical forces and the interaction between them, has been considered in Ref. [32]. This latter study allows for the pulling forces on each molecule to be different, resulting asymmetric braid structures, but this generalized case has not yet been extended beyond the ground state. In all of these studies, the molecular ends of the molecules are held apart at the same distance as the distance between the two molecular centrelines in the braid.

In the experimental setup of Refs. [1] and [3], the situation is somewhat different; the molecules are held apart a distance of the order of a micron, a length scale much larger than the Debye screening length for the electrostatic interactions. Therefore, to minimize the elastic energy, and the total energy of the system, the radius of the braid should be very much smaller than the distance between the two ends. To illustrate this, let us discuss what happens as we rotate the molecules away from their parallel configuration. Before braiding occurs, the molecules experience effectively no other forces apart from mechanical forces; when we start rotating the molecules, their unbent average centre lines come close together, at their midpoints, simply from geometry. The point where braiding actually occurs is at very slightly under half a turn of the bead. Braiding occurs when the two molecules start to experience repulsive intermolecular forces. It is these forces that make it energetically favourable for the molecules to start bending their average centre lines to accommodate a certain number of bead turns. This must always occur before half a turn, because they will eventually come into steric contact before then, if braiding has not occurred sooner. On the other hand, if the forces had an effective range of the same magnitude as the separation between the ends, we would have expected that braiding would commence immediately on turning the bead, and the braid radius would be of comparable magnitude to the separation between molecular ends. In Refs. [33] and [34], a model describing the braiding experiments of Refs. [1] and [3] was proposed that takes account of the large separation between the two sets of molecular ends. This was then extended to calculate the root mean squared fluctuations in inter-axial separation between molecules in the braid self consistently [35]. Assuming only repulsive contributions from electrostatic interactions between the two molecules, the model has been shown to fit the experimental data quite well [35], justifying its reliability.

In this study, using the model developed in Refs. [33,34,35], we now focus our attention on what might happen at ionic conditions close to those where condensation of the molecules occurs. Here, we might reasonably expect, along with the electrostatically dominated repulsive interaction, an additional attractive contribution to be important. This attractive contribution may subsequently cause a flattening or local minimum in the total interaction potential. If present, this may lead, as we shall demonstrate, to a collapse of the braid into a tighter structure as the number of turns of the bead or the pulling force is increased. Interestingly enough, once this collapse has occurred, the inter-axial separation stays roughly constant. This is due to large short range repulsive forces that make the system behave rather like the molecules were in steric contact, at this distance [28]. Because of coexistence between the collapsed and looser braid structure, the average braid radius changes in a continuous manner; there are no sudden jumps. Therefore, these effects might be able to explain extension plots in recent single molecule twisting experiments [13]. The analysis of Ref. [28] suggests that the inter-axial separation, in the plectoneme, should decrease with increasing pulling force; before then staying roughly constant at the value $\approx 30 \text{Å}$, when fitting the extension data. These experiments were done in the presence of DNA condensing agents, but in ionic conditions where condensation does not occur.

The attractive contribution may, or may not, depend on the helical structure of the DNA; it is not clear whether helix dependent forces significantly contribute to attraction or some mechanism that is independent of helical structure is mainly responsible for attraction. Nevertheless, one of the main ideas behind this work is to propose a general frame work that enables one to calculate predicted moment and extension curves, for dual molecule braiding experiments, from a microscopic model of interaction or simulations (provided that the results for the interaction energy are fitted to some empirical formula). With these points in mind, we have tried to make our calculations as general as possible, presenting results for an attractive interaction that do not depend on helix structure, as well as one that does. Indeed, both types of interaction can cause the braid to collapse. Though, while the former respects the symmetry between left and right handed braids, the latter breaks it. This symmetry breaking is seen in both the results for the extension of the molecules and the torque that needs to be applied to them to generate a number of braid turns. For the helix-dependent forces, this study goes beyond that of Ref. [33] in few different ways. The first is that we now include undulations for the braided section in all cases that we study. Secondly, we have investigated how helix geometry might limit the internal braiding torque caused by helix-specific interactions. Thirdly, we have generated results over a much larger range of parameters to look for qualitative trends.

The rest of the paper is organized in the following way. In the theory section, we start by discussing how to characterize the geometry of the braided section of the two molecules. This is then followed by a description of the various contributions to the energy functional describing the total energy of configurations of the braid. Here, we introduce expressions describing the interaction between molecules, for the various cases that we are interested in. The result for the free energy is then given, where each term is discussed. How the free energy is calculated is outlined in Appendix A. The attractive free energy contributions for both the non-helix specific and helix specific terms are given separate subsections, along with a discussion of the types of braiding states that they cause, and coexistence between these states. In this work, a large number of plots for both the extension and moment as functions of the number of braid turns has been generated; this covers a large range of parameter choices for each of the cases under study. Therefore, to lighten the main text, most of these plots are presented in Appendix B. Instead, in the main text, we present representative plots, and for these particular parameter choices we show plots of the geometric parameters of the braid. In the presentation of results, this section is divided into two parts. In the first part we consider what happens with non-helix specific attraction, while in the second part we focus on an attractive component that does depend on helix structure. In the discussion we discuss the significance of our findings and advocate new work, both theoretical and experimental.

## 2. Theory

### 2.1 General considerations

Let us consider two molecules of contour length $L$ that are rotated about each other $n$ turns away from a parallel configuration, where the molecules are initially held distance $b$ apart. The molecular ends are still held apart at distance $b$. We will also suppose that each individual molecule can relieve its torsional stress, by being either nicked or through some other mechanism. In the centre, for $|n|>1/2$ the molecules form a braid as shown in Fig. 1. To describe this system, we use the model developed in Refs. [33] and [34]. The free energy free energy is divided into two terms [33]

$$\mathcal{F}_T = \mathcal{F}_{end} + \mathcal{F}_{Braid}. \tag{2.1}$$

Here, $\mathcal{F}_{end}$ is the free energy of two unbraided end pieces for each of the two molecules and $\mathcal{F}_{Braid}$ is the free energy of the braided section. Each end piece uses up contour length $(L-L_b)/2$, while the braided section uses up length $L_b$ of the total length of each molecule. Both molecules are supposed to have the same length. We define the average distance $x$ from the ends of the braided section to the ends of the two molecules, for each of the four end pieces.

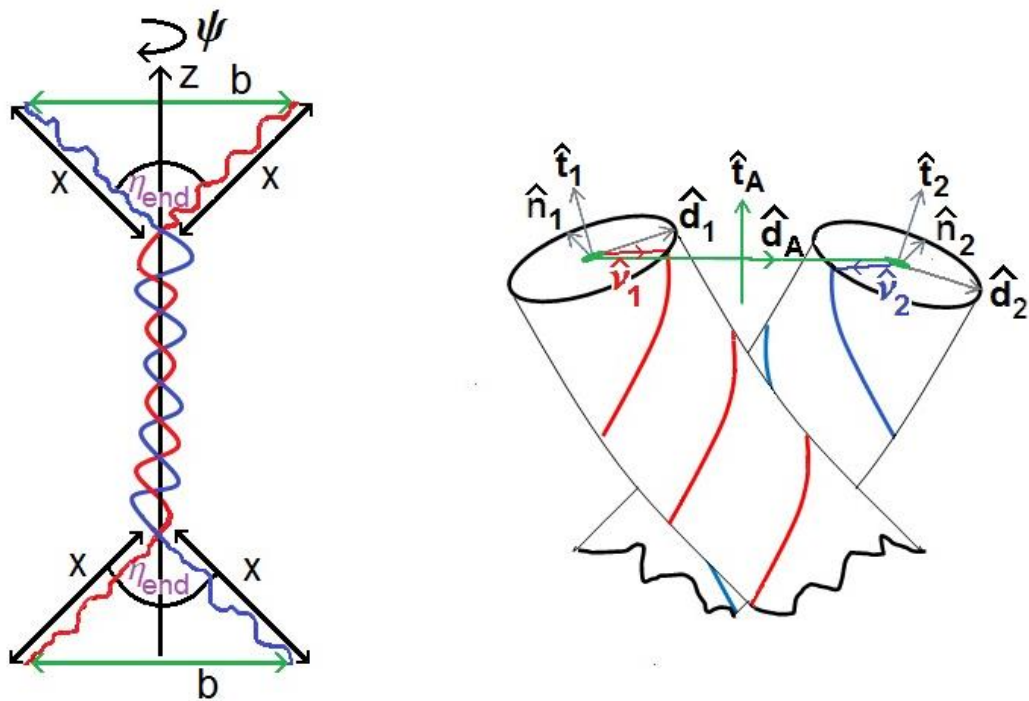

Fig.1 Schematic pictures showing the configuration of the two molecules. In the left hand figure, we show the global geometry of the system. The red (lighter) line denotes one molecule, while the blue (darker) line denotes the other one. The two sets of molecular ends are held distance $b$ apart and one set of ends are rotated an angle $\psi$ while other set remain fixed. Here $x$ is the average distance between the end of the braid and molecular ends. To generate this configuration, two of the ends may be attached to a magnetic bead and the other two ends to a substrate, as in the experiments of Ref. [3]. In the second picture the molecules are shown as being helical. We show the tangent vectors ($\hat{\mathbf{t}}_1(s)$ and $\hat{\mathbf{t}}_2(s)$) of the two molecular centre lines, the vector $\hat{\mathbf{d}}_A(s)$ that lies along a line connecting the two molecular centre lines (shown in green) and the tangent vector of the braid centre line. All of these define the local configuration of the braid. Also shown are the vectors $\hat{\mathbf{n}}_\mu(s)$ and $\hat{\mathbf{d}}_\mu(s)$, which are defined in Eq. (2.2), specifying the braid frames [32] of the two molecules. When $R'(s)=0$ we have that $\hat{\mathbf{d}}_1(s)=\hat{\mathbf{d}}_2(s)=\hat{\mathbf{d}}(s)$. The orientations of the two helices (or other simple structure type) are characterized, in their braid frames, through the vectors $\hat{\mathbf{v}}_1(s)$ and $\hat{\mathbf{v}}_2(s)$, which are given by Eq. (2.3) within the braid frames.

## 2.2 The Braided Section

Along the braided section we may define an arc length coordinate $s$ that runs from $-L_b/2$ to $L_b/2$. As we move along the braid, the axis of the thermally averaged braid is considered to be straight. This axis is defined as the axis of rotation of the parts of the two molecules forming the braided section, and lies at the midpoint of lines of length $R(s)$ connecting the two molecules (see Fig. 1). Along the thermally averaged braid axis a force $F=k_BTF_R/l_p$ is applied (here, $l_p$ is the bending persistence length), and a moment $M=k_BTM_R$ is applied about it. Fluctuations of the braid axis away from the average position are characterized through lateral displacements $x_A$ and $y_A$, which lie on axes perpendicular to average braid axis within the lab frame. To further describe the local geometry of the braid we define the braid tilt angle, $\eta$. This is defined as the angle between the two molecular centre lines of molecules 1 and 2, such that $\hat{\mathbf{t}}_1.\hat{\mathbf{t}}_2 = \cos\eta$, where $\hat{\mathbf{t}}_1$ and $\hat{\mathbf{t}}_2$ are the respective tangent vectors of these lines. A positive value of $\eta$ corresponds to a left handed braid, while a negative value a right handed one.

To characterize the helical structure of the two molecules forming the braid it is useful to construct what we call braid frames [32]. Along the lines of length $R$ connecting the two centre lines a unit vector $\hat{\mathbf{d}}$ points. We then may write

$$\hat{\mathbf{n}}_\mu(s) = \frac{\hat{\mathbf{t}}_\mu(s) \times \hat{\mathbf{d}}(s)}{\left|\hat{\mathbf{t}}_\mu(s) \times \hat{\mathbf{d}}(s)\right|}, \qquad \hat{\mathbf{d}}_\mu(s) = \hat{\mathbf{n}}_\mu(s) \times \hat{\mathbf{t}}_\mu(s), \qquad (2.2)$$

where the index $\mu$ corresponds to the molecule in question { $\mu = 1, 2$ }. Then, a line may be drawn along the surface of the molecules to indicate the relative azimuthal positions of particular molecular groups forming the macro-molecules. The trajectory of this line, on the molecular surface, (see Fig. 1) may be characterized through the vectors

$$\hat{\mathbf{v}}_\mu(s) = \cos\xi_\mu(s)\hat{\mathbf{d}}_\mu(s) + \sin\xi_\mu(s)\hat{\mathbf{n}}_\mu(s). \qquad (2.3)$$

As we shall soon see, such a mathematical construction is useful as it provides a reference to describe the individual torsional stresses on each of the molecules. For DNA, a natural choice for such a line is bisecting the minor groove, so tracing out a helix.

Now, to find expressions for $\mathcal{F}_{Braid}$ we need to consider an energy functional to be able to write the partition function. This functional describes the energy of each configuration of the braid accessible by thermal fluctuations In what follows, the thermal averages of $R(s)$ and $\eta(s)$ will be given by $\langle R(s) \rangle = R_0$ and $\langle \eta(s) \rangle = \eta_0$, and we may express $R(s) = R_0 + \delta R(s)$ and $\eta(s) = \eta_0 + \delta\eta(s)$, where for the fluctuations $\langle \delta R(s) \rangle = 0$ and $\langle \delta\eta(s) \rangle = 0$. This energy functional is described (expressions for various terms derived in Ref. [36]) by following sum of terms

$$E_{braid} = E_{st} + E_B + E_{Tw} + E_W + E_{int}. \qquad (2.4)$$

The first of these terms is the steric interaction, which we model as a harmonic potential (which was originally proposed in Ref. [37]). Thus, we write

$$E_{st} = \frac{\alpha_H}{2} \int_{-L_b/2}^{L_b/2} ds \left(\delta R(s)\right)^2. \qquad (2.5)$$

We will discuss what $\alpha_H$ should be chosen to be a little later. The second of these terms is the elastic bending energy contribution. If we assume the bending of the molecules can be described by the elastic rod model we obtain [36]

$$\frac{E_B}{k_B T} \approx \frac{E_{A,B}}{k_B T} + \frac{E_{R,B}}{k_B T}. \tag{2.6}$$

The first term in Eq. (2.6), $E_{A,B}$, is the contribution due to fluctuations of the braid axis. It takes the form

$$\frac{E_{A,B}}{k_B T} = \int_{-L_b/2}^{L_b/2} ds \left( l_p \left[ \left( \frac{dx'_A(s)}{ds} \right)^2 + \left( \frac{dy'_A(s)}{ds} \right)^2 \right] \right), \tag{2.7}$$

where $l_p = B/k_B T$ is the bending persistence length of each molecule (for DNA it is given by $l_p \approx 500 \text{Å}$), and $B$ is the elastic bending rigidity. The second term in Eq. (2.6) is the elastic contribution from forming a braid and from undulations of the two molecules relative to each other

$$\begin{aligned}\frac{E_{R,B}}{k_B T} = \int_{-L_b/2}^{L_b/2} ds \big[ &\mathcal{E}_R(R''(s), R'(s), R(s), \delta\eta'(s), \delta\eta(s)) \theta(\delta R(s) - d_{min}) \theta(d_{max} - \delta R(s)) \\ &+ \mathcal{E}_R(R''(s), R'(s), R_0 + d_{max}, \delta\eta'(s), \delta\eta(s)) \theta(\delta R(s) - d_{max}) \\ &+ \mathcal{E}_R(R''(s), R'(s), R_0 + d_{min}, \delta\eta'(s), \delta\eta(s)) \theta(d_{min} - \delta R(s)) \big], \end{aligned} \tag{2.8}$$

and

$$\begin{aligned}\mathcal{E}_R&(R''(s), R'(s), R(s), \delta\eta'(s), \delta\eta(s)) = \\ &\frac{l_p}{4}\left(\frac{d^2 R(s)}{ds^2}\right)^2 + \frac{l_p}{4}\left(\frac{d\delta\eta(s)}{ds}\right)^2 - \left(\frac{dR(s)}{ds}\right)^2 \frac{l_p}{R(s)^2}\sin^2\left(\frac{\eta_0}{2}\right) \\ &+ \frac{4 l_p}{R(s)^2}\left[\sin^4\left(\frac{\eta_0}{2}\right) + \frac{\delta\eta(s)^2}{2}\left(3\cos^2\left(\frac{\eta_0}{2}\right)\sin^2\left(\frac{\eta_0}{2}\right) - \sin^4\left(\frac{\eta_0}{2}\right)\right)\right] \\ &+ \left(\frac{dR(s)}{ds}\right)\left(\frac{d\delta\eta(s)}{ds}\right)\frac{3}{R(s)}\sin\left(\frac{\eta_0}{2}\right)\cos\left(\frac{\eta_0}{2}\right). \end{aligned} \tag{2.9}$$

In writing Eq. (2.8) we have imposed the cut-offs $d_{min}$ and $d_{max}$ on $\delta R(s)$. When $\delta R(s) > d_{max}$ we replace $\delta R(s)$ with $d_{max}$, and when $\delta R(s) < d_{min}$ we replace $\delta R(s)$ with $d_{min}$. The values $d_{min}$ and $d_{max}$ are the minimum and maximum values that the two molecular axes can separate in the braid, due to the molecules coming into steric contact. This procedure prevents unphysical values of the elastic energy when the true steric interaction is replaced by the pseudo potential described by Eq.(2.5). Indeed, these unphysical values arise from the molecules interpenetrating each other. In the absence of all other intermolecular interactions, it is reasonable to suppose that the root mean square of the $\delta R(s)$ fluctuations is proportional to the space available to fluctuate in, as originally argued in Ref. [37]- in our case $d_{min} - d_{max}$. We suppose, as in Ref. [38], that we may write (in the absence of non-steric interactions) $\langle \delta R(s)^2 \rangle \approx (d_{min} - d_{max})^2 / 4$. This reasoning leads us to a value of $\alpha_H$, namely

$$\alpha_H = \frac{2}{(d_{max} - d_{min})^{8/3} (l_p)^{1/3}}. \tag{2.10}$$

We will consider two particular choices for estimating $d_{min}$ and $d_{max}$. The first choice is $d_{max} = -d_{min} = R_0 - 2a$, which may be more appropriate for a loosely braided state [34]. This choice was implicit in the works of Refs. [21] and [30]. As a second choice we still have that $d_{min} = 2a - R_0$ but with

$$d_{max} \approx \left( \frac{\pi R_0}{\tan(\eta_0 / 2)} \right)^{3/2} \frac{1}{\left( l_p^b \sqrt{2} \right)^{1/2}}, \tag{2.11}$$

which was argued in Ref. [38] to be, perhaps, a better estimate for a tightly braided state.

Next, we consider the twisting part of the elastic energy, which in the elastic rod model, may be written as

$$\frac{E_{Tw}}{k_B T} \approx \frac{l_c}{2} \int_{-L_b/2}^{L_b/2} ds \left[ \left( g_1(s) - g_1^0(s) \right)^2 + \left( g_2(s) - g_2^0(s) \right)^2 \right], \tag{2.12}$$

where $g_1(s)$ and $g_2(s)$ are the actual rates, and $g_1^0(s)$ and $g_2^0(s)$ are the intrinsic rates (in the absence of any torsional stress) at which $\hat{\mathbf{v}}_1(s)$ and $\hat{\mathbf{v}}_2(s)$ precess about molecular centre line tangents $\hat{\mathbf{t}}_1(s)$ and $\hat{\mathbf{t}}_2(s)$, respectively. In general, they can be calculated through the expressions

$$g_\mu(s) = \left( \hat{\mathbf{t}}_\mu(s) \times \hat{\mathbf{v}}_\mu(s) \right) \cdot \frac{d\hat{\mathbf{v}}_\mu(s)}{ds}. \tag{2.13}$$

The parameter $l_c$ is the effective thermal persistence length for torsional fluctuations. This may also contain a contribution from stretching fluctuations [39], and therefore can be quite different from the mechanical torsional persistence length (that measured for a stretched molecule). Therefore, particular care must be taken with dealing with the mechanical response of the molecules with respect to a torque that provides twisting. In this case, how the molecular contour length changes with stretching fluctuations should be considered explicitly. As we deal with only nicked molecules, and so only twisting thermal fluctuations, Eq. (2.12) should suffice. For DNA, we estimate the value $l_c \approx 400\text{Å}$ [38]. We can write Eq. (2.12) approximately as

$$\frac{E_{Tw}}{k_B T} \approx \frac{l_c}{2} \int_{-L_b/2}^{L_b/2} ds \left[ \left( \frac{d\xi_1(s)}{ds_0} - \frac{\sin \eta(s)}{R} - g_1^0(s) \right)^2 + \left( \frac{d\xi_2(s)}{ds_0} - \frac{\sin \eta(s)}{R} - g_2^0(s) \right)^2 \right]. \tag{2.14}$$

For molecules which can be simply considered as uniform linear chains, simply $g_1^0(s) = g_2^0(s) = 0$. For molecules that form ideal, identical helices $g_1^0(s) = g_2^0(s) = g_0$. To describe randomly distorted helices we may write

$$g_\mu^0(s) = \bar{g}_0 + \Delta g_\mu^0(s), \tag{2.15}$$

where $\bar{g}_0$ is the average value of $g_\mu^0(s_0)$, which we suppose, here, to be the same for both molecules. We have that $\bar{g}_0 = 2\pi/H$, where the $H$ is the average helical pitch; for DNA it is $H \approx 33.8\text{Å}$. The function $\Delta g_\mu^0(s_0)$ is assumed to be a random field with a Gaussian distribution (for a justification see Ref. [40]), so that

$$\left\langle \Delta g_1^0(s) \Delta g_1^0(s') \right\rangle_{g_1} = \left\langle \Delta g_2^0(s) \Delta g_2^0(s') \right\rangle_{g_2} = \frac{1}{\lambda_c^{(0)}} \delta(s-s'), \tag{2.16}$$

$$\left\langle \Delta g_1^0(s) \right\rangle_{g_1} = \left\langle \Delta g_2^0(s) \right\rangle_{g_2} = 0. \tag{2.17}$$

Here the subscripts $g_1$ and $g_2$ refer to ensemble averaging over all the realizations of $\Delta g_1(s)$ and $\Delta g_2(s)$. Indeed, for DNA the helix is actually a distorted one due to the imperfect stacking of base pairs (see Refs. [5], [40]). For DNA $\lambda_c^{(0)}$ is estimated to be $\lambda_c^{(0)} \approx 150\text{Å}$ [40].

We can further rewrite Eq. (2.14) as

$$\frac{E_{Tw}}{k_B T} \approx \frac{l_c}{4} \int_{-L_b/2}^{L_b/2} ds \left[ \left( \frac{d\Delta\Phi(s)}{ds} - \Delta g^0(s) \right)^2 + \left( \frac{d\bar{\Phi}(s)}{ds_0} - \frac{2\sin\eta(s)}{R} - \bar{g}^0(s) \right)^2 \right], \tag{2.18}$$

where $\Delta\Phi(s) = \xi_1(s) - \xi_2(s)$, $\Delta g^0(s) = \Delta g_1(s) - \Delta g_2(s)$, $\bar{\Phi}(s) = \xi_1(s) + \xi_2(s)$ and $\bar{g}^0(s) = 2\bar{g} + \Delta g_1(s) + \Delta g_2(s)$.

Now we consider the work term $E_W$ that constrains both $N$ and $z_b$, the length that the braided section makes along the $z$-axis. This term can be written as (See Refs. [36])

$$\frac{\tilde{E}_W}{k_B T} = -\frac{F_R}{l_p} \int_{-L_b/2}^{L_b/2} ds \left[ \cos\left(\frac{\eta_0}{2}\right) - \frac{1}{8}\cos\left(\frac{\eta_0}{2}\right)\delta\eta(s)^2 - \frac{1}{2\cos\left(\frac{\eta_0}{2}\right)} \left[ \left(\frac{dx_A(s)}{ds}\right)^2 + \left(\frac{dy_A(s)}{ds}\right)^2 \right] \right]$$

$$+ M_R \int_{-L_b/2}^{L_b/2} ds \left[ \sin\left(\frac{\eta_0}{2}\right) - \frac{1}{8}\sin\left(\frac{\eta_0}{2}\right)\delta\eta(s)^2 - \frac{1}{8}\left(\frac{dR(s)}{ds}\right)^2 \sin\left(\frac{\eta_0}{2}\right)^{-1} \right] \left\{ \frac{2}{R(s)} \theta(\delta R(s) - d_{\min}) \right. \tag{2.19}$$

$$\left. \theta(d_{\max} - \delta R(s)) + \frac{2}{R_0 + \delta_{\min}} \theta(\delta R(s) - d_{\min}) + \frac{2}{R_0 + \delta_{\max}} \theta(d_{\max} - \delta R(s)) \right\} - 2\pi M_R Wr_b.$$

Again, in writing Eq. (2.19), we have imposed the cut-offs $d_{min}$ and $d_{max}$ on $\delta R(s)$ to prevent unphysical values. The first term constrains the average braid extension $\langle z_b \rangle$, for a given reduced pulling force $F_R$. The second term that multiplies $M_R$ is the braid twist, the number of times the two molecules rotate about the braid axis. Last of all $M_R$ couples to braid writhe which is defined as

$$Wr_b = \frac{1}{4\pi} \int_{-L_A/2}^{L_A/2} d\tau \int_{-L_A/2}^{L_A/2} d\tau' \frac{(\mathbf{r}_A(\tau) - \mathbf{r}_A(\tau')) \cdot \hat{\mathbf{t}}_A(\tau) \times \hat{\mathbf{t}}_A(\tau')}{|\mathbf{r}_A(\tau) - \mathbf{r}_A(\tau')|^3}, \qquad (2.20)$$

where $\tau$ is the unit arc length coordinate, along the braid axis that has length $L_A$. For the straight axis configuration, indeed, we have that $L_A = z_b$. The vector $\mathbf{r}_A(\tau)$ is the position of a point along line of the fluctuating braid axis, and $\hat{\mathbf{t}}_A(\tau)$ is the tangent vector associated with it, which can be written as

$$\hat{\mathbf{t}}_A(\tau) = \frac{dx_A(\tau)}{d\tau} \hat{\mathbf{i}} + \frac{dy_A(\tau)}{d\tau} \hat{\mathbf{j}} + \sqrt{1 - \left(\frac{dx_A(\tau)}{d\tau}\right)^2 - \left(\frac{dy_A(\tau)}{d\tau}\right)^2} \hat{\mathbf{k}}. \qquad (2.21)$$

In Eq. (2.19), $M_R$ constrains the average value $\langle N \rangle$, where $N$ should be defined as the number of rotations of one end of the braid relative to the other end. Expressions for both $\langle z_b \rangle$ and $\langle N \rangle$ can then be got through the relations

$$\langle z_B \rangle = -\frac{1}{k_B T} \frac{\partial \mathcal{F}_{Braid}}{\partial F_R} \quad \text{and} \quad \langle N \rangle = \frac{1}{2\pi k_B T} \frac{\partial \mathcal{F}_{Braid}}{\partial M_R}. \qquad (2.22)$$

In the thermodynamic limit where $L \to \infty$ we can neglect fluctuations in both $z_B$ and $N$. Therefore, we can write for $n$ (when $R_0 \ll b$)

$$n \approx N + \text{sgn}(n)/2 \approx \langle N \rangle + \text{sgn}(n)/2, \qquad (2.23)$$

and also write for $z$, the total end to end distance of the two molecules

$$z \approx z_b + \sqrt{x^2 - b^2} \approx \langle z_b \rangle + \sqrt{x^2 - b^2}. \qquad (2.24)$$

Note that positive values of both $N$ and $n$ correspond to left handed braids.

Now, we will consider the interaction energy as the sum of a repulsive term and an attractive term such that

$$E_{int} = E_{rep} + E_{attr}. \qquad (2.25)$$

In considering $E_{attr}$ we will consider two possible sets of cases. One case $E_{attr}$ is chosen to be non-helix dependent and in the other $E_{attr}$ is helix specific, i.e. depending on helix structure. In all cases we will consider the same repulsive term $E_{rep}$.

## 2.3 Repulsive interaction terms

For the repulsive interaction term we have

$$E_{rep} = \int_{-L_b/2}^{L_b/2} ds \Big[ \big( \mathcal{E}_{dir}(R(s)) + \mathcal{E}_{img}(R(s)) \big) \theta(\delta R(s) - d_{min}) \theta(d_{max} - \delta R(s))$$
$$+ \big( \mathcal{E}_{dir}(R_0 + d_{min}) + \mathcal{E}_{img}(R_0 + d_{min}) \big) \theta(d_{min} - \delta R(s)) \qquad (2.26)$$
$$+ \big( \mathcal{E}_{dir}(R_0 + d_{max}) + \mathcal{E}_{img}(R_0 + d_{max}) \big) \theta(\delta R(s) - d_{max}) \Big].$$

Here, $\mathcal{E}_{dir}(R(s))$ is a term due to electrostatic repulsion from interactions between the charges of one molecule and the other, and is given by

$$\frac{\mathcal{E}_{dir}(R)}{k_B T} = \frac{2 l_B (1-\theta)^2}{l_e^2} \frac{K_0(R \kappa_D)}{[a \kappa_D K_1(a \kappa_D)]^2}, \qquad (2.27)$$

where $a$ is the effective radius of the molecules; $l_e = e/\eta$, where $e$ is the unit charge and $\eta$ is the linear charge density; $\kappa_D$ is the inverse Debye screening length, and $l_B$ is Bjerrum length (we take $l_B \approx 7\text{Å}$). For DNA molecules we take $a \approx 11.2\text{Å}$ and $l_e \approx 1.7\text{Å}$. The parameter $\theta$ is the fraction of the molecular charge which is neutralized by either Manning condensed or bound counter-ions.

The term $\mathcal{E}_{img}(R(s))$ is the contribution due to one molecule interacting with its image charge reflection at the surface of the other molecule (see Ref.[41]). Here, we will now suppose that both molecules have a DNA like helical charge distribution and write down (according to Ref. [41])

$$\frac{\mathcal{E}_{img}(R)}{k_B T} = -\frac{2 l_B}{l_e^2} \sum_{n=-\infty}^{\infty} \sum_{j=-\infty}^{\infty} \xi_n^2 \frac{K_{n-j}(R \kappa_n) K_{n-j}(R \kappa_n)}{[a \kappa_n K_n(a \kappa_n)]^2} \frac{I'_j(a \kappa_n)}{K'_j(a \kappa_n)} \qquad (2.28)$$

where

$$\xi_n = \delta_{n,0} \theta - \cos(n \phi_s) + (1 - \delta_{n,0})(f_1 + f_2 (-1)^n) \theta \qquad (2.29)$$

and $\kappa_n = \sqrt{\kappa_D^2 + n^2 \bar{g}_0^2}$. Here, we have assumed that $\bar{g}_0 \lambda_c^{(0)} \gg 1$, so that we can set $\bar{\Phi}'(s) = 2 \sin \eta(s) / R + \bar{g}^0(s)$ and $\kappa_n$ can be considered independent of $s$ (for arguments why this should be see Ref [36] or the supplemental material of Ref. [38]). Here $\phi_s \approx 0.4\pi$ is half-width of the minor groove of the DNA. The parameters $f_1$ and $f_2$ are the fractions of condensed and bound-ions that are localized in the minor and major grooves respectively.

For molecules that may be simply described as uniform charged rods we simply keep the $n=0$ term, thus setting $\xi_n = \delta_{n,0}(\theta-1)$. For molecules described by a single helix and no localized counter-ions ($f_1 = f_2 = 0$) we set $\xi_n = \delta_{n,0}\theta - 1$. Other choices of $\xi_n$ may correspond to other types of charged helical molecules. Also, in the general framework, we are not restricted in our choice of $\mathcal{E}_{img}(R)$, any other short ranged repulsive potential could, in fact, be used. However, we will focus on electrostatics for a DNA like helical structure, and to make progress we choose Eq. (2.28). This choice seems to fit the experimental data of [3] quite well [35], although it could be argued that $\mathcal{E}_{dir}(R)$ is only important in this case, due to the relatively large inter-axial separations.

### *2.4 Considering a non-helix specific attractive interaction term*

Here, we suppose that attractive helix specific forces are not important and include a non-helix specific contribution. This supposes that any attractive helix specific forces are, for the most part, washed out by thermal and structural distortions of the molecular helices, as well as perhaps other factors. As the repulsive terms (Eq. (2.28)) are relatively insensitive to $f_1$ and $f_2$, to reduce the number of parameters, it makes sense to set the parameters $f_1 = f_2 = 0$. That is to say, we consider the condensed counter-ions as being uniformly smeared about the molecules.

For the non-helix specific attractive term, we write

$$E_{attr} = \int_{-L_b/2}^{L_b/2} ds \Big[ \mathcal{E}_{attr,NH}(R(s))\theta(\delta R(s) - d_{min})\theta(d_{max} - \delta R(s)) \qquad (2.30)$$
$$+ \mathcal{E}_{attr,NH}(R_0 + d_{min})\theta(d_{min} - \delta R(s)) + \mathcal{E}_{attr,NH}(R_0 + d_{max})\theta(\delta R(s) - d_{max}) \Big],$$

where we will suppose that

$$\frac{\mathcal{E}_{attr,NH}(R)}{k_B T} = -\frac{2 l_B f_N (1-\theta)^2}{l_e^2} \frac{K_0(R\kappa_{NH})}{\left[ a\kappa_0 K_1(a\kappa_0) \right]^2}. \qquad (2.31)$$

Here, $f_N$ is the effective strength of these interactions measured in proportion to $\mathcal{E}_{dir}(R)$. In writing Eq. (2.31) we have assumed, for cylindrical geometry, an exponentially decaying interaction with effective decay length $\kappa_{NH}^{-1}$.

Non-helix specific attraction may be caused by correlation forces. These forces come about from the fact that the densities of condensed ions about the molecules fluctuate. Fluctuations where the density of ions is less than the average value about one molecule may correlate, through screened electrostatic forces, with fluctuations where the ion density is more than average near the other molecule, and visa versa. If the average charge densities of condensed counter-ions near each molecule are assumed uniform, and therefore unaffected by the helical structure, these forces are helix non-specific. In the case where the average densities are affected helix structure there is still a helix non-specific contribution from their spatial average [42]. For such non-helix specific correlation forces, one may naively expect that $\kappa_{NH} = 2\kappa_D$. A simple argument for this is that the amount of the correlations between the fluctuations about each molecule, which determines the effective pattern of fluctuations in charge densities on each molecule, should decay with characteristic length $\kappa_D^{-1}$. This is because the strength of the correlations is governed by screened electrostatics. Then, on top of this, the interaction between the correlated charge fluctuations should decay off with distance as $\kappa_D^{-1}$. Therefore, combining these two factors together suggests that the combined decay length should indeed be $(2\kappa_D)^{-1}$. A more sophisticated analysis of correlation forces between charged cylinders [42] suggests a more complicated picture, when it comes to actually trying to calculate these forces from some microscopic model. However, the much simpler choice $\kappa_{NH} = 2\kappa_D$ may be sufficient in qualitatively describing such interactions. Therefore, we choose it as one possible choice for $\kappa_{NH}$.

As a second choice, we suppose that $\kappa_{NH}$ is independent of salt concentration and choose a value $(\kappa_{NH})^{-1} = 4.8\text{Å}$. This is a decay range suggested by an empirical law used to describe the osmotic pressure versus inter molecular separation in polyamine condensed DNA fibres [43], also used in Ref. [44]. A salt independent decay length suggests that screened electrostatic interactions do not have a role in determining the degree of attraction. We may think of these two choices as two extremes for the non-chiral attraction.

*2.5 Considering helix specific attractive interaction terms*

For the helix specific forces we write down

$$E_{attr} = \int_{-L_b/2}^{L_b/2} ds \Big[ \mathcal{E}_{attr,H}(R(s),\eta(s),\Delta\Phi(s))\theta(\delta R(s) - d_{\min})\theta(d_{\max} - \delta R(s))$$

$$+ \mathcal{E}_{attr,H}(R_0 + d_{\min},\eta(s),\Delta\Phi(s))\theta(d_{\min} - \delta R(s)) + \mathcal{E}_{attr,NH}(R_0 + d_{\min},),\eta(s),\Delta\Phi(s))\theta(\delta R(s) - d_{\max}) \Big].$$

(2.32)

We will consider now possible forms for $\mathcal{E}_{attr,H}(R(s),\eta(s),\Delta\Phi(s))$. Provided that $R(s)$, $\eta(s)$ and $\Delta\Phi(s)$ vary slowly enough, over length scales much larger than the characteristic decay length(s) of the interaction, one can use a form for $\mathcal{E}_{attr,H}$, calculated for $R$, $\eta$ and $\Delta\Phi$ being constant with respect to $s$. One may simply now replace constant values of $R$, $\eta$ and $\Delta\Phi$ with functions of $s$ in such a result. For a particular example how this approximation may come about, and what possible corrections to it may arise, see Ref. [45].

Now, it should be possible to write some density function associated with of a molecule described by an ideal helical structure (where the helix is considered as being continuous and where $\hat{\mathbf{v}}_\mu(s)$ may be considered as an axis of symmetry in the molecular cross-section)

$$\rho_\mu(r,\xi_\mu(s)) = \sum_{n=0}^{\infty} \rho_{\mu,n}(r)\cos\left(n\xi_\mu(s)\right) = \sum_{n=0}^{\infty} \rho_{\mu,n}(r)\cos\left(n(\phi_\mu - \overline{g}_0 s)\right), \quad (2.33)$$

where $r$ is the distance away from the molecular centre line in a plane perpendicular to the tangent vector $\hat{\mathbf{t}}_\mu(s)$. Here, $\phi_\mu$ is the azimuthal orientation of the helix $\mu$ at $s=0$, i.e. which way $\hat{\mathbf{v}}_\mu(0)$ points. Now let us consider an interaction that depends on $\rho_\mu(r,\xi_\mu(s))$, for each molecule. If the molecules are parallel, for continuous helices it should depend only on the phase difference $\Delta\Phi = \xi_1(s) - \xi_2(s)$, and not on the phase sum $\overline{\Phi}(s) = \xi_1(s) + \xi_2(s)$, through symmetry arguments (provided that the molecules are much longer than the characteristic decay length(s) of the interaction). This suggests that (for continuous helices) we may write

$$\mathcal{E}_{attr,H}(R,\eta=0,\Delta\Phi) = \sum_{n=1}^{\infty} a_{n,0}(R)\cos n\Delta\Phi. \quad (2.34)$$

Now in general, where $\eta \neq 0$ we should be able to approximate the interaction energy as

$$\mathcal{E}_{attr,H}(R,\eta,\Delta\Phi) \approx \sum_{n=1}^{\infty}\sum_{m=0}^{\infty} a_{n,m}(R)\sin^m\eta\cos n\Delta\Phi, \quad (2.35)$$

where we have supposed that there is no explicit dependence on the phase sum $\overline{\Phi}(s)$. This should be valid provided that angle $\eta$ is not too large, which we assume to be the case. In writing Eq. (2.35), the interaction energy has been expanded out as a power series in $\sin\eta$, as well as the Fourier series in $\Delta\Phi$. Any additional contribution from harmonics arising from the discreteness of molecular groups making up the helix has been neglected. This is because for DNA the average distance between base pairs is $\langle h \rangle \approx 3.4\text{Å}$ whereas the average helical pitch is $H \approx 33.8\text{Å}$. This difference in magnitude between these two scales suggests that any dependence of the interaction on the base pair spacing should decay more rapidly with $R$ than the continuous helix contribution to Eq. (2.35) [5]. Also, the contribution from discreteness to Eq. (2.35) is supressed much more by thermal fluctuations [5].

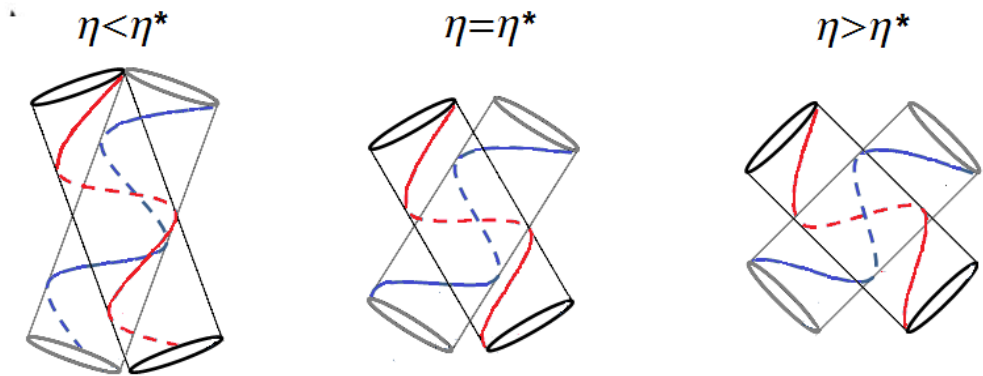

Fig.2. Schematic picture of one of the crossings of two helical molecules forming the braid. The black cylinder with a red helix is the molecule in front, while the grey cylinder with blue helix is the molecule behind. The figure illustrates that there is an optimal angle $\eta^*$ where the part of the red helix, at the back of the front molecule (dashed line), fits best between the part blue helix on the front of the molecule behind (solid line).

We will now approximate Eq. (2.35) by firstly truncating the Fourier sum. We give three arguments for truncating the Fourier sum in Eq. (2.35). Each value of $n$ arises from a density function contribution $\rho_{\mu,n}(r)\cos\left(n(\phi_\mu - 2\pi s/H)\right)$ for each molecule (see Eq.(2.33)). Along the direction of the molecular centre line, each contribution has a period $H/n$, and so oscillates more rapidly with increasing $n$. This suggests that the effective decay length of each harmonic should decrease as $n$ increased. This is indeed the case when electrostatics is considered. Therefore, one should be able to truncate the sum if $R$ is sufficiently large. The second argument is that there may be some smearing of the helices due to the finite size of molecular groups and thermal fluctuations. Then Debye-Waller factors that go as $\exp(-n^2\alpha)$ effectively truncate the sum, where $\alpha$ is a factor determined by these effects. Finally, large $n$ contributions to Eq.(2.35) are supressed more by distortions away from the ideal helix structure (see subsection 2.8 below). We truncate the Fourier series at $n=2$ (this valid for the results arising from the mean-field KL theory [39] presented below), although more terms in the Fourier series can be retained.

We now truncate the power series in $\sin\eta$. In one approximation we retain only the zeroth and first order terms in $\sin\eta$. This yields the following expression

$$\mathcal{E}_{attr,H}(R(s),\eta(s),\Delta\Phi(s)) \approx \left(a_{1,1}(R)+a_{1,2}(R)\sin\eta\right)\cos\Delta\Phi + \left(a_{2,1}(R)+a_{2,2}(R)\sin\eta\right)\cos 2\Delta\Phi. \quad (2.36)$$

However, a problem with this approximation is that it may miss some important physics, if helix specific forces are indeed important. A consideration of the geometry of the molecular helices, in the braid, suggests that there should be, in the absence of a bending rigidity, an optimum value for $\sin\eta$ at $\eta = \eta^*$. Illustrated in Fig.2, this optimum angle is where helix of one molecule fits best in between the helix the other molecule, on the sides of the molecules facing each other. When $\sin\eta > \sin\eta^*$ or $\sin\eta < \sin\eta^*$, there may be unfavourable crossings between helices (see Fig. 2), which make the interaction less attractive. Geometrical arguments suggest that this optimum angle, for DNA, should be taken to be roughly $\sin\eta^* \approx 2H/(2\pi a) \approx 0.96$ [46]. Unfortunately, retaining terms up to first order in $\sin\eta$ does not capture this effect, as one simply obtains the torque that wants to drive $\eta$ away from $\eta = 0$. Thus, neglecting this additional effect due to helix geometry overestimates the effect of helix specific attractive forces. To try an capture this limitation on the size of $\sin\eta$, one can indeed try truncate Eq. (2.35) to second order in $\sin\eta$. However, in practice, this may be a laborious process, unless Eq. (2.35) is fitted empirically to simulation data. Instead, we may take a second approach. We can estimate this effect, simply, from these geometrical considerations. We write an interaction energy density of the form:

$$\mathcal{E}_{attr,H}(R,\eta,\Delta\Phi) \approx \left(a_{1,1}(R)+a_{1,2}(R)\left(\sin\eta - \frac{\sin^2\eta}{2\sin\eta^*}\right)\right)\cos\Delta\Phi$$
$$+ \left(a_{2,1}(R)+a_{2,2}(R)\left(\sin\eta - \frac{\sin^2\eta}{2\sin\eta^*}\right)\right)\cos 2\Delta\Phi. \quad (2.37)$$

The expression in Eq. (2.37) is constructed so that in the absence of bending terms it is minimized at $\sin\eta = \sin\eta^*$. Also, when we neglect the additional $\sin^2\eta$ terms we simply recover back the approximation given by Eq. (2.36).

The forms of Eqs. (2.36) and (2.37) are still quite general, and may still apply to any possible interaction potential depends on the helix structure of the molecules. However, to go further and generate results, we need to specify a specific model for the helix-specific interactions to obtain expressions for the coefficients $a_{n,m}(R)$. We choose the mean-field KL theory [41] for which we have [6],[45]

$$a_{n,0}(R) = \frac{4l_B}{l_e^2} \frac{\xi_n^2 (-1)^n K_0(\kappa_n R)}{(\kappa_n a K_n'(\kappa_n a))^2}, \quad (2.38)$$

$$a_{n,1}(R) = \frac{4l_B n^2 a \bar{g}_0}{l_e^2} \frac{(-1)^n \xi_n^2 K_1(\kappa_n R)}{(K_n'(\kappa_n a))^2 (\kappa_n a)^3}. \quad (2.39)$$

Regardless of the choice of interaction model, we expect the qualitative features in the results presented below to remain the same, if sufficiently strong helix specific forces are indeed present. If correlation forces are important, the calculations of Refs. [42] and [47] suggest that there might be a helix specific contribution to these forces, but perhaps somehow this could be washed out in real systems. It is also suggested that hydration forces would also have helix dependence [41]. In the case of the mean-field electrostatics utilized by the KL theory, to get an idea of how laborious calculating exactly the coefficients of the $\sin^2 \eta$ terms is, we refer the reader to Ref. [45]. In Ref. [45] a general frame work for calculating the Debye screened electrostatic energy for generalized braid configurations formed of two charged helices is presented [48].

Here, to limit the parameter space, we will suppose that $f_1 = 0.4 f_H$ and $f_2 = 0.6 f_H$. That is to say we fix the ratio of ions localized in the major to the minor groove at a particular ratio. The parameter $f_H$ is a measure of how localized the counter-ions are to the grooves. This means that we choose $\xi_n = \delta_{n,0} \theta - \cos(n\phi_s) + (1-\delta_{n,0})(0.4 f_H + f_H(-0.6)^n) \theta$. As we shall see, $f_H$ and $\theta$ are control parameters for the size of the contribution from helix dependent forces.

### 2.6 The general form for the braiding free energy

In Appendix A we will summarize the main steps used in obtaining the free energy, $\mathcal{F}_{Braid}$, most notably the variational approximation that is used. Full step by step details of such a calculation of can be found in Ref. [36]. The braid free energy is then to be minimized over $\langle (R(s) - R_0)^2 \rangle \approx d_R^2$ as well as $\theta_R^2 = \langle R'(s)^2 \rangle$, which control the size of undulations of the molecules relative to each other within the braid. In the main text, we present the final result, by first writing the braid free energy as

$$\mathcal{F}_{Braid} = L_b k_B T f_{braid} = L_b k_B T (f_{conf} + f_{bend} + f_{x,y} + f_W + f_{rep} + f_{attr}). \quad (2.40)$$

The terms $f_{conf}, f_{bend}, f_{x,y}$ and $f_W$ have already been discussed in Ref. [35], for the case $d_{max} = -d_{min} = R_0 - 2a$. The first of these is

$$f_{conf} = \frac{d_R^2}{(d_{max} - d_{min})^{8/3} (l_p)^{1/3}} + \frac{1}{4l_p \theta_R^2} + \frac{l_p \theta_R^4}{4 d_R^2} + \frac{\alpha_\eta^{1/2}}{2^{1/2} l_p^{1/2}}. \quad (2.41)$$

The term $f_{conf}$ is the entropic free energy cost required to confine the molecules such that $\langle (R(s) - R_0)^2 \rangle \approx d_R^2$, which includes the contribution from steric interactions (Eq.(2.5)) that depends on $d_{max}$ and $d_{min}$. Eq. (2.41) favours $d_R = (d_{max} - d_{min})/2$, as this maximizes the entropy. This can be seen by minimizing Eq. (2.41) with respect to both $d_R$ and $\theta_R$. However, the thermally averaged interaction energy causes $d_R$ to be smaller. This is because average interaction terms in the free energy, also depend on $d_R$. Therefore, increasing $d_R$ causes the average interaction energy to become more repulsive. The parameter $d_R$ is therefore determined self-consistently by minimization of the total free energy, which includes all these contributions. The term $\alpha_\eta$ is the effective spring constant that determines the size of fluctuations in $\eta$. As it depends on the form of the interaction, we will present expressions for it within the next two subsections.

The next contribution is

$$f_{bend} = \frac{\tilde{f}_1(R_0, d_R, R_0 - 2a, -(R_0 - 2a)) l_p}{R_0^2} \left( 4 \sin^4 \left( \frac{\eta_0}{2} \right) - \theta_R^2 \sin^2 \left( \frac{\eta_0}{2} \right) \right), \quad (2.42)$$

where

$$\tilde{f}_1(R_0, d_R, d_{max}, d_{min}) = \frac{R_0^2}{d_R \sqrt{2\pi}} \int_{d_{min}}^{d_{max}} \frac{dx}{(R_0 + x)^2} \exp\left( -\frac{x^2}{2 d_R^2} \right)$$
$$+ \frac{1}{2} \left( \frac{R_0^2}{(R_0 + d_{min})^2} \left( 1 - \mathrm{erf}\left( -\frac{d_{min}}{d_R \sqrt{2}} \right) \right) + \frac{R_0^2}{(R_0 + d_{max})^2} \left( 1 - \mathrm{erf}\left( \frac{d_{max}}{d_R \sqrt{2}} \right) \right) \right). \quad (2.43)$$

The term $f_{bend}$ is the contribution to the free energy from the thermally averaged bending energy. Here, the second term in Eq. (2.42) comes from the term that depends on $R'(s)^2$ in Eq. (2.9), the expression for the bending energy. The contribution $f_{x,y}$ is given by

$$f_{x,y} = \frac{1}{l_p} \left( \frac{F_R}{2 \cos(\eta_0 / 2)} \right)^{1/2}. \quad (2.44)$$

This contribution comes from undulations of the braid centre line, the size of these fluctuations is limited by the elastic energy functional contribution Eq. (2.7) as well as a contribution dependent on $F$ coming from Eq. (2.19), which constrains $\langle z_b \rangle$. It is essentially the free energy due to the entropy loss caused by straightening out the braid centre line away from a coiled configuration, through applying the pulling force $F$. Also, we have a term that includes other contributions from Eq. (2.19), namely

$$f_W = -\frac{F_R}{l_p}\cos\left(\frac{\eta_0}{2}\right) + \frac{M_R \tilde{f}_2(R_0, d_R, R_0 - 2a, -(R_0 - 2a))}{R_0}\left(2\sin\left(\frac{\eta_0}{2}\right) - \frac{\theta_R^2}{4}\sin\left(\frac{\eta_0}{2}\right)\right)^{-1}$$
$$-\frac{M_R^2}{16 l_p}\left(\frac{1}{2 l_p F_R}\right)^{1/2}\frac{1}{\cos\left(\frac{\eta_0}{2}\right)^{7/2}}.$$

(2.45)

The first of these terms is simply the contribution from the work done elongating the braid in the absence of thermal fluctuations. The second term is the contribution from the work done in creating a number of braid turns through moment $M_R$. The last term is due to the coupling of $M_R$ to the braid writhe (described by Eq. (2.20)). When the average braid axis is straight one would expect $\langle Wr_b \rangle = 0$. However, strictly speaking this is only the case in the limit $M_R \to 0$. There are corrections due to fluctuations in $Wr_b$, which correct the free energy and also lead to a small non-zero value of $\langle Wr_b \rangle$ that depends on $M_R$. This correction in the free energy is taken account of (to leading order) in the last term of Eq. (2.45). As was noted in Refs. [34,35], this last term is similar to what was originally derived in Ref. [31], now taking account of an average braid structure so that $\tau \neq s$. This is to say that the arc length coordinate of the braid axis is different from that of the molecular centre lines, as for the molecules to precess around the braid axis they need to travel further.

The contribution from the repulsive interactions to the free energy is given by

$$f_{rep} = \frac{2 l_B (1 - \theta_c)^2}{l_e^2 (a\kappa_D)^2 K_1(a\kappa_D)^2} g_0(\kappa_D R_0, \kappa_D d_R, d_{max}/d_R, d_{min}/d_R)$$
$$-\frac{2 l_B}{l_e^2}\sum_{n=-\infty}^{\infty}\frac{\xi_n^2}{(\kappa_n a K_n'(\kappa_n a))^2} g_{img}(n, \kappa_n R_0, \kappa_n d_R, d_{max}/d_R, d_{min}/d_R; a),$$

(2.46)

where

$$g_j(\kappa R_0, \kappa d_R, d_{max}/d_R, d_{min}/d_R) = \frac{1}{\sqrt{2\pi}}\int_{d_{min}/d_R}^{d_{max}/d_R} dy K_j(\kappa R_0 + y\kappa d_R)\exp\left(-\frac{y^2}{2}\right)$$
$$+\frac{1}{2}K_j(\kappa(R_0 + d_{min}))\left[1 - \mathrm{erf}\left(-\frac{1}{\sqrt{2}}\frac{d_{min}}{d_R}\right)\right] + \frac{1}{2}K_j(\kappa(R_0 + d_{max}))\left[1 - \mathrm{erf}\left(\frac{1}{\sqrt{2}}\frac{d_{max}}{d_R}\right)\right],$$

(2.47)

and

$$g_{img}(n,\kappa R_0,\kappa d_R,d_{max}/d_R,d_{min}/d_R;a) = \frac{1}{\sqrt{2\pi}}\sum_{j=-\infty}^{\infty}\int_{d_{min}/d_R}^{d_{max}/d_R}dy K_{n-j}(\kappa R_0+y\kappa d_R)K_{n-j}(\kappa R_0+y\kappa d_R)$$

$$\frac{I'_j(\kappa a)}{K'_j(\kappa a)}\exp\left(-\frac{y^2}{2}\right)+\frac{1}{2}\sum_{j=-\infty}^{\infty}K_{n-j}(\kappa(R_0+d_{min}))K_{n-j}(\kappa(R_0+d_{min}))\frac{I'_j(\kappa a)}{K'_j(\kappa a)}\left[1-\text{erf}\left(-\frac{1}{\sqrt{2}}\frac{d_{min}}{d_R}\right)\right]$$

$$+\frac{1}{2}\sum_{j=-\infty}^{\infty}K_{n-j}(\kappa(R_0+d_{max}))K_{n-j}(\kappa(R_0+d_{max}))\frac{I'_j(\kappa a)}{K'_j(\kappa a)}\left[1-\text{erf}\left(\frac{1}{\sqrt{2}}\frac{d_{max}}{d_R}\right)\right].$$

(2.48)

Now, the forms of the contributions $f_{attr}$ are quite new. Again, because of their dependence on the type of interaction, and how they affect the results that we have, we will discuss them in the next two subsections.

### 2.7 The contribution from non-helix specific attractive interaction terms

For the non-helix specific attractive interaction term given by Eqs. (2.30) and (2.31), we may write the free energy contribution. This is simply this term averaged over the undulations; it reads as

$$f_{attr} = -\frac{2l_B f_N(1-\theta_c)^2}{l_e^2(a\kappa_D)^2 K_1(a\kappa_D)^2}g_0(\kappa_{NH}R_0,\kappa_{NH}d_R,(R_0-2a)/d_R,-(R_0-2a)/d_R).\quad(2.49)$$

We also ahve the effective spring constant, $\alpha_\eta$ for fluctuations in $\eta$, given here by

$$\alpha_\eta = \frac{4l_p\tilde{f}_1(R_0,d_R,d_{max},d_{min})}{R_0^2}\left(3\cos^2\left(\frac{\eta_0}{2}\right)\sin^2\left(\frac{\eta_0}{2}\right)-\sin^4\left(\frac{\eta_0}{2}\right)\right)+\frac{F_R}{4}\cos\left(\frac{\eta_0}{2}\right)$$

$$-\frac{M_R\tilde{f}_2(R_0,d_R,d_{max},d_{min})}{2R_0}\sin\left(\frac{\eta_0}{2}\right).$$

(2.50)

The form for $\alpha_\eta$ given by Eq. (2.50) is the same as for in Ref. [35], as the non-helix specific interactions do not depend on $\eta(s)$. In this case we minimize over $R_0$ and $\eta_0$ in addition to $d_R$ and $\theta_R$.

If $f_N$ is not large enough, as we shall see in the results section, we have only one braiding state with a continuous change in the braid radius and tilt angle, as we change the value of the moment $M$. If we increase the amount of attraction by increasing $f_N$, the braided configuration separates into two states. There exists a loose state and a tighter collapsed state, with a transition between the two at a critical value of $M=M_c$. This does not need condensation to occur; we do not require a minimum in the interaction potential to become negative. For the collapse to happen all we require is flattening of the potential, or the existence of a local minimum, at some point in free energy surface. To illustrate how we have a transition between the two states, in Fig 3 we present a schematic picture of what happens, as one increases the value of $M$, to $\mathcal{F}_{Braid}$ as a function of $R_0$. Here we have supposed a flattening of $\mathcal{F}_{Braid}$ at $M=0$ due to an attractive component. As we increase $M$ (for positive values), this develops into a local minimum in $\mathcal{F}_{Braid}$. At the value $M_c$, we have two minima for the two states with the same energy. Then above this the minima for the tightly braided state, with the smaller value of $R_0$, is favoured.

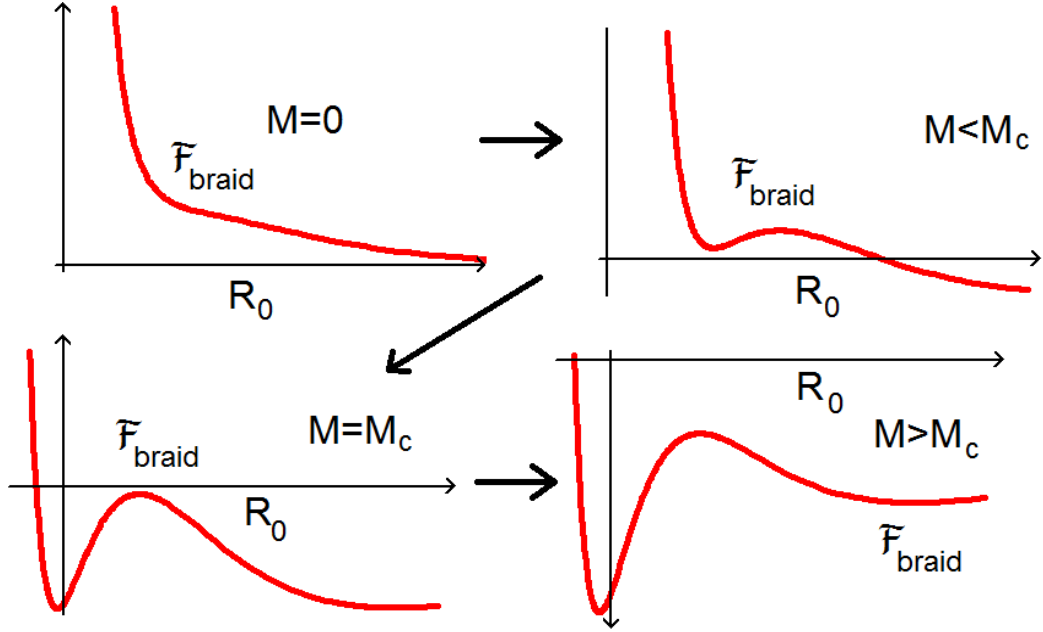

Fig.3 Schematic picture of what happens to the free energy of the braid $\mathcal{F}_{Braid}$, as a function of $R_0$, when the applied moment $M$ is changed, in a case where braid collapse occurs. The black arrows indicate the direction of increasing $M$ (here $M$ is considered positive). As we increase $M$, we see that $\mathcal{F}_{Braid}$ develops into a local minimum. At the critical moment $M = M_c$ we two minima become equal. When $M > M_c$, the free energy minimum with the smaller value of $R_0$ is favoured.

At $M_c$, the two states that minimize the free energy have different sets of values for both $\eta$ and $R$, minimize the free energy. We can write these two sets of values as $\{\eta_c^T, R_c^T\}$ and $\{\eta_c^L, R_c^L\}$ for the tight and loose braiding states, respectively. At this critical value, the purely tightly braided state has a value of $N$, which we define as $N_c^T$, whereas the loosely braided state has a different value $N_c^L$. This is due to the difference between $\{\eta_c^T, R_c^T\}$ and $\{\eta_c^L, R_c^L\}$. Note that the transition is symmetric with respect to $N \rightarrow -N$ so that there is another transition at $-M_c$, below which one is in the tightly braided state. This symmetry is due to the fact that the free energy is invariant under $\eta_0 \rightarrow -\eta_0$ and $M \rightarrow -M$. Thus, there has to be two coexistence regions for which $|N_c^T| < |N| < |N_c^L|$, where $M$ stays constant at the critical values $\pm M_c$. In the coexistence region, both states contribute to the braid. Here, the contour lengths $L^T(N)$ for the tightly braided state and $L^L(N)$ for the loosely braided state are given by

$$L^T(N) = \frac{(|N_c^L| - |N|) L_b(N_c^T)}{(|N_c^L| - |N_c^T|)}, \qquad L^L(N) = \frac{(|N| - |N_c^T|) L_b(N_c^L)}{(|N_c^L| - |N_c^T|)}, \qquad (2.51)$$

and also the braid extension $z_b$ is given by

$$z_T(N) \approx -L^T(N)\left(\frac{df^T_{Braid}}{dF}\right)\bigg|_{M_R=M_c} - L^L(N)\left(\frac{df^L_{Braid}}{dF}\right)\bigg|_{M_R=M_c} + \sqrt{x^2-b^2}, \qquad (2.52)$$

where $f^T_{Braid}$ and $f^L_{Braid}$ are $f_{Braid}$ taken at the sets of values $\{\eta^T_c, R^T_c\}$ and $\{\eta^L_c, R^L_c\}$, minimizing the free energy at $M_c$, respectively.

For the loosely braided state we have assumed that $d_{max} = -d_{min} = R_0 - 2a$, for the tightly braided state close to the transition we have assumed this again. However, for the tightly braided state away from the transition ($|M| > |M_c|$) we have also made the choice $-d_{min} = R_0 - 2a$ with $d_{max}$ given by Eq. (2.11) (for $\theta = 0.8$).

## 2.8 The contribution from helix specific interaction terms

For helix specific interactions $f_{attr}$ is rather more complicated, as the $\Delta\Phi$ degrees of freedom matter in this case due to the form of the interaction (Eq. (2.36) or Eq. (2.37)). From Eqs. (2.32), (2.36), and (2.37) we have that for $f_{attr}$ (see Appendix B for an outline of the calculation)

$$f_{attr} = \frac{(l_c + \lambda_c)^2}{16\lambda^*_h \lambda_c l_c} + \sum_{n=1}^{2} \cos(n\Delta\bar{\Phi})\exp\left(-\frac{n^2\lambda^*_h}{2\lambda_c}\right)\frac{4l_B\xi_n^2(-1)^n}{l_e^2(a\kappa_n)^2 K'_n(a\kappa_n)^2}$$
$$\left(g_0\left(\kappa_n R_0, \kappa_n d_R, \frac{d_{max}}{d_R}, \frac{d_{min}}{d_R}\right) + \frac{n^2\bar{g}_0}{\kappa_n} g_1\left(\kappa_n R_0, \kappa_n d_R, \frac{d_{max}}{d_R}, \frac{d_{min}}{d_R}\right) U(\sin\eta_0)\right), \qquad (2.53)$$

and the effective spring constant for fluctuations in $\eta$ is given by

$$\alpha_\eta = \frac{4l_p \tilde{f}_1(R_0, d_R, d_{max}, d_{min})}{R_0^2}\left(3\cos^2\left(\frac{\eta_0}{2}\right)\sin^2\left(\frac{\eta_0}{2}\right) - \sin^4\left(\frac{\eta_0}{2}\right)\right) + \frac{F_R}{4}\cos\left(\frac{\eta_0}{2}\right)$$
$$-\sum_{n=1}^{2}\frac{l_B\xi_n^2(-1)^n}{l_e^2}\frac{\cos(n\Delta\bar{\Phi})}{(a\kappa_n)^2 K'_n(a\kappa_n)^2}\frac{4n^2\bar{g}_0}{\kappa_n} g_1(\kappa_n R_0, \kappa_n d_R, d_{max}/d_R, d_{min}/d_R)\exp\left(-\frac{n^2\lambda^*_h}{2\lambda_c}\right)\frac{d^2 U(\sin\eta_0)}{d\eta_0^2}$$
$$-\frac{M_R \tilde{f}_2(R_0, d_R, d_{max}, d_{min})}{2R_0}\sin\left(\frac{\eta_0}{2}\right),$$

$$(2.54)$$

where we have the combined persistence length $1/\lambda_c = 1/\lambda_c^{(0)} + 1/l_c$. In the case where we have just have only the $\sin\eta$ terms in the interaction energy (see Eq. (2.36)) we simply have

$$U(\sin\eta_0) = \sin\eta_0. \qquad (2.55)$$

When we estimate the $\sin^2 \eta$ term (see Eq. (2.37) ), $U(\sin \eta_0)$ gets modified to

$$U(\sin \eta_0) = \sin \eta_0 - \frac{\sin^2 \eta_0}{2 \sin \eta^*}. \tag{2.56}$$

In Eqs. (2.53) and (2.54), $\Delta \bar{\Phi}$ is the average value of $\Delta \Phi(s)$ over both thermal fluctuations and the intrinsic helix distortions, $\Delta g_\mu^0(s)$. In Eqs. (2.53) and (2.54), we also have the helical adaptation length $\lambda_h^*$. This length has physical meaning, when we consider the correlation function $\mathcal{G}(s-s') = \langle (\Delta \Phi(s) - \Delta \Phi(s'))^2 \rangle$, describing the accumulation of mismatch in the phase difference. Over length scales smaller than the helical adaptation length ( $\lambda_h^*$ ) $\mathcal{G}(s-s')$ increases in a linear fashion in $s-s'$ [39], as would be expected if $\Delta \Phi'(s) = \Delta g^0(s)$ (the prime refers to derivative with respect to argument), i.e. there is no contribution from Eqs. (2.53) and (2.54). However, $f_{attr}$ (Eq. (2.53)) enables the molecules adapt themselves torsionally, so that the phases $\xi_1(s)$ and $\xi_2(s)$ are strongly correlated with each other. This occurs over large length scales, so to minimize the cost of entropy and twisting elastic energy (the first term in Eq.(2.53)), but also to lower the average interaction energy (rest of equation (2.53)). Therefore, at length scales larger than $\lambda_h^*$, this adaptation causes $\mathcal{G}(s-s')$ to saturate to a constant value $\lambda_h^* / \lambda_c$ [39]. This constant value indicates an imperfect alignment of the two helices, the size of which is a measure of how imperfect the alignment is. This imperfect alignment of helices manifests itself through factors $\exp(-n^2 \lambda_h^* / (2\lambda_c))$ that weaken the average helix specific interaction terms in $f_{attr}$. The free energy is now minimized with respect to $\Delta \bar{\Phi}$ and $\lambda_h^*$ in addition to $d_R$, $\theta_R$, $R_0$ and $\eta_0$. The free energy contribution, Eq. (2.53), allows for multiple solutions that correspond to different types of braiding states.

One solution that always minimizes Eq. (2.53) is $\lambda_h^* = \infty$. Here, the contribution from $\mathcal{E}_{attr,H}(R(s), \eta(s), \Delta \Phi(s))$ is almost washed out by thermal and intrinsic helix distortions. In this case the correlation function $\mathcal{G}(s-s')$ keeps accumulating as $|s-s'|$ over the whole length of the molecules. However, as was discussed in Ref. [35], there may still be weak correlations between the phases $\xi_1(s)$ and $\xi_2(s)$ on the two molecules. However, a preferred optimum difference between the two phases $\Delta \bar{\Phi}$ cannot be maintained along the braid due to $\lambda_h^* = \infty$. Nevertheless, in small fluctuating regions along the braid $\xi_1(s)$ and $\xi_2(s)$ are correlated to reduce the average interaction energy for helix specific forces, creating a small amount of attraction. This may lead to a weak contribution from $\mathcal{E}_{attr,H}(R(s), \eta(s), \Delta \Phi(s))$ for this state. Currently however, in the interests of simplicity, we do not try and incorporate this second order effect, which is indeed likely to be very small when $R_0 - 2a$ is of a few or more Debye screening lengths. This is because the effective screening length for such correlation forces is halved, very much in analogy to the already discussed correlation forces caused by ions. However, as discussed in Ref. [35], it might indeed be responsible for some slight asymmetry seen in some of the experimental data of Refs. [1] and [3], when the molecules are forced close enough together. Indeed, its inclusion to what was considered in Ref. [35] may fit the experimental data even better. For the $\lambda_h^* = \infty$ state we assume that $d_{\max} = -d_{\min} = R_0 - 2a$.

If $\theta$ and $f_H$ are made large enough, there are two solutions that are allowed when $\lambda_h^*$ is finite. One is where $\Delta\bar{\Phi} = 0$ and another is where $\Delta\bar{\Phi} \approx \pi/2$. Both of these solutions form tighter braids than those formed for the $\lambda_h^* = \infty$ state. The tightest braided configuration (the one with the smallest $R_0$) is the $\Delta\bar{\Phi} \approx \pi/2$ state. The former state can now be favoured (has the lowest free energy) for both left and right handed braids. This is unlike in Ref. [33], where for the parameters examined, it only seemed to be preferred for only right handed braids. The reason for this difference is the addition of $f_{conf}$ in the free energy, which takes account of an entropy cost needed to confine the molecules within a braided configuration. The term $f_{conf}$ always favours the most the loosest braided configuration, which happens to be $\lambda_h^* = \infty$; therefore, this state is favoured more across the parameter space with this term's inclusion. This term favours also the $\Delta\bar{\Phi} = 0$ over the $\Delta\bar{\Phi} \approx \pi/2$ state, again due to the looser braided structure of the former state. Therefore, there can be cases where $\Delta\bar{\Phi} = 0$ is energetically favoured for $\eta_0 > 0$, and the region where the $\Delta\bar{\Phi} \approx \pi/2$ state occurs is markedly reduced (see Results section) from what was seen in Ref [33]. As always, the $\Delta\bar{\Phi} \approx \pi/2$ state is only favoured for left handed braids. For the $\Delta\bar{\Phi} = 0$ state we assume that $d_{\max} = -d_{\min} = R_0 - 2a$, but for the tightly braided $\Delta\bar{\Phi} \approx \pi/2$ state $-d_{\min} = R_0 - 2a$ with $d_{\max}$ given by Eq. (2.11), as $R_0 - 2a$ is rather small.

As we move from large negative values of $N$ to large positive ones, Eq. (2.53) allows for a few different transitions and coexistence regions, provided that both $\theta$ and $f_H$ are sufficiently large. Though, no longer do we have the transition between the $\lambda_h^* = \infty$ state for $\eta < 0$ and the $\Delta\bar{\Phi} \approx \pi/2$ state for $\eta > 0$, as there is insufficient attraction to outweigh the confinement entropy cost for the latter state. As we start to increase $N$, the first possible transition that may occur (for $N < 0$) is the transition between the $\Delta\bar{\Phi} = 0$, $\lambda_h^*$ finite state and the $\lambda_h^* = \infty$ state. This transition occurs at some critical moment $M_c^{(1)}$. Then, when we move to positive $N$, at a certain point, we can either get a transition back to the state where $\Delta\bar{\Phi} = 0$ or to the $\Delta\bar{\Phi} \approx \pi/2$ state. This depends on the choice of the parameter values $\theta$ and $f_H$. For the former transition, we may define a critical moment $M_c^{(2)}$ and for the latter a different critical moment $M_c^{(3)}$. Finally, as we further increase $N$, if the $\Delta\bar{\Phi} = 0$ state has been favoured for $\eta_0 > 0$, we get a transition, at $M_c^{(4)}$, between it and the $\Delta\bar{\Phi} \approx \pi/2$ state. At all of these critical moments $M_c^{(1)}$, $M_c^{(2)}$, $M_c^{(3)}$ and $M_c^{(4)}$, we have coexistence regions, between the two states on either side of the particular transition in question. This is because, always, the two states on either side of the transition give different values of $R_0$ and $\eta_0$ that minimize the free energy, at the critical moments $M_c^{(j)}$.

To get a clearer picture of these coexistence regions, let's look at the one at $M_c^{(4)}$. At this value, the set of values $\{\eta_c^{\Delta\Phi=0}, R_c^{\Delta\Phi=0}\}$ minimize the free energy for the $\Delta\Phi = 0$ state, and $\{\eta_c^{\Delta\Phi\approx\pi/2}, R_c^{\Delta\Phi\approx\pi/2}\}$ for the $\Delta\bar{\Phi} \approx \pi/2$ state. These sets of geometric parameters yield the critical numbers of braid turns

$N = N_c^{\Delta\bar{\Phi}=0}$ and $N = N_c^{\Delta\bar{\Phi}\approx\pi/2}$, for the two pure braiding states, respectively. Therefore, our coexistence region between the two states is when $N_c^{\Delta\bar{\Phi}=0} < N < N_c^{\Delta\bar{\Phi}\approx\pi/2}$. Within this region, similar to Eq. (2.51), we have that

$$L^{\Delta\bar{\Phi}\approx\pi/2}(N) = \frac{\left(\left|N_c^{\Delta\bar{\Phi}=0}\right| - |N|\right) L_b(N_c^{\Delta\bar{\Phi}\approx\pi/2})}{\left(\left|N_c^{\Delta\bar{\Phi}=0}\right| - \left|N_c^{\Delta\bar{\Phi}\approx\pi/2}\right|\right)}, \qquad L^{\Delta\bar{\Phi}=0}(N) = \frac{\left(|N| - \left|N_c^{\Delta\bar{\Phi}\approx\pi/2}\right|\right) L_b(N_c^{\Delta\bar{\Phi}=0})}{\left(\left|N_c^{\Delta\bar{\Phi}=0}\right| - \left|N_c^{\Delta\bar{\Phi}\approx\pi/2}\right|\right)}, \qquad (2.57)$$

where $L^{\Delta\bar{\Phi}\approx\pi/2}(N)$ and $L^{\Delta\bar{\Phi}=0}(N)$ are the contributions from the $\Delta\bar{\Phi} \approx \pi/2$ and $\Delta\bar{\Phi} = 0$ states to $L_b$, respectively. Also, we have that (similar to Eq. (2.52))

$$z_T(N) \approx -L^{\Delta\bar{\Phi}\approx\pi/2}(N)\left(\frac{df_{Braid}^{\Delta\bar{\Phi}\approx\pi/2}}{dF}\right)\Bigg|_{M_R = M_c} - L^{\Delta\bar{\Phi}=0}(N)\left(\frac{df_{Braid}^{\Delta\bar{\Phi}=0}}{dF}\right)\Bigg|_{M_R = M_c} + \sqrt{x^2 - b^2}, \qquad (2.58)$$

where $f_{Braid}^{\Delta\bar{\Phi}\approx\pi/2}$ and $f_{Braid}^{\Delta\bar{\Phi}=0}$ are $f_{Braid}$ taken at the minimum value sets $\{\eta_C^{\Delta\bar{\Phi}\approx\pi/2}, R_C^{\Delta\bar{\Phi}\approx\pi/2}\}$ and $\{\eta_C^{\Delta\bar{\Phi}=0}, R_C^{\Delta\bar{\Phi}=0}\}$, respectively. Similar coexistence regions, as well as relations describing these regions (similar to Eqs. (2.57) and (2.58)), can be determined for the other transitions.

## 2.9 The end pieces

In the model we still have yet to give expressions for both $\mathcal{F}_{end}$ and $x$. As the end sections are described by the worm like chain model [49], we find the following form for $\mathcal{F}_{end}$ [33,34]

$$\frac{\mathcal{F}_{end}}{k_B T(L - L_b)} \approx -\frac{F_R b}{l_p \left|\sin\left(\frac{\eta_{end}}{2}\right)\right|}\left(\cos\left(\frac{\eta_{end}}{2}\right) - \left(\frac{1}{2F_R \cos\left(\frac{\eta_{end}}{2}\right)}\right)^{1/2}\right), \qquad (2.59)$$

where $\eta_{end} = \eta(L_b/2) = \eta(-L_b/2)$, the angle between the two molecular centrelines at the start and end of the braided section. The distance $x$ is determined to be [33]

$$2x = (L - L_b)(1 - \sqrt{\cos(\eta_{end}/2)/2F_R}). \qquad (2.60)$$

Since $x$ can also be specified through $2x\sin(\eta_{end}/2) = b$ this allows us to write [33,34]

$$L_b \approx \left[L - \left(1 + \left(\frac{\cos\left(\frac{\eta_{end}}{2}\right)}{2F_R}\right)^{1/2}\right) \frac{b}{\left|\sin\left(\frac{\eta_{end}}{2}\right)\right|}\right] \theta\left(L - \left(1 + \left(\frac{\cos\left(\frac{\eta_{end}}{2}\right)}{2F_R}\right)^{1/2}\right) \frac{b}{\left|\sin\left(\frac{\eta_{end}}{2}\right)\right|}\right), \qquad (2.61)$$

as well as

$$z \approx \langle z_B \rangle + \sqrt{(L-L_b)^2 \left(1 - \sqrt{\frac{\cos(\eta_{end}/2)}{F_R}}\right)^2 - b^2}. \qquad (2.62)$$

We can write for the length dependence of the free energy of the braid, when the braid is sufficiently large, $\mathcal{F}_{Braid} = k_B T L_b f_{Braid}$. Therefore we have assumed that $f_{Braid}$ is effectively decoupled from $\eta_{end}$. More elaborate numerical calculations, presented in Ref. [33], indeed suggest that this is indeed true for a sufficiently long braid. Subsequently, we can now simply minimise $\eta_{end}$ using Eqs. (2.1), (2.59) and (2.61), obtaining the expression [33]

$$\cos\left(\frac{\eta_{end}}{2}\right) \approx -\frac{F_R}{l_p f_{Braid}} - \left(\frac{2}{F_R}\right)^{1/2}\left(-\frac{F_R}{l_p f_{Braid}}\right)^{3/2}, \qquad (2.63)$$

for sufficiently large pulling force, $F_R \gg 1$. For a braid to form, we always require that $\eta_{end} > \eta_{min}$ (for $\eta > 0$) and $\eta_{end} < -\eta_{min}$ (for $\eta < 0$), where $\eta_{min}$ and $-\eta_{min}$ are values of $\eta_{end}$ that solve Eq. (2.61) when $L_b = 0$. There are values of the moment $M_{min}^+$ and $M_{min}^-$ at which both $\eta_{min}$ and $-\eta_{min}$ occur. These two moment values are finite, as to form a braid we must do mechanical work against the pulling force $F$ when we rotate from parallel molecules ( the $n = 0$ position) to where molecules just to form about to form a braid (at $n = \pm 1/2$), as the end to end distance is shorter in the latter case. As always we have the value $L_b = 0$, at the point of braid formation, we must always have $M_{min}^- = -M_{min}^+$.

## 3. Results

We consider four possible cases for attractive interactions between the molecules in the braided state: i.) where $f_{attr}$ and $\alpha_\eta$ are given by Eqs. (2.49) and (2.50), where we set $\kappa_{NH} = 2\kappa_D$; ii.) where $f_{attr}$ and $\alpha_\eta$ are still given by Eqs. (2.49) and (2.50), but we set $\kappa_{NH}^{-1} = 4.8\text{Å}$; iii.) where $f_{attr}$ and $\alpha_\eta$ are given by Eqs. (2.53), (2.54) and (2.55); and iv.) where $f_{attr}$ and $\alpha_\eta$ are given by Eqs. (2.53), (2.54) and (2.56), where the helical geometry limits $\eta_0$.

In all numerical calculations we have set the parameters $L = 55000\text{Å}$ and $b = 7000\text{Å}$ [50]. There is a large number of plots illustrating qualitative trends in the numerical data for both $M_R$ and $z$ as functions of $N$. Therefore, for all the four cases considered, we have chosen to present all of this numerical data in Appendix B. However, in the main text we will focus on particular examples, as well as discussing trends seen in the curves presented in Appendix B. For the particular examples shown in the main text we use a Debye screening length of $\kappa_D^{-1} = 15.5\text{Å}$ (again, for the rationale behind this choice see [50]). In Appendix B, we do investigate changing $\kappa_D$ by also picking values $\kappa_D^{-1} = 11.7\text{Å}$ and $7.75\text{Å}$ [50] to calculate both $M_R$ and $z$ as functions of $N$. This section is divided into two subsections; the first for non-helix specific attractive terms, the second is for helix specific ones.

### 3.1 With non-helix specific attractive interactions

Let us first consider non-helix specific attractive interactions. In Fig 4 we show a couple of results from Appendix B for the moment (or torque) $M$ that needs to be applied to the braid, as well as the end to end distance $z$, as a function of the number of braid turns. These have been generated with two choices of parameter values $\theta = 0.6$, $f_{NH} = 0.3$ and $\theta = 0.7$, $f_{NH} = 0.7$. In these plots, we also compare the case where we choose $\kappa_{NH}^{-1} = 4.8\text{Å}$ with that where we set $\kappa_{NH} = 2\kappa_D$.

The overall trends, with increasing pulling force, are for both $|M|$ and $z$ to increase. The latter trend is explained by the fact that as we increase pulling force $F$ we elongate the end pieces, and reduce $\eta_{end}$, therefore increasing $z$. This maximizes the mechanical work that is performed against the internal forces, i.e. electrostatic and steric repulsion. There is a discontinuity in $M$, at $N=0$ ($n=\pm 1/2$), as to produce a braid, we must first perform mechanical work against the pulling force. This is to reduce the extension as we first rotate from $n=0$ to $n=\pm 1/2$. The size of the discontinuity in $M$ increases with pulling force as the amount of mechanical work that we need to do against it increases. With increasing $|N|$, the trend is for $z$ to decrease, as both $L_b$ and $\eta_0$ must increase to accommodate the growing number of braid turns. Indeed, as the length of the braid $L_b$ increases, the magnitude of $\eta_{end}$ must increase through Eq. (2.61), and so affecting $z$. Also, as we increase $|N|$, the magnitude of the moment $|M|$ increases. This is because as we further decrease $z$ we must do even more mechanical work against the pulling force; also this trend increasing as $F$ grows. Note as interaction between molecules is not chiral, we have the symmetries $z(N) = z(-N)$ and $M(N) = -M(-N)$.

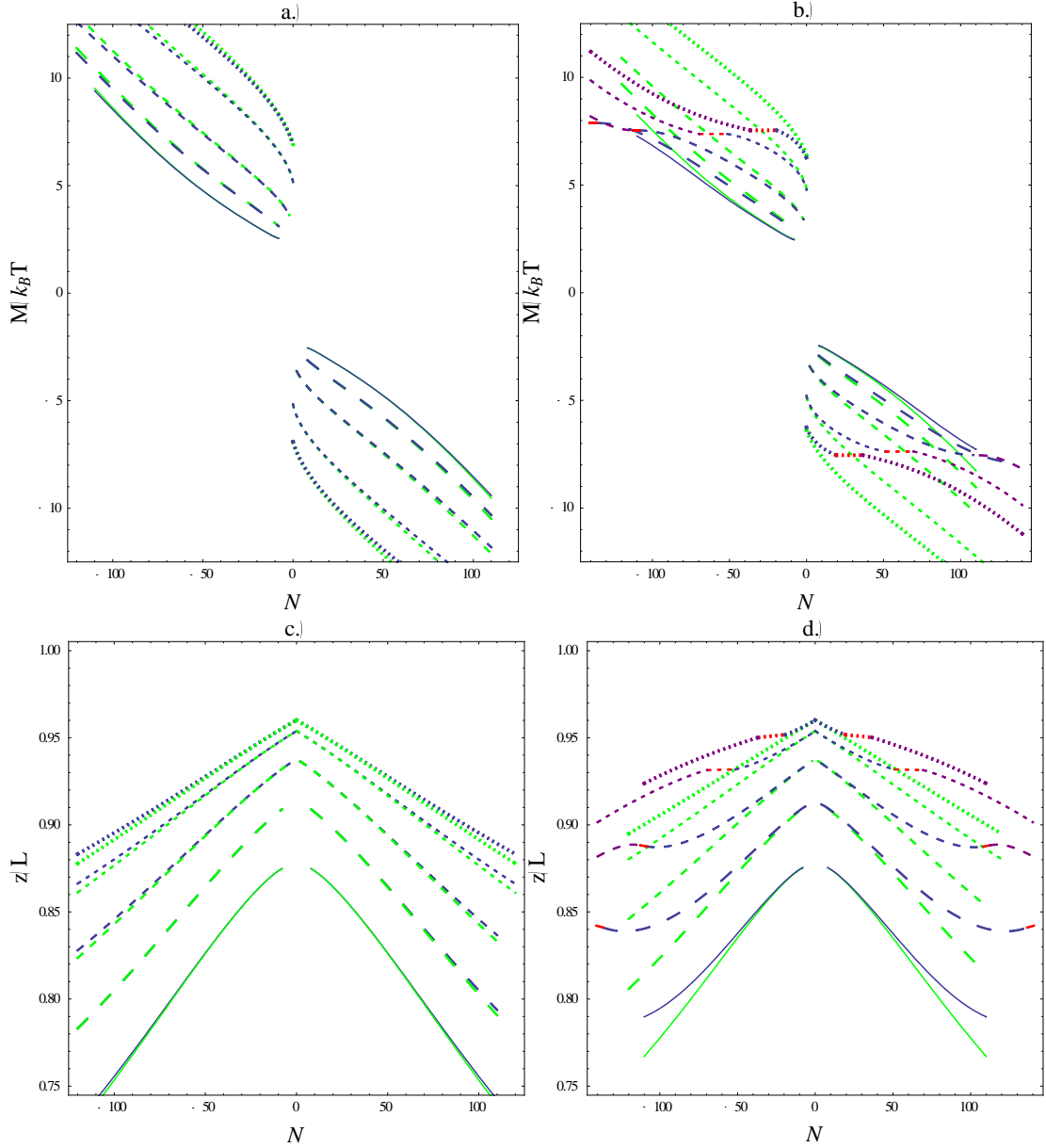

Fig.4. The moment $M$ and the extension $z/L$ as functions of the number of braid turns $N$ for non-helix specific attractive forces. Shown in a.) and b.) are plots for $M$, and in c.) and d.) plots for $z/L$. In a.) and c.) the parameter values of $\theta = 0.6$ and $f_N = 0.3$ is used, while in b.) and d.) the parameter values $\theta = 0.7$ and $f_N = 0.7$, are used. All calculations are done with a total molecular contour length of $L = 55000\text{Å}$ and separation between ends $b = 7000\text{Å}$. Here, a Debye screening length of $\kappa_D^{-1} = 15.5\text{Å}$ is used. The green (light grey) curves correspond to the choice of $\kappa_{NH}^{-1} = 4.8\text{Å}$ for the decay length of the attractive interaction, while all the other colours correspond to the choice $\kappa_{NH} = 2\kappa_D$. For this second choice, purple curves correspond to a collapsed braided state (occurring when $|M| > |M_c|$), the blue curves correspond to the looser braided state (occurring when $|M| < |M_c|$), and the red curves (straight lines at $|M| = |M_c|$, which are flat for the moment curves) are coexistence between the two states. In all curves we have used the choice $-d_{min} = d_{max} = R_0 - 2a$. The solid, long dashed, medium dashed, short dashed and dotted lines correspond to the force values $F = 3.5\,pN, 7\,pN, 14\,pN, 28\,pN$ and $40\,pN$, respectively.

When we set $\theta = 0.6$ and $f_N = 0.3$, the plots for the two choices of $\kappa_{NH}^{-1} = 4.8\text{Å}$ and $\kappa_{NH} = 2\kappa_D$ lie close together. Slight differences start only to appear when we increase the pulling force, thus bringing the molecules closer together (see Fig. 5 below). This is because the attractive term, Eq. (2.49), contributes little to the overall strength of the interaction. However, things change dramatically when we go to $\theta = 0.7$ and $f_N = 0.7$. Though, little has changed for the fixed value of $\kappa_{NH}^{-1} = 4.8\text{Å}$, for the choice $\kappa_{NH} = 2\kappa_D$ the extension curves have very different shape. Now, we see collapse into a tighter braiding state, with the appearance of the coexistence region discussed in Subsection 2.7. Here, some parts of the braid are now in a tightly braided state, while other parts are in the looser braided state. In the extension plots, before we move into the coexistence region, as we increase $|N|$, the gradient of $z$ starts to flatten for $\kappa_{NH} = 2\kappa_D$, making these curves deviate away from those calculated for $\kappa_{NH}^{-1} = 4.8\text{Å}$. This effect may be due to the fact that when $N$ increases $R_0$ decreases much more rapidly for $\kappa_{NH} = 2\kappa_D$ than for $\kappa_{NH}^{-1} = 4.8\text{Å}$, but both $L_b$ and $\eta_0$ change much less. This may be caused by substantially reduced repulsion between the two molecules when $\kappa_{NH} = 2\kappa_D$. This makes the braid very much softer with respect to compression; the reduction in $R_0$. Because of this, the system reduces $R_0$ to increase the number of turns $N$. This is opposed to doing work purely against the pulling force in reducing $z$ (that controls both the values of $\eta_0$ and $L_b$) to produce the same number of turns. This relative ease in compressing $R_0$ also manifests itself in a flattening of the moment curves, as one increases $N$ before reaching the coexistence region. When $|N|$ is increased above the coexistence region, we move into the pure tightly braided state. In this state, $R_0$ is much smaller, causing both $M$ and $z$ to change much less with changing $N$ than in the looser state. This is because one gains more turns from changing $L_b$ and so doing mechanical work against the pulling force.

We now discuss some more of the trends seen in the Graphs of Appendix B. In our complete set of plots (Appendix B), we looked at nine parameter values corresponding to the choices $\theta = 0.6, 0.7$ and $0.8$, in combination with the choices $f_N = 0.3, 0.5$ and $0.7$. Clearly by increasing both $\theta$ and $f_N$, we increase the importance of the attractive term, which is indeed seen in the plots for $\kappa_{NH} = 2\kappa_D$ over this range. However, for the case of $\kappa_{NH}^{-1} = 4.8\text{Å}$, not much happens, as the decay range is too short to affect things. The attractive term has to compete with the short ranged repulsive interaction modelled by $\mathcal{E}_{img}(R)$ (Eq.(2.28)). In both cases, at the value $\theta = 0.6$, for the values of $f_N$ considered, no collapse is seen, as the attractive component is not strong enough. However, as we increase $f_N$, we start to see an increase in the reduction of the magnitude of the gradients of both $M$ and $z$, with increasing $|N|$, for the choice $\kappa_{NH} = 2\kappa_D$. As was discussed in the previous paragraph, this is probably an indication of a significant weakening of repulsion between the two molecules. Both for $\theta = 0.7$ and $\theta = 0.8$ coexistence occurs when $f_N$ is sufficiently high, for the $\kappa_{NH} = 2\kappa_D$ case. For values of $f_N$ just below where this coexistence occurs, an inflexion is sometimes seen in the curves. This is probably due to the rate of change in $R_0$ with respect to increasing $N$, after changing rapidly, suddenly becoming much slower when one approaches the inter-axial separation $R \approx 30\text{Å}$. This sudden change with increasing $N$ is because of strong short range repulsion that in contained in Eq. (2.28). Once the molecules become sufficiently close, this term causes the braid to suddenly become much more rigid with respect to $R_0$ as $N$ is changed. In Appendix B we do not present any curves for the choice $\theta = 0.8$, $f_N = 0.7$ and $\kappa_{NH} = 2\kappa_D$, as we have a negative interaction energy, and so one may have molecular condensation. For the choice $\theta = 0.8$, $f_N = 0.5$ and $\kappa_{NH} = 2\kappa_D$, as $R_0 - 2a$ becomes particularly small we have considered both the choices where $-d_{min} = d_{max} = R_0 - 2a$, and where $-d_{min} = R_0 - 2a$ with $d_{max}$ given by Eq. (2.11) (for $\theta = 0.8$). Both curves seem to match well in their region of overlap, suggesting that for the particular choice, at these values of $N$, the steric term is not that important. This is because the steric interaction, if important, should depend more on this choice than other interactions; the other interactions contribution to $d_R$ seem to be much more important over these ranges of values. For all other curves, the choice $-d_{min} = d_{max} = R_0 - 2a$ was used.

We further investigate the effect of changing the salt concentration, and so the value of $\kappa_D$. We present plots for both $M$ and $z$ for the Debye screening length values $\kappa_D^{-1} = 11.7\text{Å}$ and $\kappa_D^{-1} = 7.75\text{Å}$, in addition to $\kappa_D^{-1} = 15.5\text{Å}$. Here we consider the pulling force values $F = 3.5\,\text{pN}$ and $F = 40\,\text{pN}$. Increasing $\kappa_D$ does indeed decrease the long range repulsion. This reduction manifests itself in a reduction in the moment $M$ required to produce a certain number of braid turns and an increase in $z$, for the case where $\kappa_{NH}^{-1} = 4.8\text{Å}$. Also for $\kappa_{NH}^{-1} = 4.8\text{Å}$, we see that the graphs depend little on $f_N$; the attractive contribution is not that important. For the case where $\kappa_{NH} = 2\kappa_D$ the situation is rather more complicated, sometimes leading to non-monotonic behaviour with increasing $\kappa_D$. As we change from $\kappa_D^{-1} = 15.5\text{Å}$ to $\kappa_D^{-1} = 11.7\text{Å}$, the collapse to a more tightly braided state goes away. This is because we reduce the range of the attractive component in the intermolecular interaction, thus any local minimum or flat region in the potential as a function of $R_0$ goes away, due the short ranged repulsive interactions. The upshot is that both $R_0$ and $\eta_0$ can no longer jump between two values for a loosely and tightly braided state. Therefore, if $\theta$ and $f_N$ are sufficiently large, the degree of repulsion actually increases as the range of the attractive component is reduced. Where this happens, instead we have an increase in $M$ and a decrease in $z$. On the other hand, when we change $\kappa_D^{-1} = 11.7\text{Å}$ to $\kappa_D^{-1} = 7.75\text{Å}$, we reduce the amount of repulsion, by reducing the long range repulsion in all cases. Therefore, here, we indeed see a decrease in $M$ and an increase in $z$. In addition, when we increase $\kappa_D$, the difference between $\kappa_{NH}^{-1} = 4.8\text{Å}$ and $\kappa_{NH} = 2\kappa_D$ case becomes less. This is hardly surprising, as the values of two decay lengths become closer to each other.

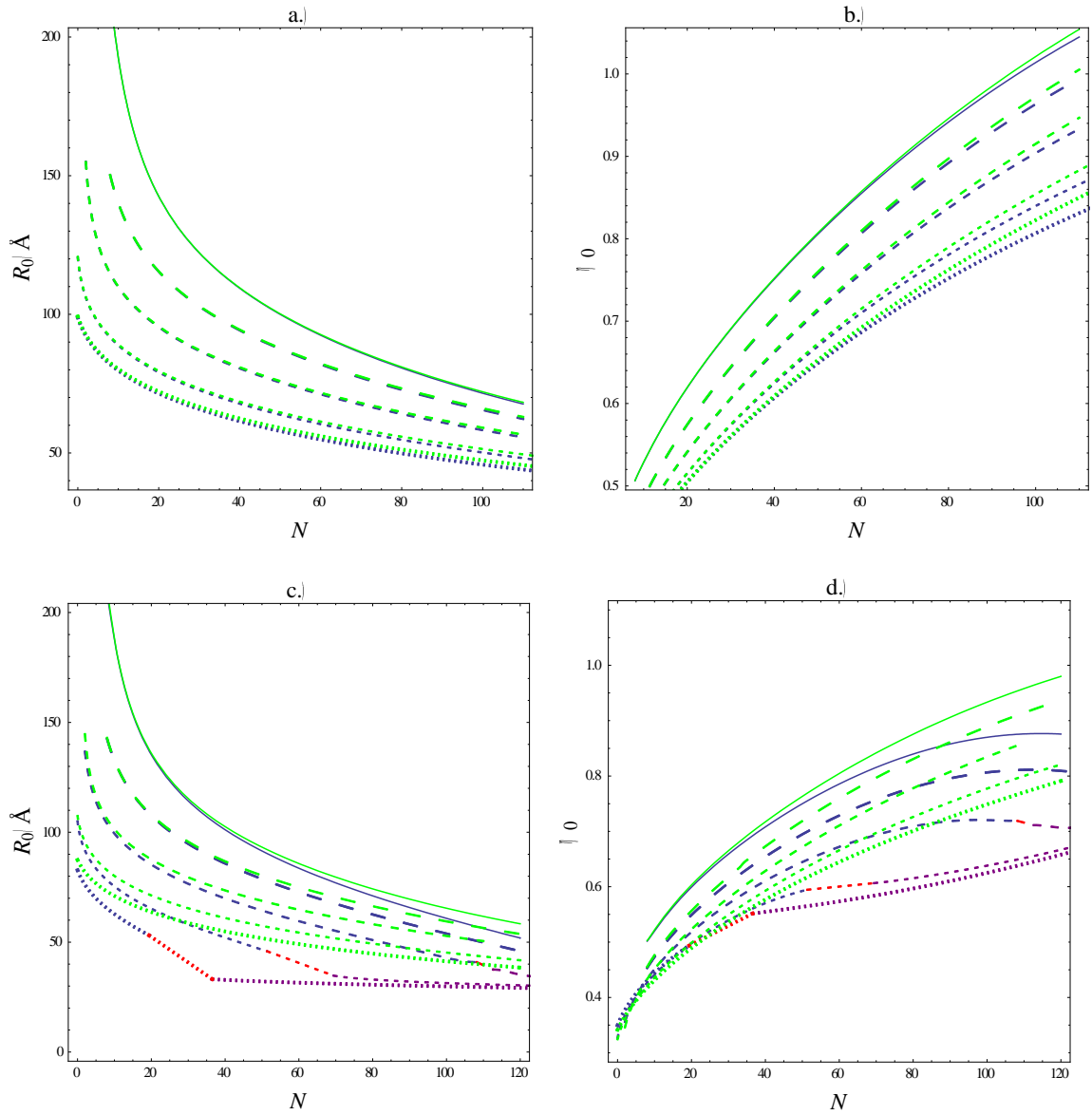

Fig.5. The average inter-axial separation $R_0$ and the average tilt angle $\eta_0$ as functions of the number of braid turns $N$ for non-helix specific attractive forces. In a.) and c.) $R_0$, and in b.) and d.) $\eta_0$, is plotted. In both a.) and b.) the parameter choice $\theta = 0.6$ and $f_N = 0.3$ is used, while in c.) and d.) the values $\theta = 0.7$ and $f_N = 0.7$ are used. All calculations are done with a total molecular contour length of $L = 55000\text{Å}$ and separation between ends $b = 7000\text{Å}$. Here, a Debye screening length of $\kappa_D^{-1} = 15.5\text{Å}$ is used. The green (light grey) curves correspond to the choice of $\kappa_{NH}^{-1} = 4.8\text{Å}$ for the decay length of the attractive interaction, while all the other colours correspond to the choice $\kappa_{NH} = 2\kappa_D$. For this second choice, the red curves (straight lines) mark coexistence between the tightly braided state and the looser braided state. The purple curves (curves joining on to the right of the straight lines at large $N$) correspond to a collapsed braided state, the loose state is denoted by the blue curves (all other curves). The solid, long dashed, medium dashed, short dashed and dotted lines correspond to the force values $F = 3.5 pN, 7 pN, 14 pN, 28 pN$ and $40 pN$, respectively. Due to the left-right handed symmetries $R_0(N) = R_0(-N)$ and $\eta_0(N) = -\eta_0(-N)$, we show only positive $N$ in these plots.

We now examine the mean geometric parameters of the braid $R_0$ and $\eta_0$ for $\kappa_D^{-1} = 15.5\text{Å}$, and show plots in Fig. 5. As we increase the pulling force we reduce both the size of $R_0$ and the magnitude of $\eta_0$. This enables the force to extend the vertical length of the braided section, and so the total extension of the molecules, whilst maintaining a fixed value of $N$. To understand why this should be, it is useful to think about the ground state of the system. In this case we have that

$$N = \pi L_b \sin(\eta_0/2)/R_0 \quad \text{and} \quad z = L_b \cos(\eta_0/2) + \sqrt{(L-L_b)^2 - b^2} \;. \tag{3.1}$$

Increasing the value of $z$ this is achieved by reducing $L_b$ and decreasing $\eta_0$, as seen in Eq. (3.1). This is what we indeed observed in Fig. 6 when thermal fluctuations are included. Therefore, it is clear that $R_0$ must decrease with increasing pulling force to maintain a fixed number of braid turns $N$ in the ground state. This argument indeed rationalizes the trends that we see. However, if $L_b$ drastically decreases with force and $R_0$ cannot change that much due to internal forces, it is quite possible for $\eta_0$ to increase, to maintain the number of braid turns. This can be the case when $b$ is sufficiently large, that the second term in the expression for $z$ dominates (Eq. (3.1)). We do not see this tendency here, though this reverse trend was seen in Ref. [34]. In the ground state, examination of Eq. (3.1) suggests that, as $N$ is increased, $R_0$ decreases and $\eta_0$ increases. When thermal fluctuations are included these trends persist, as seen in Fig.5.

We see that, for the choice of parameters $\theta = 0.6$ and $f_N = 0.3$, there is, again, little difference between setting $\kappa_{NH}^{-1} = 4.8\text{Å}$ and $\kappa_{NH} = 2\kappa_D$. There is a big difference between these choices for $\theta = 0.7$ and $f_N = 0.7$. Let us examine this difference by first looking at $R_0$. Indeed we see that $R_0$ is significantly larger for the choice of $\kappa_{NH}^{-1} = 4.8\text{Å}$ than for $\kappa_{NH} = 2\kappa_D$. This is due to the lesser degree of attraction in the former when compared with the latter. For the latter case, when the pulling force gets sufficiently large, we see collapse into the tightly braided state. This collapse can also be triggered by increasing $N$, as seen in Fig 5c. In the tightly braided state $R_0$ stays roughly constant at $R_0 \approx 30\text{Å}$, due large short ranged repulsive interactions arising from $\mathcal{E}_{img}(R)$. Shown in Fig. 5c is also the coexistence region where the spatial average of $R_0$ is plotted (in red). In the coexistence region ($|N_c^L| < N < |N_c^T|$) this is given by

$$\bar{R}(N) = \frac{\left(|N_c^L| - |N|\right) R_0(N_c^T)}{\left(|N_c^L| - |N_c^T|\right)} + \frac{\left(|N| - |N_c^T|\right) R_0(N_c^L)}{\left(|N_c^L| - |N_c^T|\right)} . \tag{3.2}$$

In fact, here, $R_0$ is either of the two values, $R(N_c^T)$ and $R(N_c^L)$, favoured by the two braiding states at the critical moment $M_c$. For values of $|N|$ smaller than at where this coexistence occurs we do indeed see that $R_0$ is changing more rapidly for the $\kappa_{NH} = 2\kappa_D$ choice than for the $\kappa_{NH}^{-1} = 4.8\text{Å}$ choice, indeed suggesting the flattening of the moment and extension graphs (see Fig. 4) discussed previously.

We now examine the differences in the plots of $\eta_0$ for the two choices of screening length. For the parameter choice $\theta = 0.6$ and $f_N = 0.3$, we again see little difference between the two choices. However, when $\theta = 0.7$ and $f_N = 0.7$, we see that the values of $\eta_0$ are much less when $\kappa_{NH} = 2\kappa_D$. Also, the gradient with respect to $N$, becomes much less for $\kappa_{NH} = 2\kappa_D$ than the $\kappa_{NH}^{-1} = 4.8\text{Å}$ case, as $N$ is increased. This is indeed in line with the fact that $R_0$ is reduced, when $\kappa_{NH} = 2\kappa_D$, and drops more rapidly with increasing $N$, as we discussed previously. Also again, for $\kappa_{NH} = 2\kappa_D$, we have the coexistence region. In this region we plot the spatial average value of $\eta_0$

$$\bar{\eta}(N) = \frac{\left(\left|N_c^L\right| - |N|\right)\eta_0(N_c^T)}{\left(\left|N_c^L\right| - \left|N_c^T\right|\right)} + \frac{\left(|N| - \left|N_c^T\right|\right)\eta_0(N_c^L)}{\left(\left|N_c^L\right| - \left|N_c^T\right|\right)}. \tag{3.3}$$

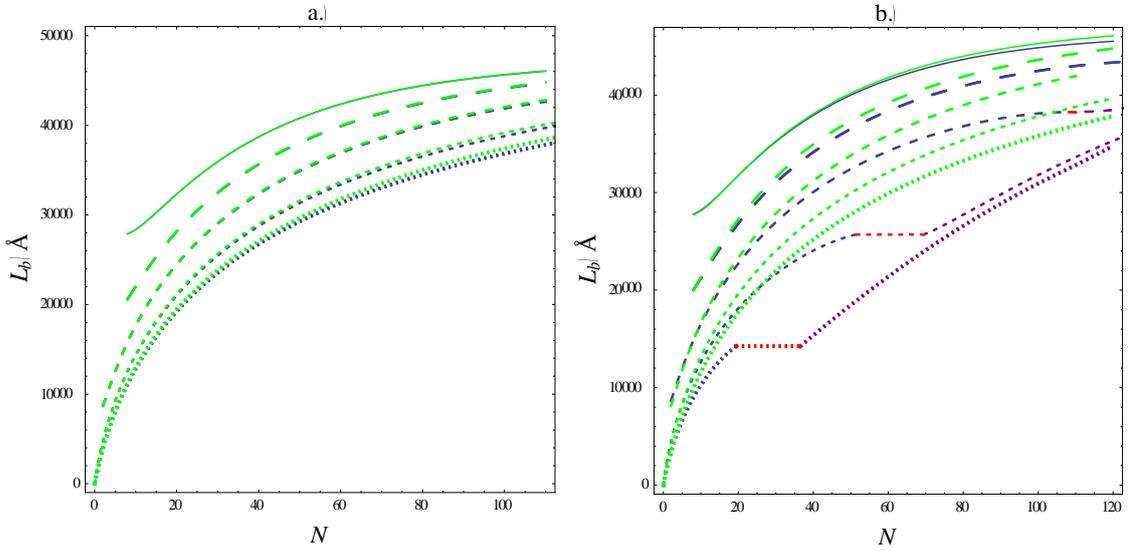

Fig.6. The contour length of the braided section $L_b$ as a function of the number of braid turns $N$ for non-helix specific attractive forces. In a.) the parameter values $\theta = 0.6$ and $f_N = 0.3$ are used, whereas in b.) $\theta = 0.7$ and $f_N = 0.7$. The values of the other parameters and colour coding are the same as in the previous two figures. Most notably, the green curves correspond to the choice of $\kappa_{NH}^{-1} = 4.8\text{Å}$ for the decay length of the attractive interaction, while all the other colours correspond to the choice $\kappa_{NH} = 2\kappa_D$. The solid, long dashed, medium dashed, short dashed and dotted lines correspond to the force values $F = 3.5\,pN, 7\,pN, 14\,pN, 28\,pN$ and $40\,pN$, respectively. Due to the left-right handed symmetry $L_b(N) = L_b(-N)$ we show only positive $N$ in these plots.

Last of all we present plots for $L_b$. Already we have discussed that, with increasing force, $L_b$ should decrease. This is indeed what is seen in Fig.6. Again, there is little difference between the two choices of screening length for the attractive component when $\theta = 0.6$ and $f_N = 0.3$, and a big difference when $\theta = 0.7$ and $f_N = 0.7$. At large force values, the values of $L_b$ are considerably smaller at fixed $N$ for $\kappa_{NH} = 2\kappa_D$ than for $\kappa_{NH}^{-1} = 4.8\text{Å}$. This reduction in $L_b$ can be accommodated due to reduced repulsion, that allows for $R_0$ to take a smaller value, thereby conserving the value of $N$. This smaller value $L_b$ is favoured by the pulling force that wants to increase the extension as much as the other forces will allow. In the coexistence region, $L_b$ stays constant as $L_b(N_c^T) = L_b(N_c^L)$, as the energy must be the same for both states and so $\eta_{end}$ (see Eq. (2.63)), therefore simply
$L_b(N) = L^T(N) + L^L(N) = L_b(N_c^L)$.

## 3.2 With helix specific attractive interactions

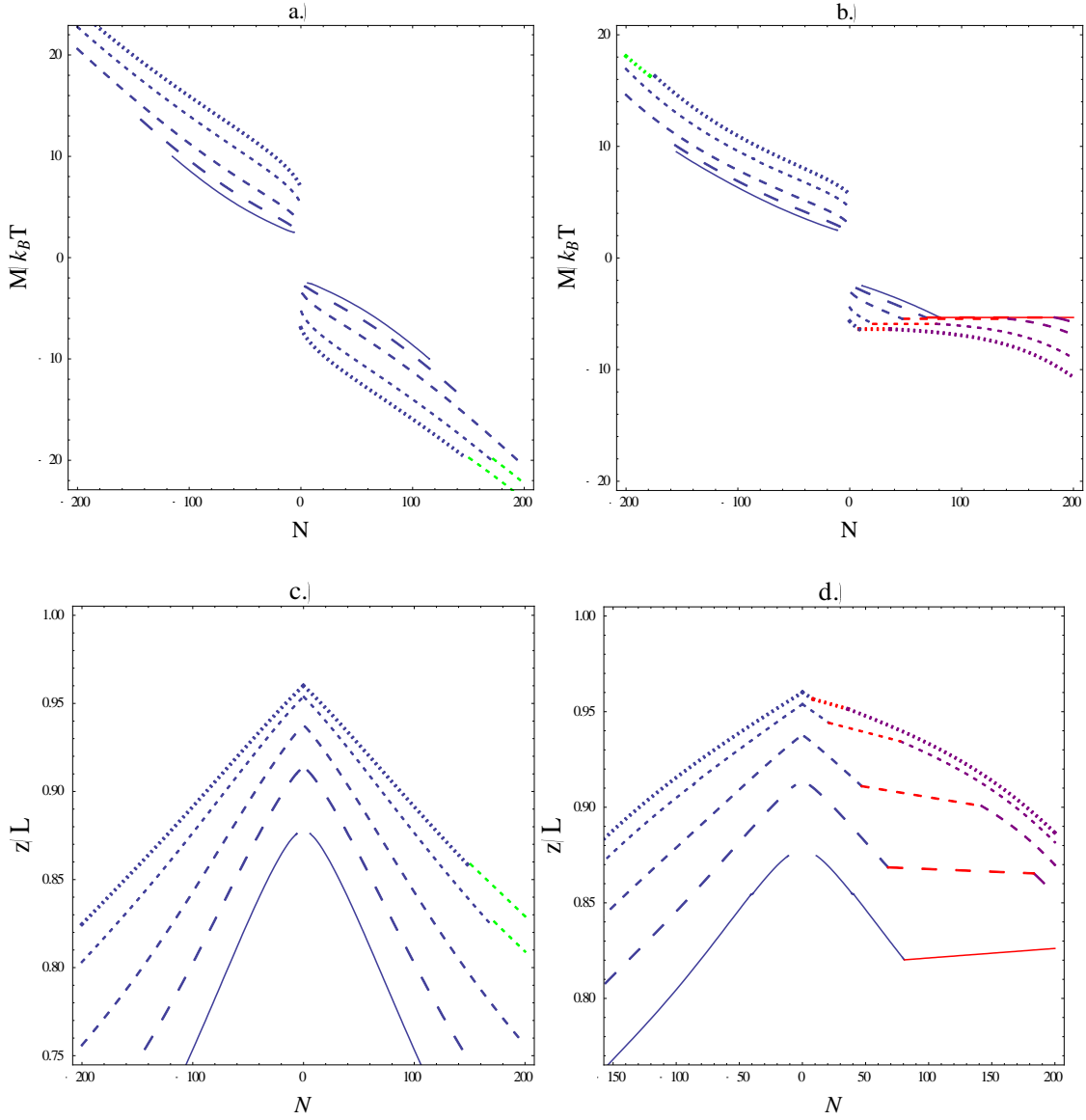

Fig.7. The moment $M$ and the extension $z/L$ as functions of the number of braid turns for helix specific attractive forces. These graphs are generated with the forms $f_{attr}$ and $\alpha_\eta$ in the free energy given by Eqs.(2.53), (2.54) and (2.55). In all plots the parameter values $L = 55000\text{Å}$, $b = 7000\text{Å}$ and $\kappa_D^{-1} = 15.5\text{Å}$ are used. Both a.) and b.) show the plots for applied moment $M$, while c.) and d.) show plots for $z/L$. For a.) and c.), the parameter values $\theta = 0.6$ and $f_H = 0.5$ are used, and for b.) and d.) the parameter values $\theta = 0.8$ and $f_H = 0.9$. Here the blue (dark grey) curves correspond to the $\lambda_h^* = \infty$ state, green (light grey) curves to the finite $\lambda_h^*$, $\Delta\bar{\Phi} = 0$ state, and the purple curves (to the right of the coexistence region) to the finite $\lambda_h^*$, $\Delta\bar{\Phi} \approx \pi/2$ state. There is a coexistence

region between the $\lambda_h^* = \infty$ state and the $\lambda_h^*$, $\Delta\bar{\Phi} \approx \pi/2$ state, which is shown in red (straight or flat lines) in both b.) and d.). Since the other coexistence regions are very narrow, we do not delineate them with red lines. For a discussion of these states see Subsection 2.8. The solid, long dashed, medium dashed, short dashed and dotted lines correspond to the force values $F = 3.5 pN, 7 pN, 14 pN, 28 pN$ and $40 pN$, respectively.

Now we examine what happens when one has helix specific attractive interactions. Firstly, due to the chirality of the molecular helices, the left-right handed symmetry is broken, when these interactions are not washed out by thermal fluctuations and intrinsic helix distortions. This washing out of chiral forces is supposed for the $\lambda_h^* = \infty$ state in the current study; but, as discussed previously, even this state may exhibit weak chirality. All plots are truncated at $\eta_{end} \approx 2$, where we might expect our model to break down, if not slightly before then as discussed previously in Refs [33,35]. Strictly speaking, for the expressions used, $\eta_{end}, \eta_0 \leq \pi/2$, and this condition must always be the obeyed for two rods in steric contact [32]. In Fig. 7 we present numerical results for both $M$ and $z$, for the case where we do not limit $\eta_0$ through the helix geometry (i.e. using Eq. (2.55) for the free energy). We show results, in the main text, for two choices of the parameters; $\theta = 0.6$, $f_H = 0.5$ and $\theta = 0.8$, $f_H = 0.9$. Firstly, in both cases, we see the same general trends for $M$ and $z$ with increasing $|N|$ and pulling force $F$, as was seen for non-helix specific attraction (see previous subsection).

For the parameter choice $\theta = 0.6$, $f_H = 0.5$, the right-left handed symmetry is broken only at very large values of $|N|$. This slight breaking of symmetry is through a transition, at positive $N$, between the $\lambda_h^* = \infty$ state and the finite $\lambda_h^*$ state where $\Delta\bar{\Phi} = 0$. Indeed, for this choice of interaction parameters, helix specific forces are sufficiently weak for the $\lambda_h^* = \infty$ state to dominate over a large range of values of $N$. The finite $\lambda_h^*$, $\Delta\bar{\Phi} = 0$ state is only favoured for left handed braids, when the two molecules can be brought sufficiently close for the helix specific interactions to be sufficiently strong. To bring the two molecules together, as discussed previously, one can increase either the pulling force or $N$, the number of braid turns. Increasing the latter will eventually cause the braid axis to buckle, whereas increasing the former is likely to supress this tendency. The coexistence region between these two states is very narrow, as there is little difference between the two sets of values of $\eta_0$ and $R_0$ for the two states at the critical moment for the transition.

When look at the plots for $\theta = 0.8$, $f_H = 0.9$, the situation drastically changes. The left-right handed braid symmetry is broken to a much greater degree. Firstly, we see the existence of large coexistence regions for positive $N$, where $M$ stays constant at $M_C^{(3)}$, the critical moment for the transition between the $\lambda_h^* = \infty$ state and the finite $\lambda_h^*$, $\Delta\bar{\Phi} \approx \pi/2$ state. These coexistence regions are much larger than those seen for $\theta = 0.6$, $f_H = 0.5$, because the $\Delta\bar{\Phi} \approx \pi/2$ state has a much smaller value of $R_0$ than the finite $\lambda_h^*$, $\Delta\bar{\Phi} = 0$ state. This new state allows for the molecules to come very close due to strong helix specific attraction. However, in the $\lambda_h^*$, $\Delta\bar{\Phi} \approx \pi/2$ state, the system must pay a much higher entropy cost to confine the molecules within this tight braid structure. Therefore, as both $\theta$, $f_H$ are reduced, weakening the attractive contribution, the looser $\Delta\bar{\Phi} = 0$ state starts to become favoured in left handed braids. As the pulling force is increased, the value of $N$ at which the transition between the $\lambda_h^* = \infty$ state and the finite $\lambda_h^*$, $\Delta\bar{\Phi} \approx \pi/2$ state occurs is reduced. This is again because increasing the pulling force brings the molecules closer together; the mechanical forces are sufficiently strong to overcome the entropy cost and so favour the finite $\lambda_h^*$, $\Delta\bar{\Phi} \approx \pi/2$ state.

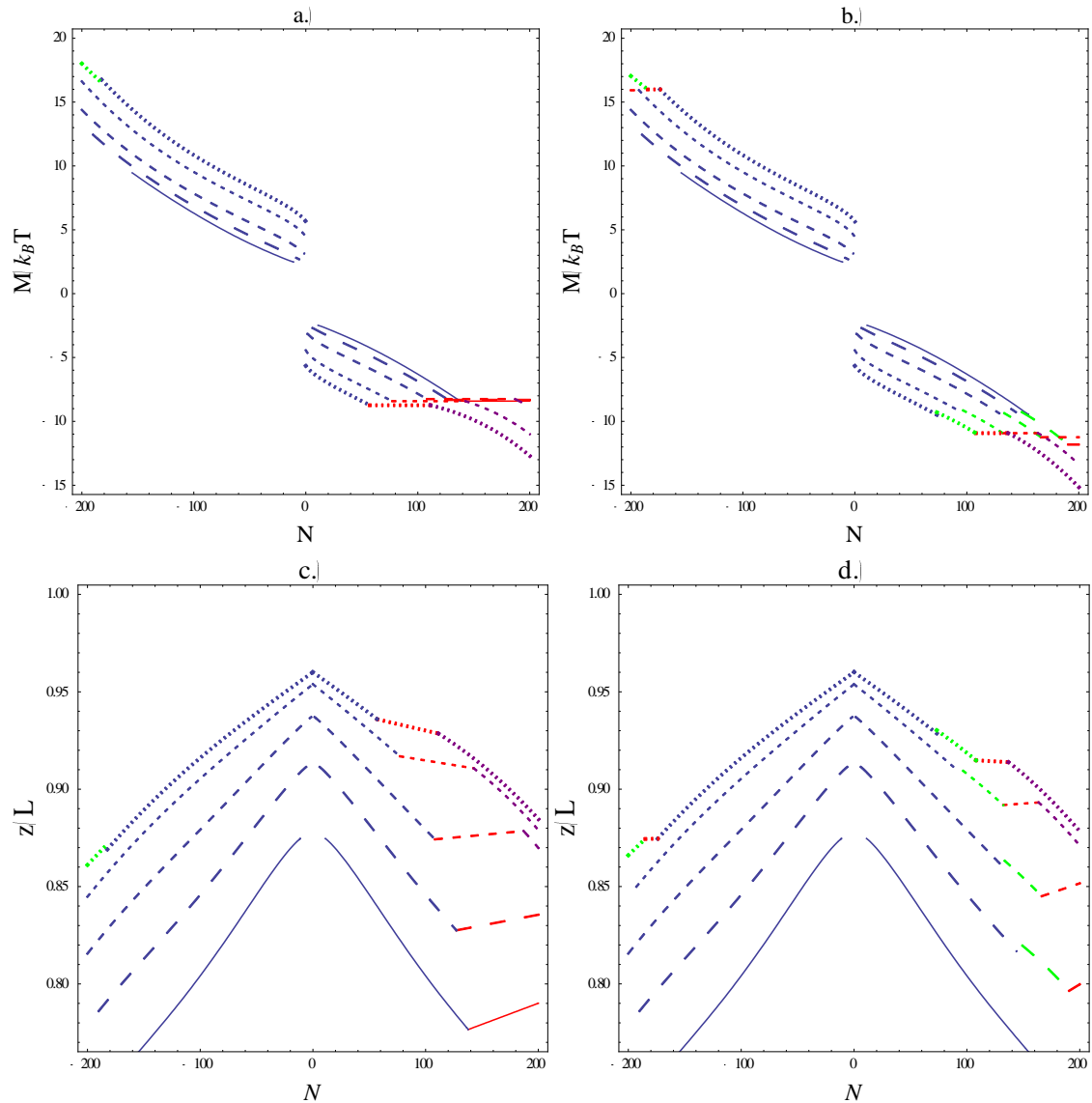

Fig.8. Comparison of moment and extension plots generated by using Eq. (2.55) with those using the Eq. (2.56), which estimates the restriction on $\eta_0$ due to helix geometry. These graphs are generated with the helix specific forms $f_{attr}$ and $\alpha_\eta$ in the free energy given by Eqs.(2.53) and (2.54). In all plots the parameter values $L = 55000\text{Å}$, $b = 7000\text{Å}$, $\kappa_D^{-1} = 15.5\text{Å}$, $\theta = 0.8$ and $f_H = 0.5$ are used. Plots for the moment are given in a.) and b.) and the extension in c.) and d.). Both a.) and c.) use Eq. (2.55), whereas both b.) and d.) use Eq. (2.56). Here, the blue (dark grey) curves correspond to the $\lambda_h^* = \infty$ state, green (light grey) curves to the finite $\lambda_h^*$, $\Delta\bar{\Phi} = 0$ states and the purple curves (to the right of the coexistence region) to the finite $\lambda_h^*$, $\Delta\bar{\Phi} \approx \pi/2$ state. There are coexistence lines between the states at the endpoints, shown in red (as flat or straight lines). Note that that in a.) and c.), for negative $N$, the coexistence region is so narrow that we do not show coexistence lines. For a discussion of these states and the coexistence regions see Subsection 2.8. The solid, long dashed, medium dashed, short dashed and dotted lines correspond to the force values $F = 3.5\,pN, 7\,pN, 14\,pN, 28\,pN$ and $40\,pN$, respectively.

We now examine what happens when we include an additional $\sin^2 \eta_0$ term (see Eq. (2.56)), estimated by the helix geometry, which limits the size of $\eta_0$. Here in the main text, we make the parameter choice $\theta = 0.8$ and $f_H = 0.5$. Looking at Fig. 8, when we include the $\sin^2 \eta_0$ term, we no-longer have one single transition between states for this choice of parameters, at positive $N$. Instead, as we increase $N$, we have a transition between the $\lambda_h^* = \infty$ state and the finite $\lambda_h^*$, $\Delta\bar{\Phi} = 0$ state, followed by a transition between this state and the finite $\lambda_h^*$, $\Delta\bar{\Phi} \approx \pi/2$ state. What is responsible for this seems to be a reduction in amount of attraction between the two molecules, due to a reduction in the energy benefit from making $\eta_0$ positive. Here, again, the looser $\Delta\bar{\Phi} = 0$ state is favoured over the much tighter $\Delta\bar{\Phi} \approx \pi/2$ state, due to a smaller reduction in the entropy due to confinement. This is the case until the mechanical forces become sufficiently strong to cause collapse into the tighter braid state. The final transition to $\Delta\bar{\Phi} \approx \pi/2$ state occurs at a larger value of $N$ than when Eq. (2.55) is used, which supposes no geometric limitation on $\eta_0$.

In Appendix B we present plots for both $M$ and $z$ for when either Eqs. (2.55) or (2.56) are used. These plots are for combinations of the values of $\theta = 0.6, 0.7, 0.8$ with the values $f_H = 0.5, 0.7, 0.9$. We first present plots with varying pulling force for the Debye screening length of $\kappa_D^{-1} = 15.5\,\text{Å}$. Indeed, as we increase both $\theta$ and $f_H$ we move from the situation that we have for $\theta = 0.6$, $f_H = 0.5$ to that we have for $\theta = 0.8$, $f_H = 0.9$. There are couple of things that we should comment on. Firstly, we see that the values of $N$ at which the transition from $\lambda_h^* = \infty$ state to the finite $\lambda_h^*$, $\Delta\bar{\Phi} = 0$ state occurs seems to depend little on the value of $f_H$. On the other hand, the region where the finite $\lambda_h^*$, $\Delta\bar{\Phi} \approx \pi/2$ state occurs depends strongly on $f_H$, and if $f_H$ is too small it does not occur at all. Secondly, only the case where we choose $\theta = 0.8$, when Eq. (2.55) is used, are the attractive interactions are sufficiently strong for there to be a collapse from the $\lambda_h^* = \infty$ state straight into the finite $\lambda_h^*$, $\Delta\bar{\Phi} \approx \pi/2$ state. If we use Eq. (2.56), such a collapse is only seen for the parameter choice $\theta = 0.8$, $f_H = 0.9$, for all other choices with $\theta = 0.8$ we see to two transitions for positive $N$.

We have investigated the dependence of both $M$ and $z$ on salt concentration through changing $\kappa_D$. We use the same values of $\kappa_D$ for helix specific interactions as those that we have used for non-helix specific interactions. In Appendix B, we present plots for both pulling force values $F = 3.5\,\text{pN}$ and $F = 40\,\text{pN}$ for varying $\kappa_D$. We see that the effect of increasing $\kappa_D$ is to allow for transitions between states to occur at lower values of $N$. This is because, when we increase $\kappa_D$, the long range electrostatic repulsion described by $\mathcal{E}_{dir}(R)$ is reduced most by a significant reduction in its decay range. Thus, it becomes much easier to bring the molecules close enough together for helix specific forces to be strong to cause a transition between braiding states. At $F = 40\,\text{pN}$, if Eq. (2.55) is used, when $\theta = 0.8$, $f_H = 0.9$ we reach a situation where when we increase $N$ from $N = 0$, we immediately start forming a braid in the collapsed finite $\lambda_h^*$, $\Delta\bar{\Phi} \approx \pi/2$ state. This is more of less what was seen in Ref. [33], at sufficiently high forces for the strong helix specific interaction case. Always, when Eq. (2.56) is used, attraction is weakened, and the transitions occur at larger values of $N$ for all parameters choices.

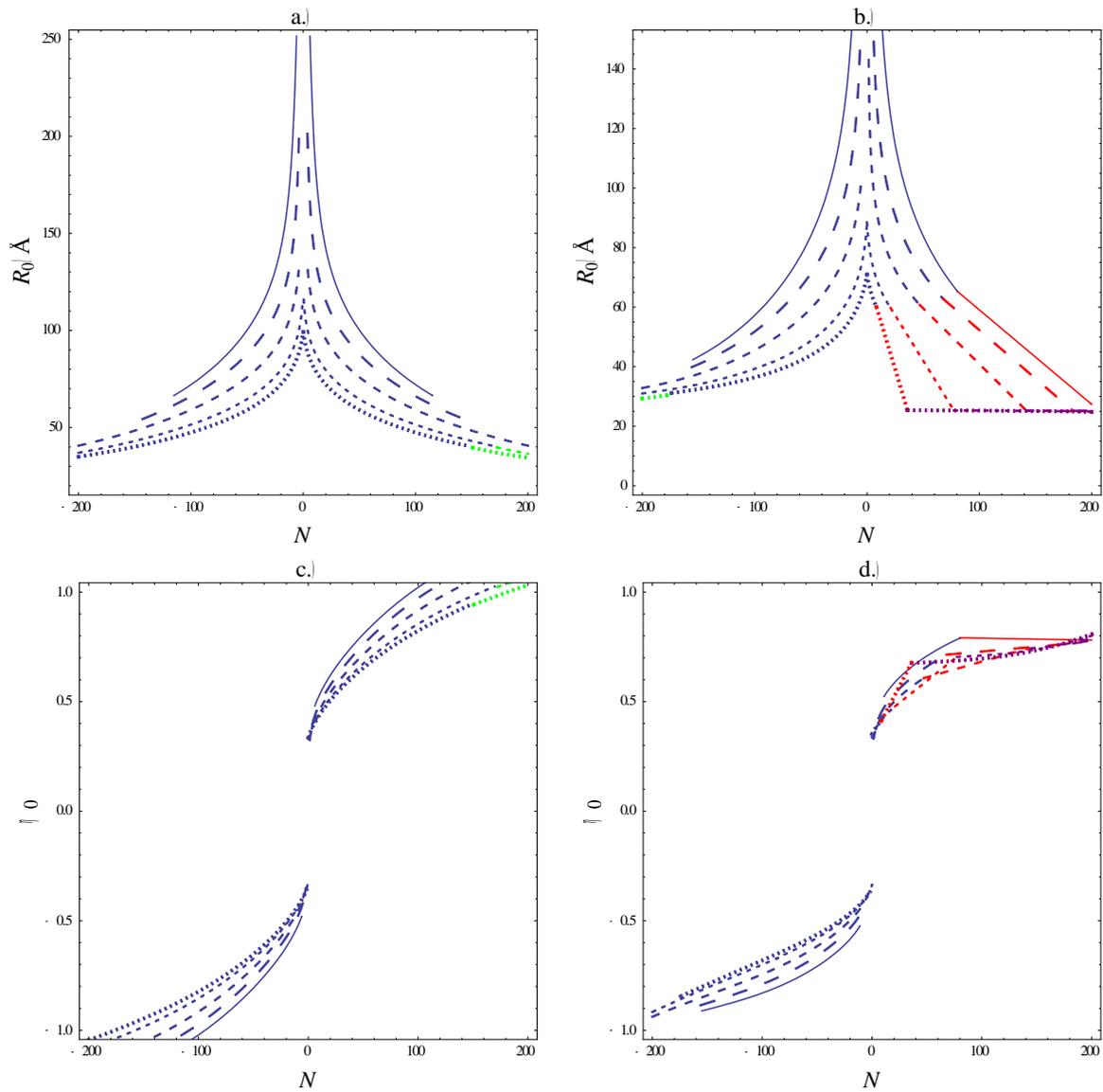

Fig.9. The average inter-axial separation $R_0$ and the average tilt angle $\eta_0$ as functions of the number of braid turns $N$. Here we consider helix specific attractive forces, where we use the forms for $f_{attr}$ and $\alpha_\eta$ given by Eqs.(2.53) (2.54) and (2.55) in the free energy. Both a.) and b.) contain plots of $R_0$, while in c.) and d.) plots of $\eta_0$ are presented. In both a.) and c.), the parameter values $\theta = 0.6$ and $f_H = 0.5$ are used, and in b.) and d.) the parameter values $\theta = 0.8$ and $f_H = 0.9$ are used. In all plots the parameter values $L = 55000\text{Å}$, $b = 7000\text{Å}$ and

$\kappa_D^{-1} = 15.5\text{Å}$ are used. The same colour coding as in Fig. 7 denotes the various braiding states and coexistence regions. The solid, long dashed, medium dashed, short dashed and dotted lines correspond to the force values $F = 3.5\,pN, 7\,pN, 14\,pN, 28\,pN$ and $40\,pN$, respectively.

Now, we investigate the geometric parameters $R_0$ and $\eta_0$ as functions of the number of braid turns $N$. We start by presenting results, in Fig. 9, for when we use Eq. (2.55) (thus, not restricting $\sin\eta$ in the interaction potential) in our calculations for both the parameter choice $\theta = 0.6$, $f_H = 0.5$, as well as for $\theta = 0.8$, $f_H = 0.9$. The trend with increasing pulling force is for both $R_0$ and $\eta_0$ to decrease. However, as dicussed previously, it may be possible for the trend in $\eta_0$ to reverse, if we increase $b$ [34]. The general trends with increasing $|N|$ are that $R_0$ decreases and $|\eta_0|$ increases. The origins of these trends has already been discussed in the previous subsection, and it seems to be generic. As we indeed see, the finite $\lambda_h^*$ $\Delta\Phi = 0$ state is not much different from the $\lambda_h^* = \infty$ state with respect to both $R_0$ and $\eta_0$. This indeed explains the narrowness of the respective coxistance regions. However, the finite $\lambda_h^*$ $\Delta\Phi \approx \pi/2$ state is very different. Here, $R_0 \approx 25\text{Å}$ and stays roughly constant with respect to $N$, due to the strong repulsion in $f_{rep}$ that makes the braid extremely rigid with regard to compression (reducing $R_0$). In this state, from the point of view of $R_0$, the braid can indeed effectively considered as two rods of effective radius $R_0/2 \approx 12.5\text{Å}$ in steric contact. Also in this state $\eta_0$ is much reduced; since $R_0$ is small, this can be accomadated at fixed $N$. This reduction brings down the bending energy and increases the braid extension $z_b$ in response to the pulling force.

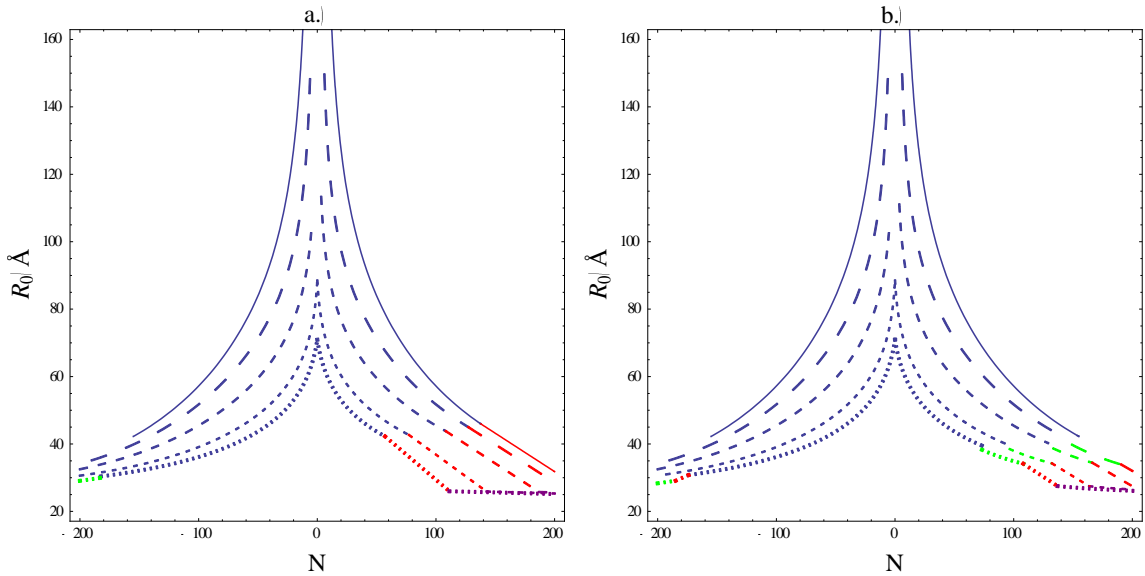

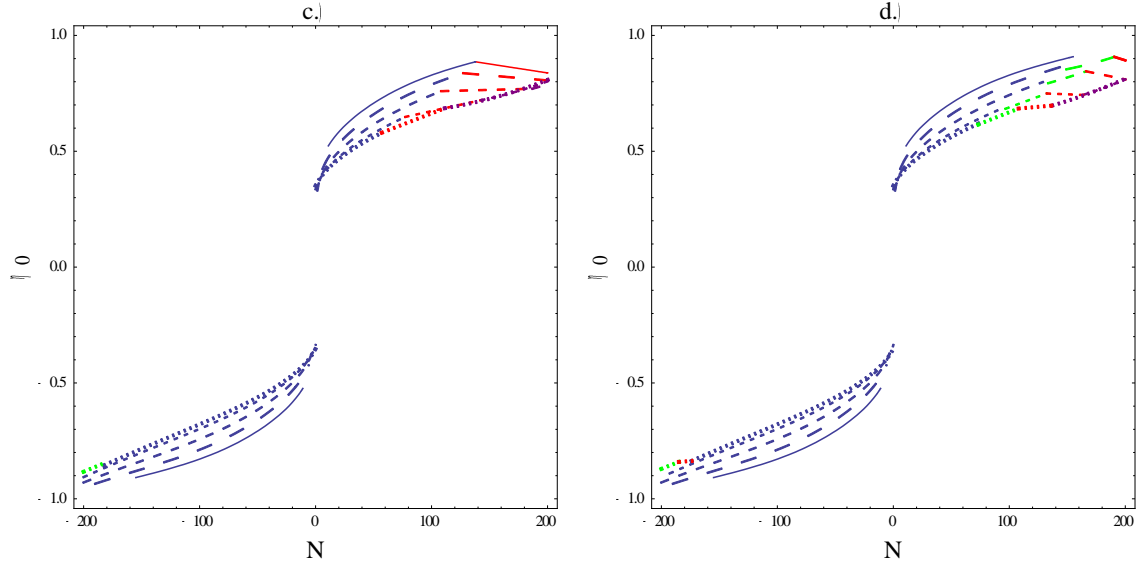

Fig.10. Comparison of plots for $R_0$ and $\eta_0$ generated by using Eq. (2.55) with those using the Eq. (2.56), which estimates the restriction on $\eta_0$ due to helix geometry. In all plots the parameter values $L = 55000\text{Å}$, $b = 7000\text{Å}$, $\kappa_D^{-1} = 15.5\text{Å}$, $\theta = 0.8$ and $f_H = 0.5$ are used. The plots of $R_0$ are given in a.) and b.) and the plots of $\eta_0$ in c.) and d.) Both a.) and c.) use Eq. (2.55), whereas both b.) and d.) use Eq. (2.56). The same colour coding is used as in Fig.8 to denote the various braiding states and coexistence regions. The solid, long dashed, medium dashed, short dashed and dotted lines correspond to the force values $F = 3.5\,\text{pN}, 7\,\text{pN},\ 14\,\text{pN}, 28\,\text{pN}$ and $40\,\text{pN}$, respectively.

Next, in Fig. 10, we compare the results for $R_0$ and $\eta_0$ calculated with Eq. (2.56) with those of Eq. (2.55) at the parameter values $\theta = 0.8$, $f_H = 0.5$. Again, we notice, when Eq. (2.56) is used as opposed to Eq. (2.55), the appearance of the $\Delta\Phi = 0$ state for positive $N$, as well as the $\Delta\Phi \approx \pi/2$ state occuring at larger values of $N$. Again, in regards to both $R_0$ and $\eta_0$, the $\Delta\Phi = 0$ state is not much different from the $\lambda_h^* = \infty$ state. When Eq. (2.56) is used, the value of $R_0$ within the $\Delta\Phi \approx \pi/2$ state is still rather insensitive to $N$. The value of $R_0$ calculated for this state slightly larger, for Eq. (2.56), than when Eq. (2.55) is used. Note that in the coxistence regions, the spatial averages of both $R_0$ and $\eta_0$ are used, which have similar forms as Eqs. (3.2) and (3.3).

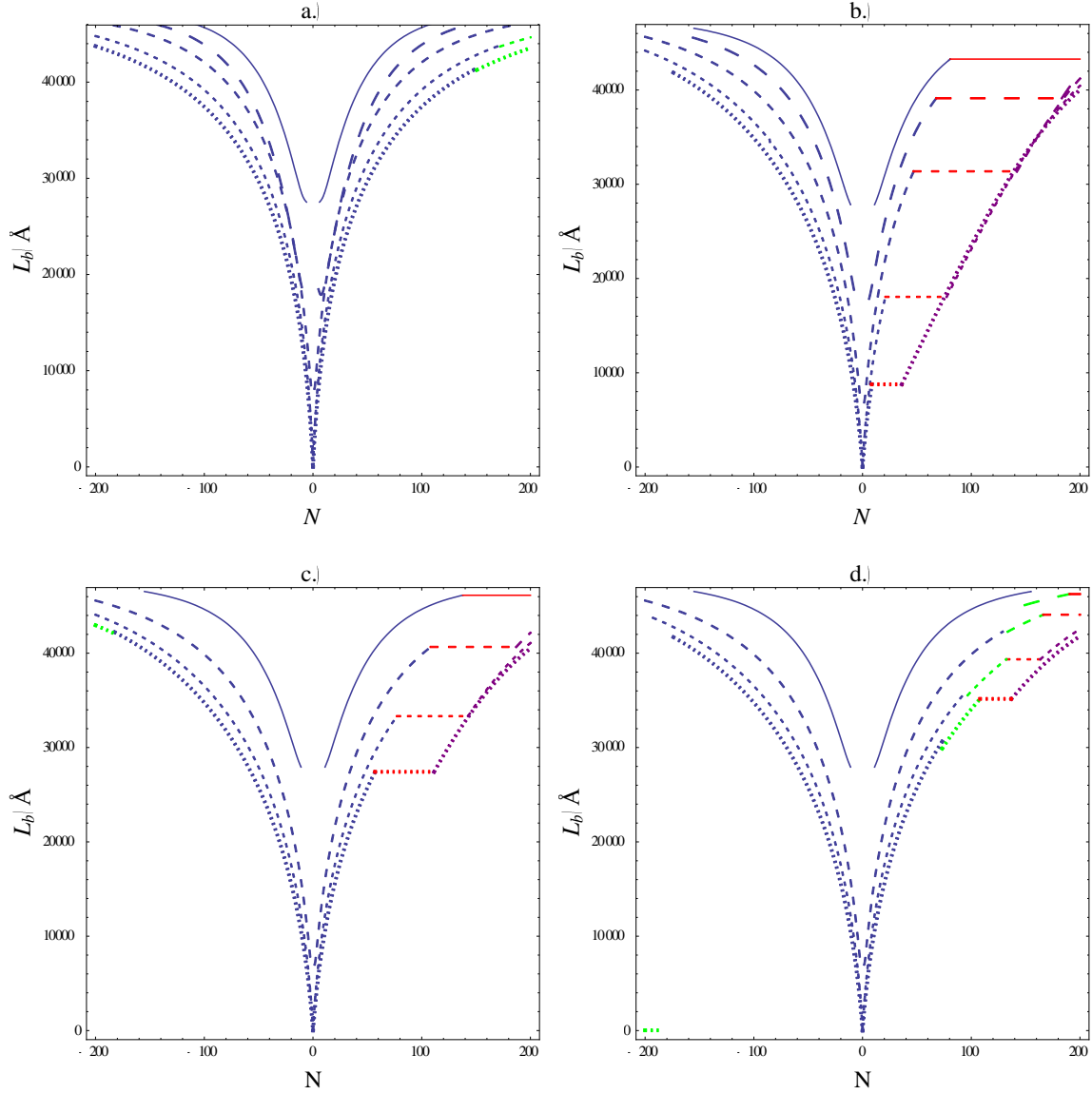

Fig.11. The contour length of the braided section as a function of the number of braid turns for helix specific attractive forces. In all plots the parameter values $L = 55000\text{Å}$, $b = 7000\text{Å}$ and $\kappa_D^{-1} = 15.5\text{Å}$ are used, as well as for a.) $\theta = 0.6$, $f_H = 0.5$, b.) $\theta = 0.8$, $f_H = 0.9$, and both c.) and d.) $\theta = 0.8$, $f_H = 0.5$. In a.), b.) and c.) Eq. (2.55) is used to generate the plots, but d.) uses Eq. (2.56), where an estimate of the restriction on $\eta_0$ due to helix geometry is used. Here, the blue (dark grey) curves correspond to the $\lambda_h^* = \infty$ state, green (light grey) curves to the finite $\lambda_h^*$, $\Delta\bar{\Phi} = 0$ states and the purple curves (to the right of coexistence lines) to the finite $\lambda_h^*$, $\Delta\bar{\Phi} \approx \pi/2$ state. There are coexistence lines between the states at the endpoints, shown in red (as flat lines). Note that some of the coexistence regions are so narrow that we do not show them. For a discussion of these states and the coexistence regions see Subsection 2.8. The solid, long dashed, medium dashed, short dashed and dotted lines correspond to the force values $F = 3.5\,pN, 7\,pN,\ 14\,pN, 28\,pN$ and $40\,pN$, respectively.

Last of all, we examine $L_b$ in Fig. 11. The general qualitative trends are the same as for non-helix specific interactions. However, these plots clearly break the left-right handed symmetry, whereas the previous ones, presented in Subsection 3.1, do not. In the first three plots of Fig. 11, we use Eq. (2.55) in our calculations. We see that for $\theta = 0.6$, $f_H = 0.5$ there is little difference between right and left handed braids with the same value of $|N|$, only a slight breaking of symmetry. However, at the parameter values of $\theta = 0.8$, $f_H = 0.5$ and $\theta = 0.8$, $f_H = 0.9$ there is a clear difference between braids of different handedness. In the $\Delta\Phi \approx \pi/2$ state, the small value of $R_0$ allows for a smaller value of $L_b$, as well as $\eta_0$ to accommodate the same number of braid turns $N$. Last of all, we see that, at $\theta = 0.8$, $f_H = 0.5$, the differences in the size of $L_b$ between the Left and Right handed braids are much less pronounced, when we use Eq. (2.56), instead of Eq. (2.55).

## 4. Discussion

We have examined what happens when we include an attractive component in the interaction potential. Indeed, a collapse of the braid from a looser braided state to a tighter one(s) is obtained. This occurs at a critical value of the moment(s) $M = M_c, M_c^{(j)}$, which needs to be applied to generate the braid. We find that, as we change the number of braid turns $N$, (and consequently $n$, which may correspond to a number of turns of a bead that the molecules are attached to) we do not get a sudden jump between the two states. Rather equilibrium thermodynamics dictates that we should have a continuous transition, in which the two states coexist in different proportions along the braided section. These phenomenona occur because we have a local minimum or flattening in the interaction potential caused by attraction. At coexistence $M = M_c, M_c^{(j)}$ and two states minimize the free energy. The size of the coexistence region in $N$ is determined by the difference in the values of average inter-axial separation $R_0$ and tilt angle $\eta_0$, for the two coexisting states.

      We may ask what conditions may be conducive to braid collapse. Firstly our study reveals that it is most likely to happen when the pulling force and the number of braid turns are sufficiently high, if we have sufficient attraction for a local minima or a flattening the interaction potential. However, condensation does not necessarily need to happen; i.e. the interaction energy becoming sufficiently negative for the molecule to collapse into a toroid state. Indeed, this phenomenon of braid collapse may occur in the presence of some condensing agent (though not causing conditions that condense DNA), or multivalent ions, to induce such an attractive term. We should point out, in the case of an attractive interaction contribution, that when the system is not in thermo-dynamical equilibrium, the coexistence may not be seen. In a dynamical situation, where we were turning the molecular ends at some finite rate, we could expect to see a jump from a local minimum state straight to the state with lowest free energy through thermal fluctuations, not a coexistence region where both states have the same energy. Also, the salt concentration is important. In the case where we had non-chiral attraction with a decay range half the Debye screening length, which ionic correlation forces might suggest, we saw that braid collapse only occurred at only the lowest salt concentration considered. In relation to this, it is interesting to see that in the results of Ref. [3], at low salt concentrations, jumps caused by braid collapse were indeed observed while rotating the bead at a constant rate. Though, the interpretation given authors of Ref. [3] was that the DNA braid had undergone some kind coalescence between two single stranded DNA from each molecule, while the other two strands had broken away. Nevertheless, we might expect to see from a dynamical model, based on this one, hysteresis curves similar to those seen in Ref. [3]. At what value of $N$ this jump would occur, and the degree of hysteresis, would depend on how quickly one turned the bead.

Based on what we have seen in our results for dual molecule braiding, we would hypothesize, in the single molecule twisting experiments of Ref. [13], that braid collapse and coexistence might have already been happening. When we look at our results for $R_0$ (c.f. Figs. 5, 9 and 10) as a function of increasing pulling force $F$ (as well as $N$), for sufficiently large values of $N$, we do see that the average value of $R_0$ starts falling until it reaches a value where it stays roughly constant. Indeed, to fit the experimental data such behaviour has been advocated in Ref. [28], flattening out to a constant value of $R_0 \approx 30 \text{Å}$ at high forces. What we obtain for the collapsed state with non-helix specific forces is about this value. For helix specific forces, the value is slightly less, but this observation does not necessarily preclude them. Firstly, it is a slightly different system; and secondly we may still be overestimating the torque that increases $\eta_0$ from these forces. Indeed, when we try and limit $\eta_0$, using geometrical arguments, the value of $R_0$ comes closer to the value of $R_0 \approx 30 \text{Å}$ in the tightly braided state. With all of this in mind, it would be interesting to do dual molecule braiding experiments in exactly the same ionic conditions as those studied in Ref. [13]. Conversely, the complete model presented here could be adapted to describe the experiments of Ref. [13]. However, if we are indeed observing braid collapse in the experiments of Ref. [13] due to attractive interactions, performing the former experiments could be even more interesting, as one should be able to investigate the chirality of such interactions.

From our results, we see that when helix specific forces are sufficiently strong, they cause large collapse and coexistence regions only in left handed braids. In contrast, for non-helix specific forces, significant collapse may occur for braids of both handedness. The curves are symmetric under $N \rightarrow -N$. Always, an indicator sign of the coexistence, at thermodynamic equilibrium, is a significant change in the rate of change of the end to end distance $z$ with respect to $N$, i.e. a change in $dz/dN$ (as seen in Figs. 4, 7 and 8). Also, the rate of change of moment should drop considerably. In the calculations presented, in the thermodynamic limit this is actually zero ($L \rightarrow \infty$). In addition to the finite size effects discussed in Ref. [33], one should also point out, as the dual molecule braid effectively is a 1-D system. Therefore, there can be no infinitely sharp (phase like) transitions. Due to thermal fluctuations we might expect some coexistence away from the value(s) $M = M_c, M_c^{(j)}$ allowing for a smoothing out of the sharp edges seen in our result curves (for instance Figs. 4, 7 and 8) moving into and away from the coexistence region at $M = M_c, M_c^{(j)}$. This refinement requires more careful treatment of the relevant local minima in the free energy, with appropriate Boltzmann weighting. This yet has yet to be considered, but it should not affect these qualitative indicators of braid collapse.

If spontaneous braiding, as discussed in Ref. [6], does indeed occur, we would expect features of the results, for strong helix specific forces (contained in Ref. [33]), to hold below some critical value of the pulling force. The first is that below a critical force, at $M = 0$, a state is favoured for which $N \neq 0$; a braided state is formed when no torque is applied to the molecules. The second is, below this critical force, the appearance of a coexistence region between a loose right handed braided state (not favoured by helix specific forces) and a tight left handed braided state (favoured). Here, coexistence between these two states occurs at $N = 0$; once the molecules are forced together (overcoming kinetic barriers) they stick together in some combination of the two states. At some parameter values where the degree of counter-ion groove localization $f_H$ is close to $1$ spontaneous braiding might still occur [38]. This, in conjunction with these results, suggests that the critical force might be at values smaller than $3.5 \text{pN}$. This value of the pulling force may be too small for high force approximations used in this study to be used. On the other hand, undulations of the braid axis may have a role to play in further weakening helix specific forces by causing the helices of the two molecules to further fall further out of alignment. This is because in a bend of the braid centre line, one molecule has to use up more contour length, and so more helical pitches (for helical molecules), than the other molecule. This effect, if matters for spontaneous braiding, is probably not so important for a braid under sufficient tension, as these undulations are likely to be suppressed. Therefore, in the model, we have not yet included this effect with helix specific interactions. However, the complete spontaneous braiding of two DNA molecules might not be possible, due to these additional effects due to undulations, even though significant helix specific forces may be present. Still, it might be possible for very long molecules for braiding to occur in regions along them, provided that the purely braided state is still metastable. The true test of whether this is at all possible will be experiment.

Already in the experiments of Ref. [3] there may be a slight indicator of chirality [38]. Though conceivably, it might be that, for two molecules, helix specific forces are weak enough to be insignificant, and non-chiral attraction is more important. This is why, in our study, we have investigated both cases. However, in assemblies of DNA and other rod like helical molecules, we do indeed see a cholesteric state [5,8]. Therefore, surely, to produce such a state that is intrinsically chiral, the intermolecular interactions must have some chiral nature in molecular assemblies. Helix specific forces, whether they are steric [51] or ranged interactions [4,52], naturally provide this. Therefore, we suggest that it may be insightful to attempt experiments of similar kind as in Ref. [1] and [3], but with more molecules attached to both the bead and the substrate, if this can indeed be technically achieved. If insignificant for two molecules, we would expect, as we increase the number of molecules in the braid, chiral effects should become more apparent; otherwise, this leaves a mystery concerning why a cholesteric state should form at all. Theoretically, braids formed of more molecules should increase the size of helix specific effects from ranged interactions. One way this happens is by increasing the number of interactions between molecules, as in DNA assemblies. This reduces the amount of fluctuations, and so increases the effective strength of helix specific forces. A second effect is the sucking of counter-ions into multi-molecular assemblies, through Donnan equilibrium [5]. This pushes up $\theta$ and reduces the effective screening length of the electrostatic interactions [5]. We do indeed see that both effects would enhance the role of helix specific forces from the trends seen in the numerical results (also see Appendix B). It is also worth pointing out that braids formed three or more actin molecules have indeed been observed [53], in high concentrations of $Mg^{2+}$ ions; what more or less appears like spontaneous braiding. We would propose, in light of the experiments of Ref. [53], that perhaps an interesting case to start with is three actin molecules at these ionic conditions, when constructing such multi-molecular mechanical braiding experiments. If spontaneous braiding does occur we would expect interesting phenomena in these braiding experiments below a critical force, as discussed in the previous paragraph and in Ref. [33].

In regards to molecular braiding, there are many routes theory can go. As well as adapting our current models to braids formed of more than two molecules, there are many other things have yet to be investigated. One obvious one, as discussed in Refs. [33,34,35] is braid buckling, where above a particular value $|N|$, we go to a state where the average braid centre line can no longer be considered as straight. Also finite size effects, for instance treating the coexistence regions more explicitly, by looking at '1-d domain wall' solutions that minimize the (free) energy functional, as was suggested in Ref. [33], and it was postulated that the 'domain wall' energy cost may lead to some kind of torque overshoot phenomena [26]. Another direction is to consider what happens to the braid when we consider different pulling forces acting on each of the two molecules, so that the system no-longer adopts the rectangle like configuration seen in Fig. 1. Already, some work has been done in this direction for the ground state in Ref. [32], but thermal fluctuations have yet to be included here.  Also, an attempt at calculating the electrostatic energy for a more generalized braid structure has already been performed [45], which is also needed to describe such situations. The statistical mechanics, here, may present a rather tricky problem as $\eta_0(s)$ and $\Delta\Phi(s)$ (if this degree of freedom is important) can no longer be considered constant in the braid. There is also new parameter that characterizes the asymmetry [54] of the braid [32,45], which will also depend on the position along the braid. However, if these functions vary slowly enough, perhaps a WKB approximation [55] may be used in calculating Eigen functions and values needed to evaluate path integrals in the statistical mechanics. Another research path would be to consider sliding of one molecule with respect to the other in the braid; already some theoretical study has been conducted [56], but a=the braided structure has yet to be explicitly considered.  On the experimental side, to investigate these effects, four bead micro-manipulation experiments of the form suggested in [57] could be useful.

We reemphasise that dual (and multi-molecular) molecular mechanical braiding experiments may provide a useful and unique tool to investigate many aspects of the nature of intermolecular interactions between rod-like molecules.

## Acknowledgements

D.J. (O') Lee would like to acknowledge useful discussions with R. Cortini , G. King, A. Korte, A. A. Korynshev, E. L. Starostin , G.H.M. van der Heijden, G.J.L. Wuite and Mara Prentis. This work was initially inspired by joint work that has been supported by the United Kingdom Engineering and Physical Sciences Research Council (grant EP/H004319/1). He would also like to acknowledge the support of the Human Frontiers Science Program (grant RGP0049/2010-C102).

## Appendix A: an outline of the calculation used to derive the Free energy of the braided section

Our starting point, here is Eqs. (2.4)-(2.39) of the main text. We can first write down the following partition function to describe the braided section

$$Z_{braid} = \int \mathcal{D}R(s) \int \mathcal{D}\eta(s) \int \mathcal{D}\Delta\Phi(s) \int \mathcal{D}x'_A(s) \int \mathcal{D}y'_A(s) \exp\left(-\frac{E_{braid}[R(s),\eta(s),\Delta\Phi(s),x'_A(s),y'_A(s)]}{k_BT}\right),$$

(A.1)

where we have five functional integrations to evaluate over all the degrees of freedom characterized through all possible realizations of the functions $R(s)$, $\eta(s)$, $\Delta\Phi(s)$, $x_A(s)$ and $y_A(s)$. In Eq. (A.1), we have emphasised that $E_{braid}$ is a functional of these five functions through use of a square bracket. In the form of $E_{braid}$ $x_A(s)$ and $y_A(s)$ are effectively decoupled from $R(s)$, $\eta(s)$ and $\Delta\Phi(s)$ can be dealt with separately. Therefore, next, we can write

$$E_{braid}[R(s),\eta(s),\Delta\Phi(s),x'_A(s),y'_A(s)] = E_R[R(s),\eta(s),\Delta\Phi(s)] + E_{x,y}[x'_A(s),y'_A(s)],$$
(A.2)

where

$$E_{x,y}[x'_A(s), y'_A(s)] = \int_{-L_b/2}^{L_b/2} ds \left( k_B T l_p \left[ \left(\frac{dx'_A(s)}{ds}\right)^2 + \left(\frac{dy'_A(s)}{ds}\right)^2 \right] + \frac{F}{2\cos\left(\frac{\eta_0}{2}\right)}\left[x'_A(s)^2 + y'_A(s)^2\right] \right) \quad (A.3)$$

$$-2\pi M_R Wr_b[x'_A(s), y'_A(s)],$$

and $E_R$ contains all else that remains that depends on $R(s)$, $\eta(s)$, and $\Delta\Phi(s)$ in what is presented in Eqs. (2.5), (2.8), (2.9), (2.18), (2.19), (2.26)-(2.32), (2.36)-(2.39). The expression for $Wr_b[x'_A(s), y'_A(s)]$ is given by Eq. (2.20) of the main text, which makes it a functional of both $x'_A(s)$ and $y'_A(s)$. This decoupling of the degrees of freedom allows us to write

$$Z_{braid} = Z_A Z_R, \quad (A.4)$$

where

$$Z_A = \int \mathcal{D}x'_A(s) \int \mathcal{D}y'_A(s) \exp\left(-\frac{E_{x,y}[x'_A(s), y'_A(s)]}{k_B T}\right), \quad (A.5)$$

$$Z_R = \int \mathcal{D}R(s) \int \mathcal{D}\eta(s) \int \mathcal{D}\Delta\Phi(s) \exp\left(-\frac{E_R[R(s), \eta(s), \Delta\Phi(s)]}{k_B T}\right). \quad (A.6)$$

The first thing that we do is to construct variational approximations to approximate both $Z_A$ and $Z_R$, as in Ref. [36] for the case of strong intermolecular interactions. This involves writing the trial energy functionals for both sets thermal fluctuations described by Eqs. (A.5) and (A.6). These are

$$\frac{E_{0,A}[x'_A(s_0), y'_A(s_0)]}{k_B T} = \int_{-L/2}^{L/2} ds_0 \left( l_p \left[ \left(\frac{dx'_A(s_0)}{ds_0}\right)^2 + \left(\frac{dy'_A(s_0)}{ds_0}\right)^2 \right] + \left(\alpha_x x'_A(s_0)^2 + \alpha_y y'_A(s_0)^2\right) \right) \quad (A.7)$$

$$-2\pi M Wr_b[x'_A(s), y'_A(s)],$$

and

$$\frac{E_{0,R}[\delta\eta(s), \delta R(s), \delta\Phi(s)]}{k_B T} = \int_{-L_b/2}^{L_b/2} ds \left( \frac{l_p}{4}\left(\frac{d^2\delta R(s)}{ds^2}\right)^2 + \frac{\beta_R}{2}\left(\frac{d\delta R(s)}{ds}\right)^2 + \frac{\alpha_R}{2}\delta R(s)^2 \right)$$
$$+ \int_{-L_b/2}^{L_b/2} ds \left( \frac{l_p}{4}\left(\frac{d\delta\eta(s)}{ds}\right)^2 + \frac{\alpha_\eta}{2}\delta\eta(s)^2 \right) + \int_{-L_b/2}^{L_b/2} ds \left( \frac{l_p}{4}\left(\frac{d\delta\Phi(s)}{ds}\right)^2 + \frac{\alpha_\Phi}{2}\delta\Phi(s)^2 \right). \quad (A.8)$$

Then contributions to the Free energy, $F_A = -k_B T \ln Z_A$ and $F_R = -k_B T \ln Z_R$ can written approximately as

$$F_A \approx F_A^T = -k_B T \ln Z_{0,A} + \langle E_{x,y}[x'_A(s), y'_A(s)]\rangle_0 - \langle E_{0,A}[x'_A(s), y'_A(s)]\rangle_0, \quad (A.9)$$

and

$$F_R \approx F_R^T = -k_B T \ln Z_{0,R} + \langle E_R[R(s),\eta(s),\Delta\Phi(s)]\rangle_0 - \langle E_{0,R}[\delta\eta(s),\delta R(s),\delta\Phi(s)]\rangle_0, \quad (A.10)$$

where

$$Z_{0,A} \approx \int \mathcal{D}x'_A(s)\int \mathcal{D}y'_A(s)\exp\left(-\frac{E_{0,A}[x'_A(s),y'_A(s)]}{k_B T}\right), \quad (A.11)$$

$$Z_{0,R} \approx \int \mathcal{D}\delta R(s)\int \mathcal{D}\delta\eta(s)\int \mathcal{D}\delta\Phi(s)\exp\left(-\frac{E_{0,R}[\delta R(s),\delta\eta(s),\delta\Phi(s)]}{k_B T}\right), \quad (A.12)$$

$$\langle E_{x,y}[x'_A(s),y'_A(s)]\rangle_0 = \frac{1}{Z_{0,A}}\int \mathcal{D}x'_A(s)\int \mathcal{D}y'_A(s)E_{x,y}[x'_A(s),y'_A(s)]\exp\left(-\frac{E_{0,A}[x'_A(s),y'_A(s)]}{k_B T}\right), \quad (A.13)$$

$$\langle E_{0,A}[x'_A(s),y'_A(s)]\rangle_0 = \frac{1}{Z_{0,A}}\int \mathcal{D}x'_A(s)\int \mathcal{D}y'_A(s)E_{0,A}[x'_A(s),y'_A(s)]\exp\left(-\frac{E_{0,A}[x'_A(s),y'_A(s)]}{k_B T}\right), \quad (A.14)$$

$$\langle E_R[R(s),\eta(s),\Delta\Phi(s)]\rangle_0$$
$$= \frac{1}{Z_{0,R}}\int \mathcal{D}\delta R(s)\int \mathcal{D}\delta\eta(s)\int \mathcal{D}\delta\Phi(s)E_R[R(s),\eta(s),\Delta\Phi(s)]\exp\left(-\frac{E_{0,R}[\delta R(s),\delta\eta(s),\delta\Phi(s)]}{k_B T}\right), \quad (A.15)$$

and

$$\langle E_{0,R}[\delta\eta(s),\delta R(s),\delta\Phi(s)]\rangle_0$$
$$= \frac{1}{Z_{0,R}}\int \mathcal{D}\delta R(s)\int \mathcal{D}\delta\eta(s)\int \mathcal{D}\delta\Phi(s)E_{0,R}[R(s),\eta(s),\Delta\Phi(s)]\exp\left(-\frac{E_{0,R}[\delta R(s),\delta\eta(s),\delta\Phi(s)]}{k_B T}\right).$$
$$(A.16)$$

Here, recall that we have written $\eta(s) = \eta_0 + \delta\eta(s)$, $\Delta\Phi(s) = \Delta\Phi_0(s) + \delta\Phi(s)$ and $R(s) = R_0 + \delta R(s)$, where $\langle \delta\eta(s)\rangle_0 = 0$, $\langle \delta\Phi(s)\rangle_0 = 0$ and $\langle \delta R(s)\rangle_0 = 0$. The Gibbs-Bogoliubov inequality tells us always that $F_A \leq F_A^T$ and $F_R \leq F_R^T$. Therefore, our best choice for the variational parameters $\alpha_x$ and $\alpha_y$ are those that minimize $F_A^T$, and best choice for $\beta_R$, $\alpha_R$, $\alpha_\eta$ and $\alpha_\Phi$ are those that minimize $F_R^T$; the closest choices of $F_A^T$ and $F_R^T$ to $F_A$ and $F_R$, respectively. The free energy needs also to be minimized with respect to $\eta_0$, $R_0$ and $\Delta\Phi_0(s)$. We treat $-2\pi M W r_b[x'_A(s), y'_A(s)]$ as a perturbation in Eq. (A.7), thus indeed assuming that the average braid axis is straight. This allows us to write, up to the first non-vanishing correction in the perturbation series,

$$-k_B T \ln Z_{0,A} \approx -k_B T \ln Z_x - k_B T \ln Z_y - \frac{(2\pi M)^2}{2k_B T}\langle\langle W r_b^2\rangle_x\rangle_y, \quad (A.17)$$

where

$$Z_x = \int \mathcal{D}x'_A(s)\exp\left(\int_{-L_b/2}^{L_b/2} ds_0 \left(l_p\left(\frac{dx'_A(s_0)}{ds_0}\right)^2 + \alpha_x x'_A(s_0)^2\right)\right), \tag{A.18}$$

$$Z_y = \int \mathcal{D}y_A(s)\exp\left(\int_{-L_b/2}^{L_b/2} ds_0 \left(l_p\left(\frac{dy'_A(s_0)}{ds_0}\right)^2 + \alpha_x y'_A(s_0)^2\right)\right). \tag{A.19}$$

The evaluation of $\left\langle\left\langle Wr_b^2\right\rangle_x\right\rangle_y$ yields (see section 7 of Ref. [36] to see details of the calculation)

$$\left\langle\left\langle Wr_b^2\right\rangle_x\right\rangle_y = \frac{1}{8(2\pi)^2 l_p^{3/2}} \frac{L_b}{\cos\left(\frac{\eta_0}{2}\right)^4} \frac{1}{\left(\alpha_y^{1/2} + \alpha_x^{1/2}\right)}, \tag{A.20}$$

whereas $-k_B T \ln Z_x$ and $-k_B T \ln Z_y$ are found by considering their derivatives with respect to $\alpha_x$ and $\alpha_y$ namely

$$-\frac{\partial \ln Z_x}{\partial \alpha_x} = L_b \left\langle x'_A(s_0)^2\right\rangle = \frac{L_b}{4\alpha_x^{1/2} l_p^{1/2}}, \qquad -\frac{\partial \ln Z_y}{\partial \alpha_y} = L_b \left\langle y'_A(s_0)^2\right\rangle = \frac{L_b}{4\alpha_y^{1/2} l_p^{1/2}}. \tag{A.21}$$

Integrating up both expressions in Eq. (A.21) then yields

$$-\ln Z_{0,A} \approx \frac{L_b \alpha_x^{1/2}}{2l_p^{1/2}} + \frac{L_b \alpha_y^{1/2}}{2l_p^{1/2}} + \Theta_{x,y} - \frac{M^2}{(k_B T)^2} \frac{1}{16 l_p^{3/2}} \frac{L_b}{\cos\left(\frac{\eta_0}{2}\right)^4} \frac{1}{\left(\alpha_y^{1/2} + \alpha_x^{1/2}\right)}, \tag{A.22}$$

where $\Theta_{x,y}$ is an arbitrary constant of integration that may be discarded. Therefore, we may fully express

$$\frac{F_A^T}{L_b k_B T} \approx \frac{\alpha_x^{1/2}}{4l_p^{1/2}} + \frac{\alpha_y^{1/2}}{4l_p^{1/2}} - \frac{M^2}{(k_B T)^2} \frac{1}{16 l_p^{3/2}} \frac{1}{\cos\left(\frac{\eta_0}{2}\right)^4} \frac{1}{\left(\alpha_y^{1/2} + \alpha_x^{1/2}\right)} + \left(\frac{1}{8\alpha_x^{1/2} l_p^{1/2}} + \frac{1}{8\alpha_y^{1/2} l_p^{1/2}}\right) \frac{F}{k_B T \cos\left(\frac{\eta_0}{2}\right)}. \tag{A.23}$$

We can then minimize over $\alpha_x$ and $\alpha_y$. If the correction due to $\left\langle\left\langle Wr_b^2\right\rangle_x\right\rangle_y$ is small, we can neglect it in the minimization. Then, simply, we have that

$$\alpha_x = \alpha_y \approx \frac{F}{2k_B T \cos\left(\frac{\eta_0}{2}\right)}. \tag{A.24}$$

Substituting back Eq. (A.24) into (A.23) gives us the expression for $f_{x,y}$ (Eq. (2.44)) and the last term in $f_W$ (Eq. (2.45) ).

Now let us focus our attention on $F_R^T$. The contributions to $-\ln Z_{0,R}$ are evaluated by considering its derivatives with respect to $\beta_R$, $\alpha_R$, $\alpha_\eta$ and $\alpha_\Phi$. All of these can be written as

$$-\left(\frac{\partial \ln Z_{0,R}}{\partial \alpha_\eta}\right)_{\alpha_R,\beta_R,\alpha_\Phi} = \frac{L_b d_\eta^2}{2}, \quad -\left(\frac{\partial \ln Z_{0,R}}{\partial \alpha_R}\right)_{\alpha_\eta,\beta_R,\alpha_\Phi} = \frac{L_b d_R^2}{2}, \quad -\left(\frac{\partial \ln Z_{0,R}}{\partial \beta_R}\right)_{\alpha_\eta,\alpha_R,\alpha_\Phi} = \frac{L_b \theta_R^2}{2},$$

and

$$-\left(\frac{\partial \ln Z_{0,R}}{\partial \alpha_\Phi}\right)_{\alpha_R,\beta_R,\alpha_\eta} = \frac{L_b d_\Phi^2}{2}. \tag{A.25}$$

The parameters $d_\eta^2$, $d_R^2$, $\theta_R^2$ and $d_\Phi^2$ are given through the following expressions

$$d_\eta^2 = \langle \delta\eta(s)^2 \rangle_0 = \frac{1}{2\pi}\int_{-\infty}^{\infty} \frac{k^2 dk}{\frac{l_p}{2}k^4 + \alpha_\eta}, \qquad d_R^2 = \langle \delta R(s)^2 \rangle_0 = \frac{1}{2\pi}\int_{-\infty}^{\infty} \frac{dk}{\frac{l_p}{2}k^4 + \beta_R k^2 + \alpha_R},$$

$$\theta_R^2 = \left\langle \left(\frac{d\delta R(s)}{ds}\right)^2 \right\rangle_0 = \frac{1}{2\pi}\int_{-\infty}^{\infty} \frac{k^2 dk}{\frac{l_p}{2}k^4 + \beta_R k^2 + \alpha_R}, \quad d_\Phi^2 = \langle \delta\Phi(s)^2 \rangle = \frac{1}{2\pi}\int_{-\infty}^{\infty} dk \frac{1}{\frac{l_c}{2}k^2 + \alpha_\Phi}.$$

$$\tag{A.26}$$

Solving both Eqs. (A.25) and (A.26) allows us to obtain $\ln Z_{0,R}$, for details of how this is done see Section 10 of Ref. [36].

The averages $\langle E_R[R(s),\eta(s),\Delta\Phi(s)] \rangle_0$ may be calculated using a general result for Gaussian path integration

$$\left\langle F\left(X(s), \frac{dX(s)}{ds}\right) \right\rangle_0 = \frac{1}{2\pi d_X \theta_X}\int_{-\infty}^{\infty} dx \int_{-\infty}^{\infty} dx' F(x,x')\exp\left(-\frac{1}{2}\frac{x^2}{d_X^2}\right)\exp\left(-\frac{1}{2}\frac{x'^2}{\theta_X^2}\right), \tag{A.27}$$

where $X(s)$ may correspond to any of the functions $x_A(s)$, $y_A(s)$, $\delta R(s)$, $\delta\eta(s)$ or $\delta\Phi(s)$, and $F(X(s),X'(s))$ represents an completely general function of $X(s)$, as well as its derivative respect to $s$, $X'(s)$. The steps in deriving Eq. (A.27) can either be found in the supplemental material to Ref. [36] or within Appendix B of Ref. [38]. Expressions for the other averages over $R(s)$ and $\eta(s)$ are relatively straightforward from Eq. (A.27). Here, we will focus on the averages $\langle \cos n\Delta\Phi(s) \rangle_0$. Using Eq. (A.27), these averages over $\Delta\Phi(s)$ can be written as

$$\langle \cos n\Delta\Phi(s)\rangle_0 = \exp\left(-\frac{n^2 d_R^2}{2}\right)\cos n\Delta\Phi_0(s). \tag{A.28}$$

Also, note that we may write

$$\left\langle\left(\frac{d\Delta\Phi(s)}{ds}-\Delta g^0(s)\right)^2\right\rangle_0 - \left\langle\left(\frac{d\delta\Phi(s)}{ds}\right)^2\right\rangle_0 = \left(\frac{d\Delta\Phi_0(s)}{ds}-\Delta g^0(s)\right)^2, \tag{A.29}$$

since $\langle \delta\Phi'(s)\rangle_0 = 0$. Now, functional minimization over $\Delta\Phi_0(s)$ leads to a non-linear equation that cannot, in general, be solved analytically for helix specific forces [36]. One notable exception is in the case where helix specific forces are not present, i.e. the averages described by Eq. (A.28) are not present, or we have that $\Delta g^0(s) = 0$ (identical sequences in perfect alignment). In both cases the solution is rather trivial, is simply satisfies the equation:

$$\frac{d\Delta\Phi_0(s)}{ds} = \Delta g^0(s). \tag{A.30}$$

However, to consider more general cases, one can write down a trial function $\Delta\Phi_0(s)$. The form of this is based on the linear response theory, valid when $\Delta\Phi_0(s)-\Delta\bar{\Phi}$ is considered to be small, where $\Delta\bar{\Phi} = \langle \Delta\Phi_0(s)\rangle_{\Delta g}$. Here, the subscript $\Delta g$, on the averaging brackets, denotes that the average is an ensemble average over all realizations of $\Delta g^0(s)$. This trial function is

$$\Delta\Phi_0(s) \approx \Delta\bar{\Phi} + \frac{1}{2}\int_{-\infty}^{\infty}\frac{(s-s')}{|s-s'|}\Delta g^0(s')\exp\left(-\frac{|s-s'|}{\tilde{\lambda}_h}\right). \tag{A.31}$$

We substitute Eq. (A.31) into the free energy approximation $F_R^T$ and average it over the realizations of $\Delta g^0(s)$. In doing so, we may make use of the results

$$\langle \cos n\Delta\Phi_0(s)\rangle_{\Delta g} = \exp\left(-\frac{n^2\tilde{\lambda}_h}{4\lambda_c}\right)\cos n\Delta\bar{\Phi}, \tag{A.32}$$

and

$$\left\langle\left(\frac{d\Delta\Phi_0(s)}{ds}-\Delta g^0(s)\right)^2\right\rangle_{\Delta g} = \frac{l_c}{8\tilde{\lambda}_h\lambda_c^{(0)}}. \tag{A.33}$$

Eqs. (A.32) and (A.33) are derived from Eqs. (2.16) and (2.17), as well as the fact that $\Delta g^0(s)$ is Gaussian distributed. A derivation of Eq. (A.32) is given by Eq. (13.9) of Ref. [36]. We now minimize $F_R^T$ with respect to $\tilde{\lambda}_h$ and $\Delta\bar{\Phi}$, as well as $\eta_0$, $R_0$, $\beta_R$, $\alpha_R$, $\alpha_\eta$ and $\alpha_\Phi$. One may write $d_\Phi^2 = \lambda_h/l_c$, and is able to show (Section 13 of Ref [36], or supplemental material of Ref [38]) that the choice $\lambda_h = \tilde{\lambda}_h$ minimises the free

energy. Then, it more convenient to rewrite all of our expressions in terms of a combined persistence length

$$\frac{1}{\lambda_c} = \frac{1}{\lambda_c^{(0)}} + \frac{1}{l_c},$$  (A.34)

and a rescaled adaptation length

$$\frac{\lambda_h^*}{2\lambda_c} = \frac{\lambda_h}{2l_c} + \frac{\lambda_h}{4\lambda_c^{(0)}}.$$  (A.35)

Following these steps in calculating $\left\langle F_R^T \right\rangle_{\Delta g}$ then yields expressions for $f_{conf}$ (Eq. (2.41)), $f_{bend}$ (Eq.(2.42) ), $f_{rep}$ (Eq. (2.46)), $f_{attr}$ (Eq. (2.49) or Eq. (2.53)) and the reminder of $f_W$ (Eq. (2.45)). For more explicit details of this calculation, as applied to more general cases, see Ref. [36].

# Appendix B: moment and extension graphs for the complete range of parameter values investigated

All plots are calculated with the parameter choices $L = 55000\text{Å}$ and $b = 7000\text{Å}$.

## B.1 Attractive Non-helix specific forces

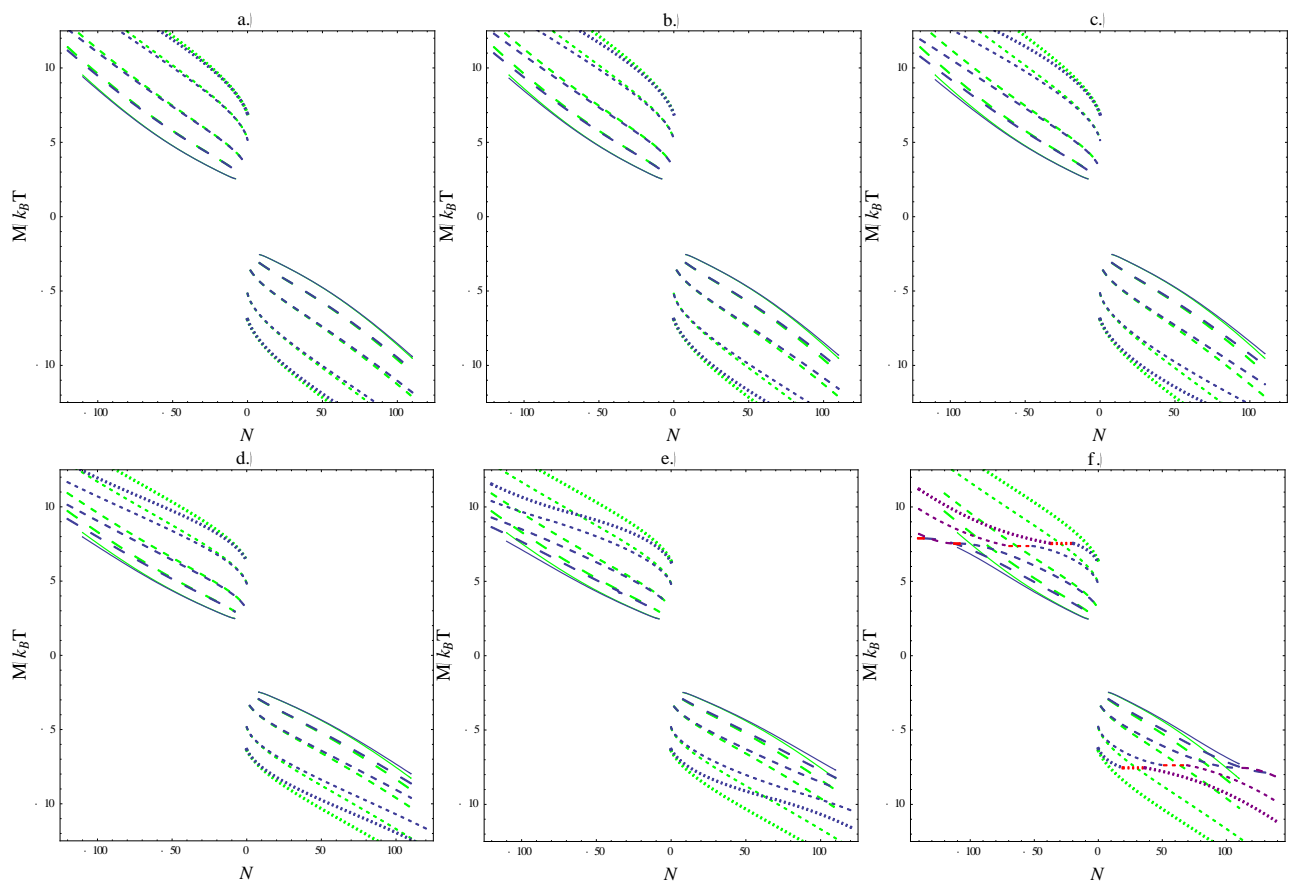

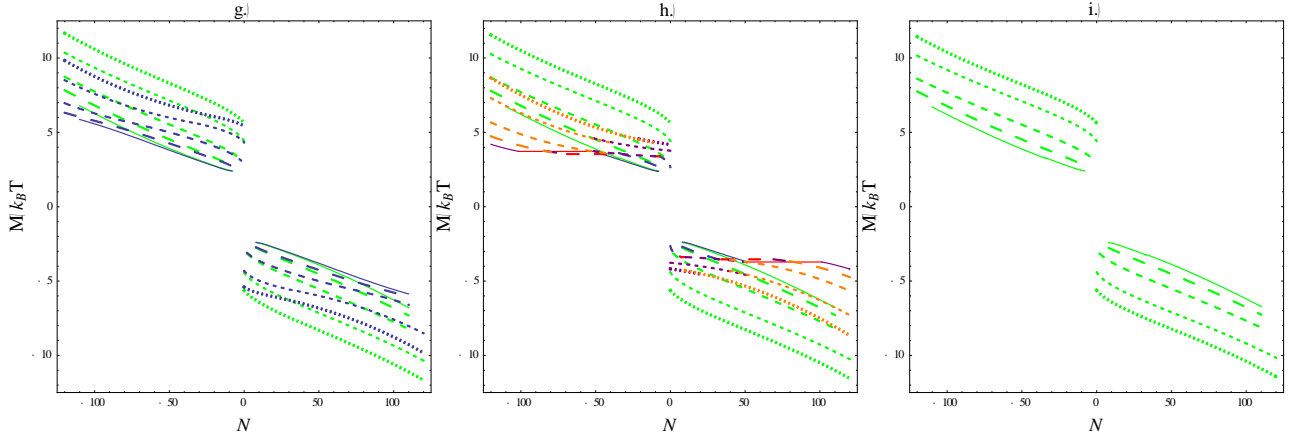

Fig. B.1 Moment curves as a function of $N$ for varying force and interaction parameter values. A non-helix specific attractive interaction term is considered here. All plots here are generated at a value of Debye screening length $\kappa_D^{-1} = 15.5\text{Å}$. The parameter values for each panel are: a.) $\theta = 0.6$, $f_N = 0.3$; b.) $\theta = 0.6$, $f_N = 0.5$; c.), $\theta = 0.6$, $f_N = 0.7$; d.) $\theta = 0.7$, $f_N = 0.3$; e.) $\theta = 0.7$, $f_N = 0.5$; f.) $\theta = 0.7$, $f_N = 0.7$; g.) $\theta = 0.8$, $f_N = 0.3$, h.) $\theta = 0.8$, $f_N = 0.5$; and i.) $\theta = 0.8$, $f_N = 0.7$. The green curves are for the choice $\kappa_{NH}^{-1} = 4.8\text{Å}$, whereas all other colours correspond to the choice $\kappa_{NH} = 2\kappa_D$. The purple and orange colours refer to a tightly braided sate, whereas the blue to a looser braided state; the latter state may collapse into the former. The difference between the purple and orange is the choice of $d_{max}$. In the purple curves it is chosen to be $d_{max} = R_0 - 2a$ for the orange curves it is given by Eq. (2.11), which was argued to be a more appropriate choice for a tightly braided state. The red part of the curves (flat lines) indicates the coexistence region between the collapsed state (calculated with the choice $d_{max} = R_0 - 2a$) and the looser state. The solid, long dashed, medium dashed, short dashed and dotted lines refer to force values $F = 3.5\,\text{pN}, 7\,\text{pN}, 14\,\text{pN}, 28\,\text{pN}$ and $40\,\text{pN}$, respectively.

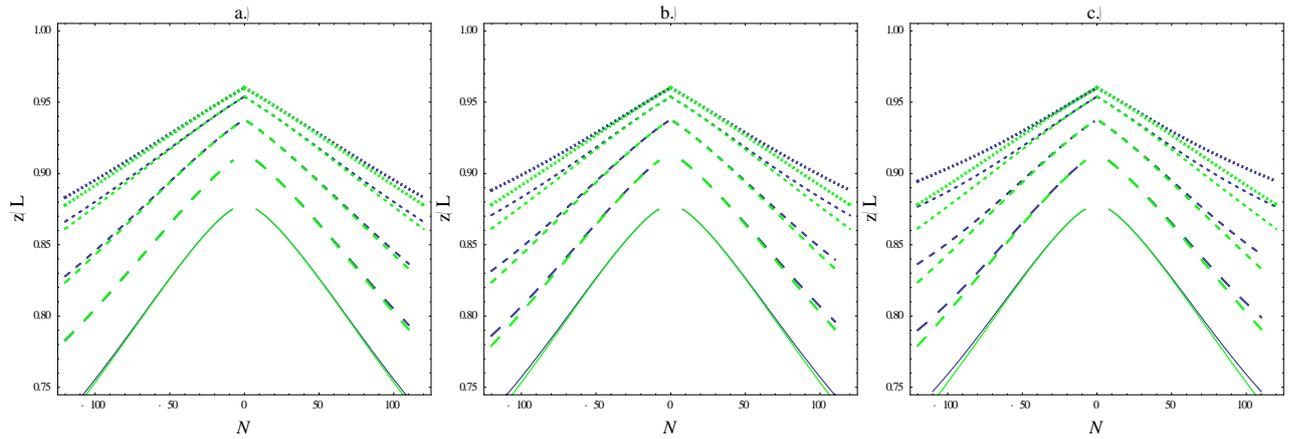

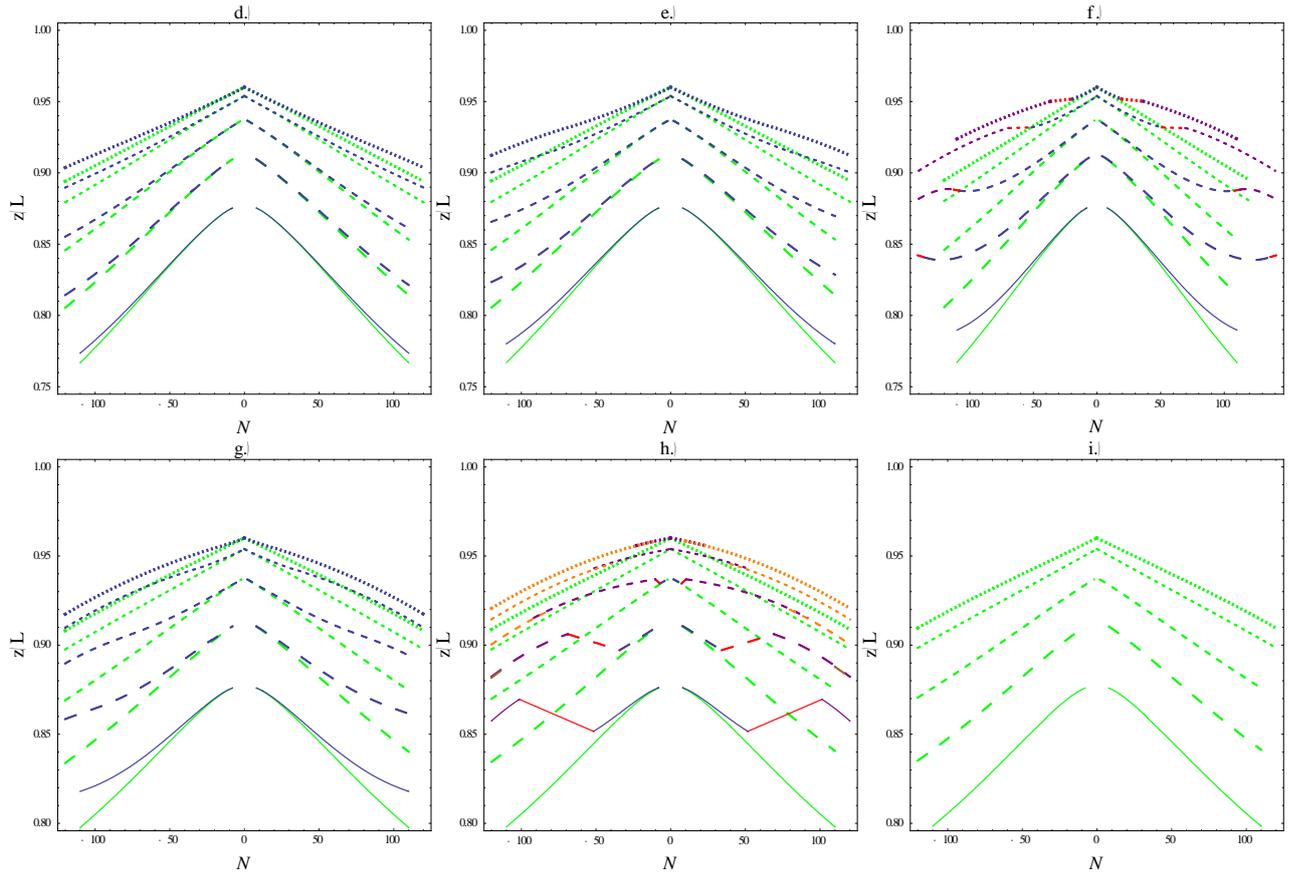

Fig. B.2 Extension curves as a function of $N$ for varying force and interaction parameter values. A non-helix specific attractive interaction term is considered here. All plots here are generated at a value of Debye screening length $\kappa_D^{-1} = 15.5\,\text{Å}$. The parameter values for each panel are: a.) $\theta = 0.6$, $f_N = 0.3$; b.) $\theta = 0.6$, $f_N = 0.5$; c.) $\theta = 0.6$, $f_N = 0.7$; d.) $\theta = 0.7$, $f_N = 0.3$; e.) $\theta = 0.7$, $f_N = 0.5$; f.) $\theta = 0.7$, $f_N = 0.7$; g.) $\theta = 0.8$, $f_N = 0.3$; h.) $\theta = 0.8$, $f_N = 0.5$; and i.) $\theta = 0.8$, $f_N = 0.7$. The green curves are for the choice $\kappa_{NH}^{-1} = 4.8\,\text{Å}$, whereas all other colours are for the choice $\kappa_{NH} = 2\kappa_D$. The purple and orange refer to a tightly braided state; whereas the blue to a looser braided state. The difference between the purple and orange is the choice of $d_{max}$. In the purple curves it is chosen to be $d_{max} = R_0 - 2a$ for the orange curves it is given by Eq. (2.11), which was argued to be more appropriate for a tightly braided state. The red part of the curves (straight lines) indicates the coexistence region between the tighter state (calculated with the choice $d_{max} = R_0 - 2a$) and the looser state. The solid, long dashed, medium dashed, short dashed and dotted lines refer to force values $F = 3.5\,\text{pN}, 7\,\text{pN}, 14\,\text{pN}, 28\,\text{pN}$ and $40\,\text{pN}$, respectively.

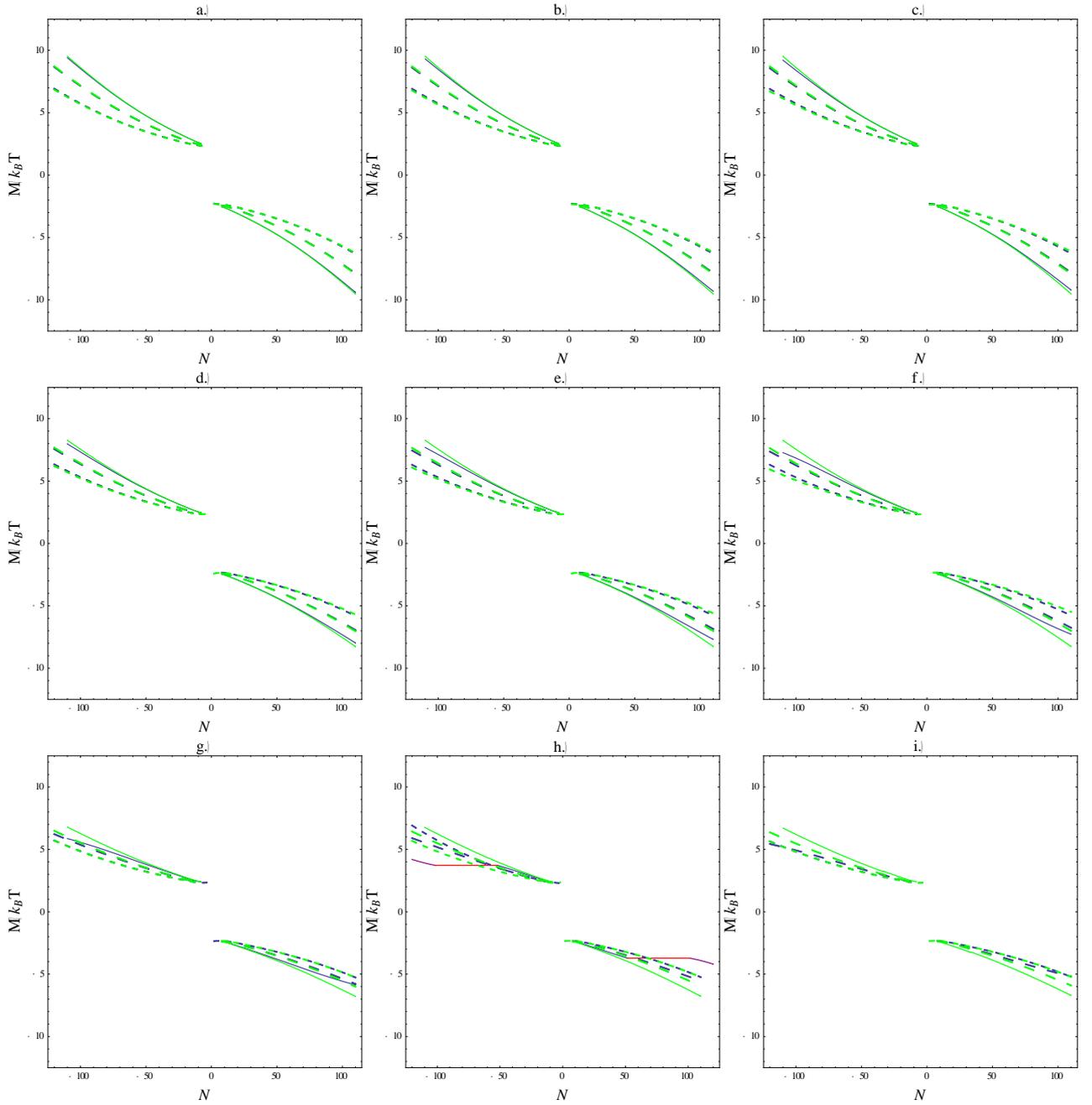

Fig. B.3 Moment curves at $F = 3.5\,\text{pN}$ as a function of $N$ for varying salt concentration and interaction parameter values. A non-helix specific interaction term is considered here. The parameter values for each panel are: a.) $\theta = 0.6$, $f_N = 0.3$; b.) $\theta = 0.6$, $f_N = 0.5$; c.) $\theta = 0.6$, $f_N = 0.7$; d.) $\theta = 0.7$, $f_N = 0.3$; e.) $\theta = 0.7$, $f_N = 0.5$; f.) $\theta = 0.7$, $f_N = 0.7$; g.) $\theta = 0.8$, $f_N = 0.3$; h.) $\theta = 0.8$, $f_N = 0.5$; and i.) $\theta = 0.8$, $f_N = 0.7$. The green curves are for the choice $\kappa_{NH}^{-1} = 4.8\,\text{Å}$, whereas all other colours correspond to the choice $\kappa_{NH} = 2\kappa_D$. The purple colour refers to a tighter braided sate, whereas the blue to a looser braided state. For the purple curves $d_{max}$ is chosen to be $d_{max} = R_0 - 2a$. The red part of the curves (flat lines) indicates the coexistence region between collapsed state and looser state. The solid, long dashed and medium dashed lines correspond to the Debye screening length values $\kappa_D^{-1} = 15.5\,\text{Å}, 11.7\,\text{Å}$ and $7.75\,\text{Å}$, respectively.

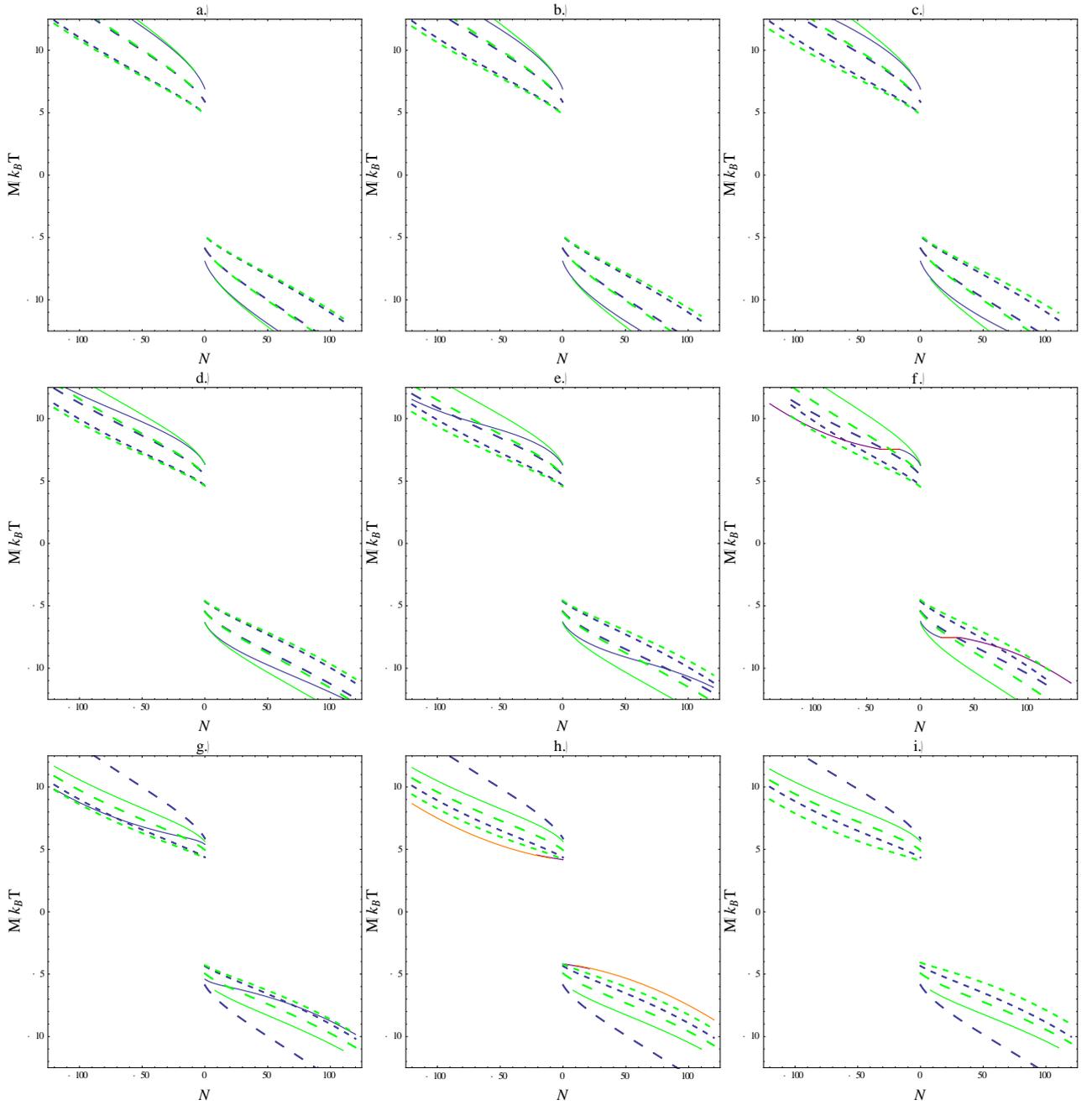

Fig. B.4 Moment curves at $F = 40\,\text{pN}$ as a function of $N$ for varying salt concentration and interaction parameter values. A non-helix specific attractive interaction term is considered here. The parameter values for each panel are: a.) $\theta = 0.6$, $f_N = 0.3$; b.) $\theta = 0.6$, $f_N = 0.5$; c.) $\theta = 0.6$, $f_N = 0.7$; d.) $\theta = 0.7$, $f_N = 0.3$; e.) $\theta = 0.7$, $f_N = 0.5$; f.) $\theta = 0.7$, $f_N = 0.7$; g.) $\theta = 0.8$, $f_N = 0.3$; h.) $\theta = 0.8$, $f_N = 0.5$; and i.) $\theta = 0.8$, $f_N = 0.7$. The green curves are for the choice $\kappa_{NH}^{-1} = 4.8\,\text{Å}$, whereas all other colours are for the choice $\kappa_{NH} = 2\kappa_D$. The purple and orange colours refer to the tightly braided sate, whereas the blue to a looser braided state. The difference between the purple and orange is the choice of $d_{\max}$. In the purple curves it is chosen to be $d_{\max} = R_0 - 2a$, for the orange curves it is given by Eq. (2.11), which was argued to be a more appropriate choice for a tightly braided state. The red part of the curves (flat lines) indicates the coexistence region between collapsed state (calculated with the choice $d_{\max} = R_0 - 2a$) and looser state. The solid, long dashed and medium dashed lines correspond to the Debye screening length values $\kappa_D^{-1} = 15.5\,\text{Å}, 11.7\,\text{Å}$ and $7.75\,\text{Å}$, respectively.

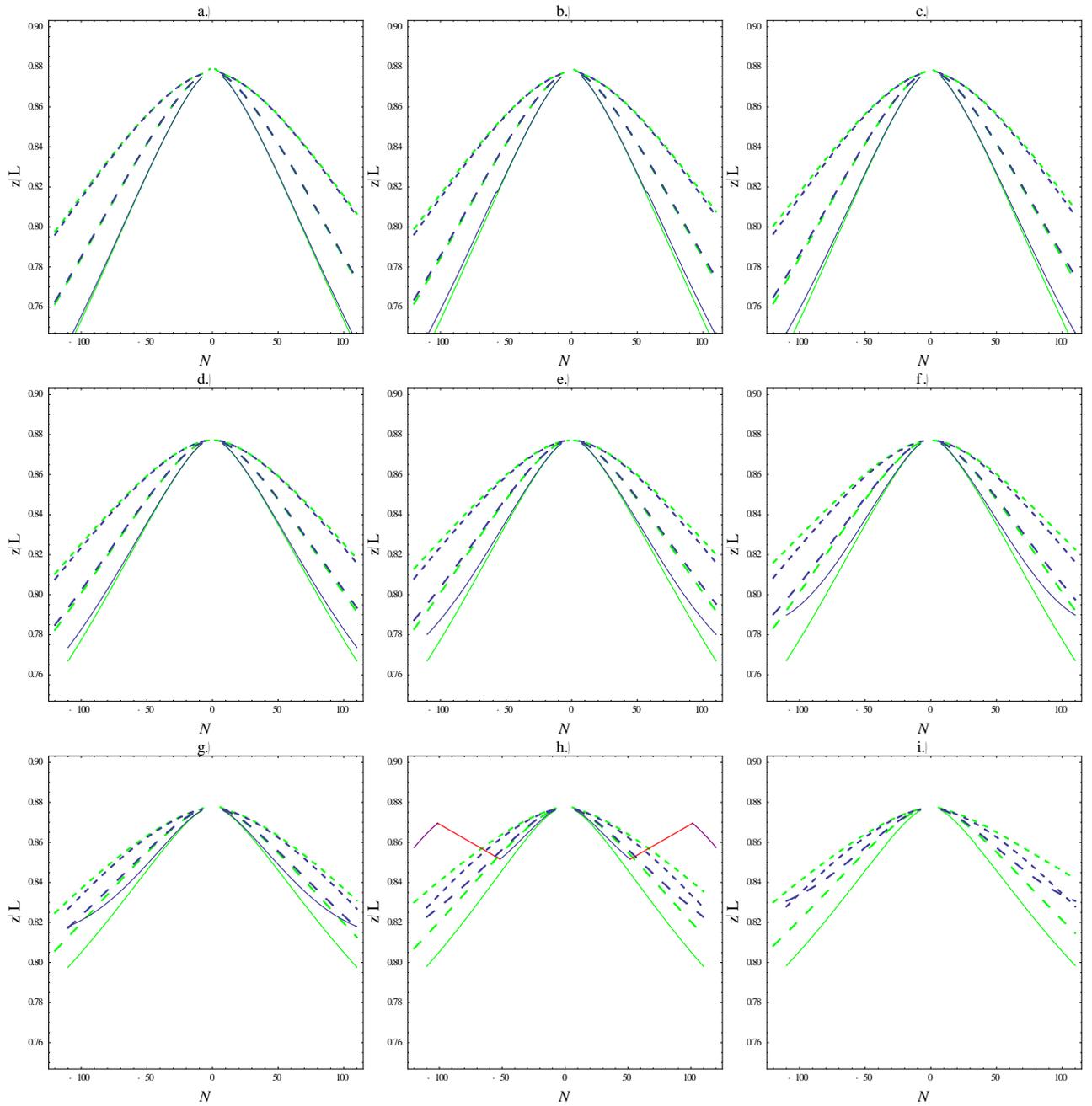

Fig. B.5 Extension curves at $F = 3.5\,\text{pN}$ as a function of $N$ for varying salt concentration and interaction parameter values. A non-helix specific attractive interaction term is considered here. The parameter values for each panel are: a.) $\theta = 0.6$, $f_N = 0.3$; b.) $\theta = 0.6$, $f_N = 0.5$; c.) $\theta = 0.6$, $f_N = 0.7$; d.) $\theta = 0.7$, $f_N = 0.3$; e.) $\theta = 0.7$, $f_N = 0.5$; f.) $\theta = 0.7$, $f_N = 0.7$; g.) $\theta = 0.8$, $f_N = 0.3$; h.) $\theta = 0.8$, $f_N = 0.5$; and i.) $\theta = 0.8$, $f_N = 0.7$. The green curves are for the decay range choice $\kappa_{NH}^{-1} = 4.8\,\text{Å}$, whereas all other colours are for the choice $\kappa_{NH} = 2\kappa_D$. The purple colour refers to a tighter braided state, whereas the blue to a looser braided state. For the purple curves $d_{\max}$ is chosen to be $d_{\max} = R_0 - 2a$. The red part of the curves (straight lines) indicates the coexistence region between tighter state and looser state. The solid, long dashed and medium dashed lines correspond to the Debye screening length values $\kappa_D^{-1} = 15.5\,\text{Å}, 11.7\,\text{Å}$ and $7.75\,\text{Å}$, respectively.

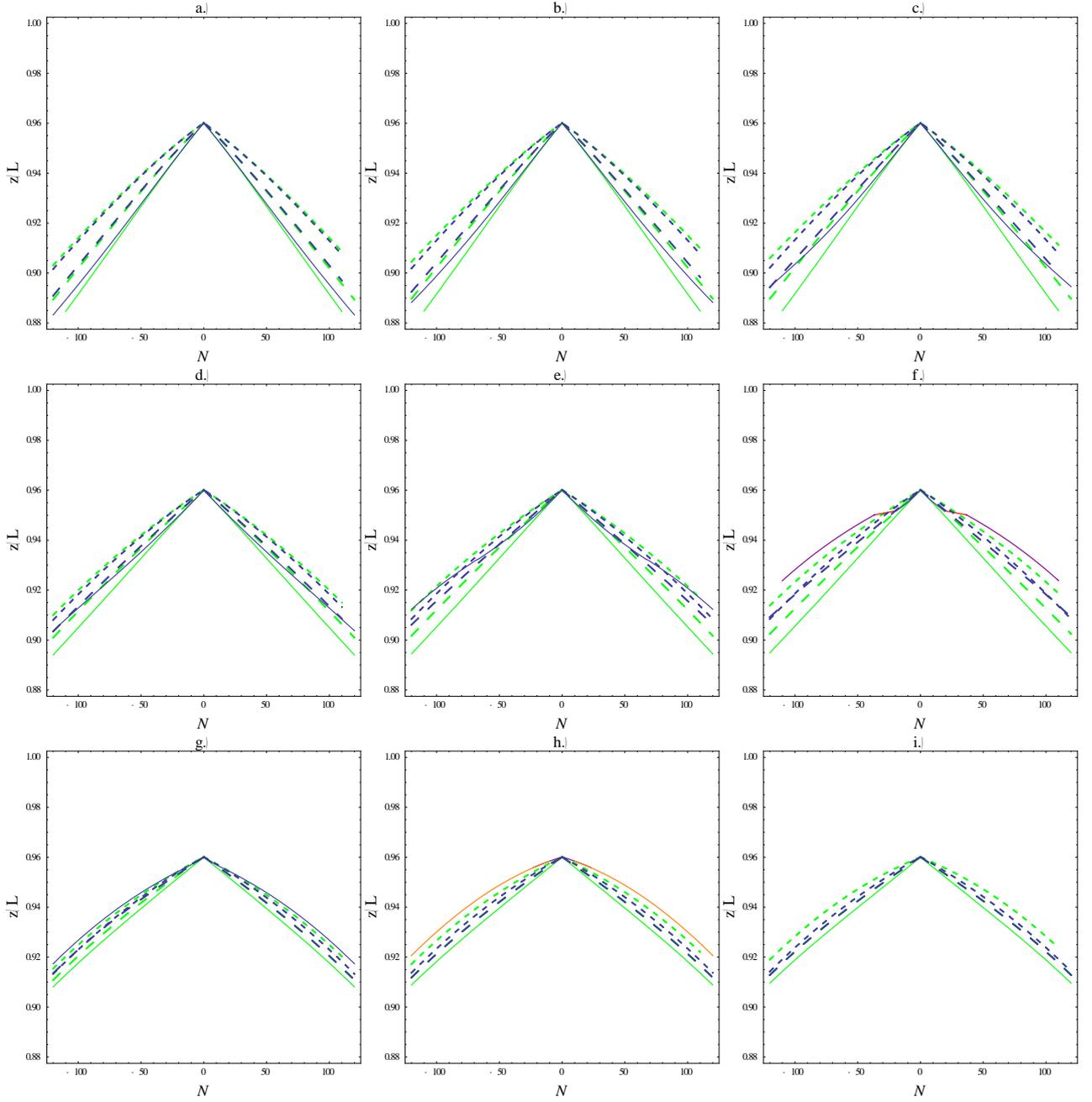

Fig. B.6 Extension curves at $F = 40\,\text{pN}$ as a function of $N$ for varying salt concentration and interaction parameter values. A non-helix specific attractive interaction term is considered here. The parameter values for each panel are: a.) $\theta = 0.6$, $f_N = 0.3$; b.) $\theta = 0.6$, $f_N = 0.5$; c.) $\theta = 0.6$, $f_N = 0.7$; d.) $\theta = 0.7$, $f_N = 0.3$; e.) $\theta = 0.7$, $f_N = 0.5$; f.) $\theta = 0.7$, $f_N = 0.7$; g.) $\theta = 0.8$, $f_N = 0.3$; h.) $\theta = 0.8$, $f_N = 0.5$; and i.) $\theta = 0.8$, $f_N = 0.7$. The green curves are for the choice $\kappa_{NH}^{-1} = 4.8\,\text{Å}$, whereas all other colours are for the choice $\kappa_{NH} = 2\kappa_D$. The purple and orange colours refer to a tighter braided state, whereas the blue to a looser braided state. The difference between the purple and orange colours is the choice of $d_{\max}$, in the purple curves it is chosen to be $d_{\max} = R_0 - 2a$ for the orange curves it is given by Eq. (2.11), which was argued to be more appropriate for a tightly braided state. The red part of the curves (straight lines) indicates the coexistence region between collapsed state (calculated with the choice $d_{max} = R_0 - 2a$) and looser state. The solid, long dashed and medium dashed lines correspond to the Debye screening length values $\kappa_D^{-1} = 15.5\text{Å}, 11.7\text{Å}$ and $7.75\text{Å}$, respectively.

## B.2 Attractive Helix specific forces

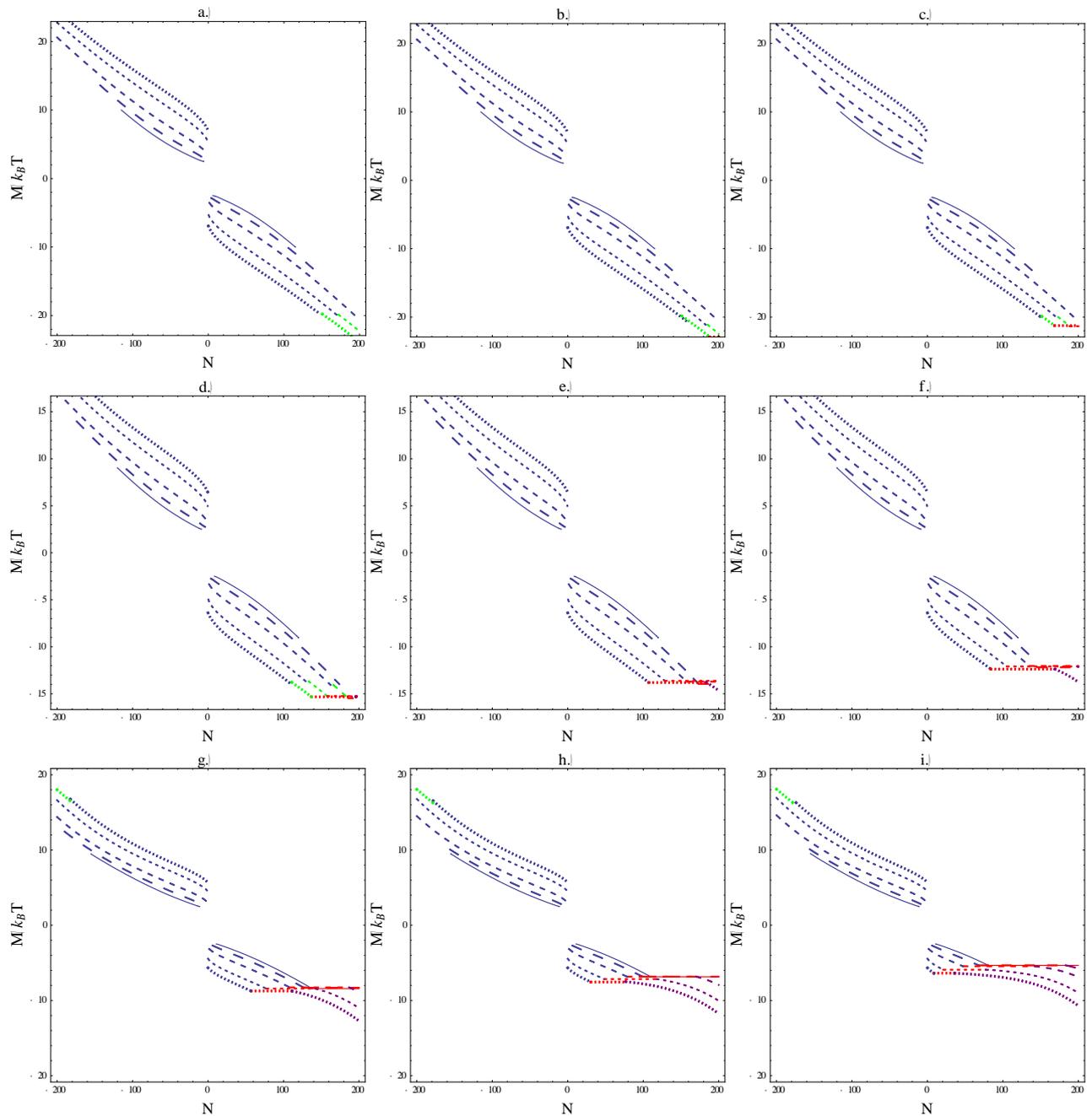

Fig. B.7 Moment curves as a function of $N$ for varying force and interaction parameter values. An attractive helix specific interaction term, calculated using Eqs. (2.53) and (2.55), is considered here. All plots here are generated at a Debye screening length value $\kappa_D^{-1} = 15.5\text{Å}$. The parameter values for each panel are: a.) $\theta = 0.6$, $f_H = 0.5$; b.) $\theta = 0.6$, $f_H = 0.7$; c.) $\theta = 0.6$, $f_H = 0.9$; d.) $\theta = 0.7$, $f_H = 0.5$; e.) $\theta = 0.7$, $f_H = 0.7$; f.) $\theta = 0.7$, $f_H = 0.9$; g.) $\theta = 0.8$, $f_H = 0.5$, h.) $\theta = 0.8$, $f_H = 0.7$ and i.) $\theta = 0.8$, $f_H = 0.9$. The blue curves correspond to the $\lambda_h^* = \infty$ state, the green curves to the finite $\lambda_h^*$, $\Delta\bar{\Phi} = 0$ state, and the purple curves to the $\lambda_h^*$, $\Delta\bar{\Phi} \approx \pi/2$ state. The red

lines correspond to the coexistence regions, where the two states at the end points coexist, and where the moment stays constant. The solid, long dashed, medium dashed, short dashed and dotted lines refer to force values $F = 3.5\,\text{pN}, 7\,\text{pN}, 14\,\text{pN}, 28\,\text{pN}$ and $40\,\text{pN}$, respectively.

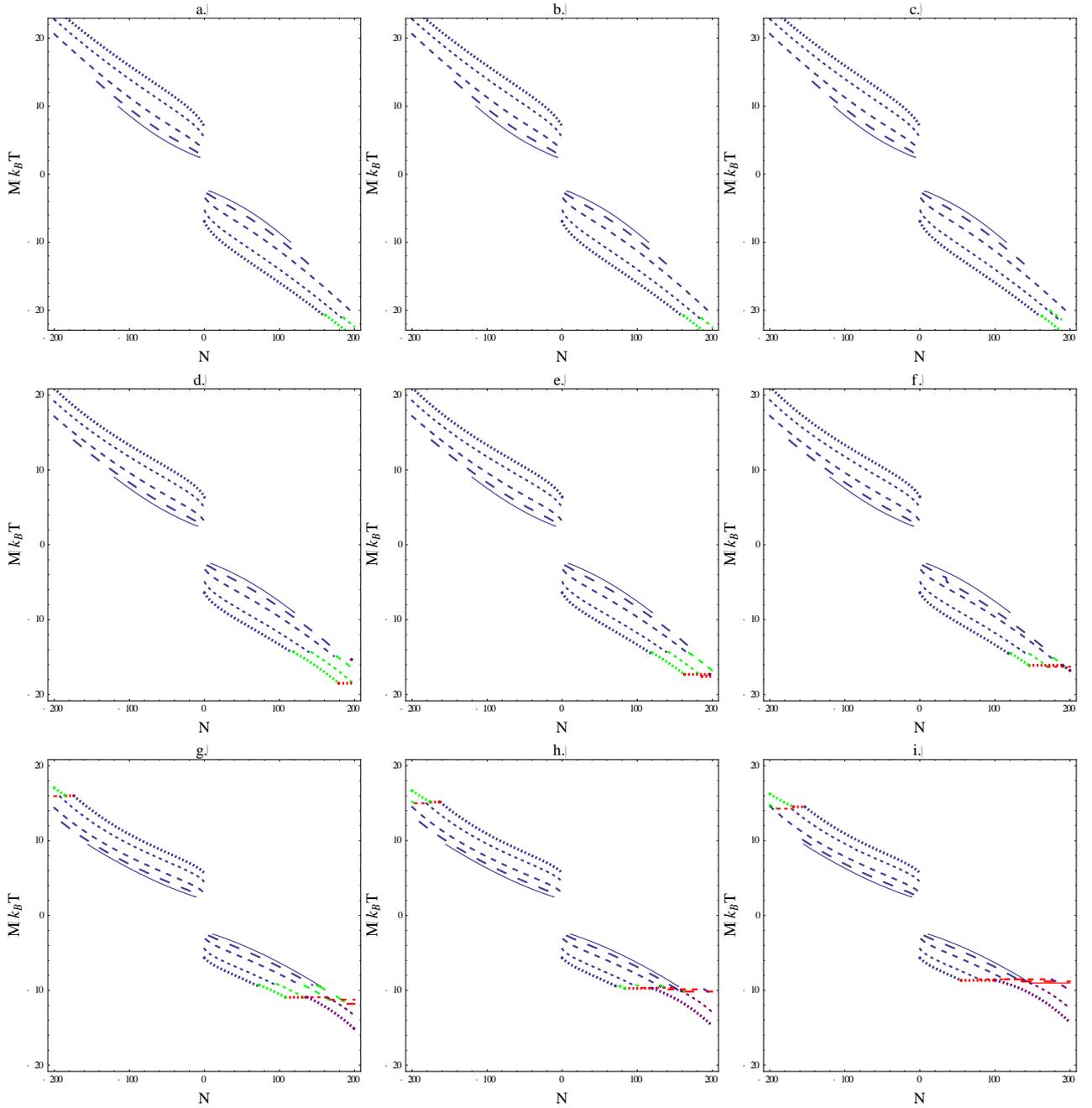

Fig. B.8. Moment curves as a function of $N$ for varying force and interaction parameter values. An attractive helix specific interaction term of the form of Eqs. (2.53) and (2.56) is considered here. This supposes that $\sin \eta(s)$ is limited by the helix geometry. All plots here are generated using a Debye screening value $\kappa_D^{-1} = 15.5\,\text{Å}$. The parameter values for each panel are: a.) $\theta = 0.6$, $f_H = 0.5$; b.) $\theta = 0.6$, $f_H = 0.7$; c.) $\theta = 0.6$, $f_H = 0.9$; d.) $\theta = 0.7$, $f_H = 0.5$; e.) $\theta = 0.7$, $f_H = 0.7$; f.) $\theta = 0.7$, $f_H = 0.9$; g.) $\theta = 0.8$, $f_H = 0.5$, h.) $\theta = 0.8$, $f_H = 0.7$ and i.) $\theta = 0.8$, $f_H = 0.9$. The blue curves correspond to the $\lambda_h^* = \infty$ state, the green curves to the finite $\lambda_h^*$,

$\Delta\bar{\Phi} = 0$ state, and the purple curves to the finite $\lambda_h^*$, $\Delta\bar{\Phi} \approx \pi/2$ state. The lines correspond to coexistence regions, where the two states at the end points coexist, and where the moment is constant. The solid, long dashed, medium dashed, short dashed and dotted lines refer to force values $F = 3.5\,\text{pN}, 7\,\text{pN}, 14\,\text{pN}, 28\,\text{pN}$ and $40\,\text{pN}$, respectively.

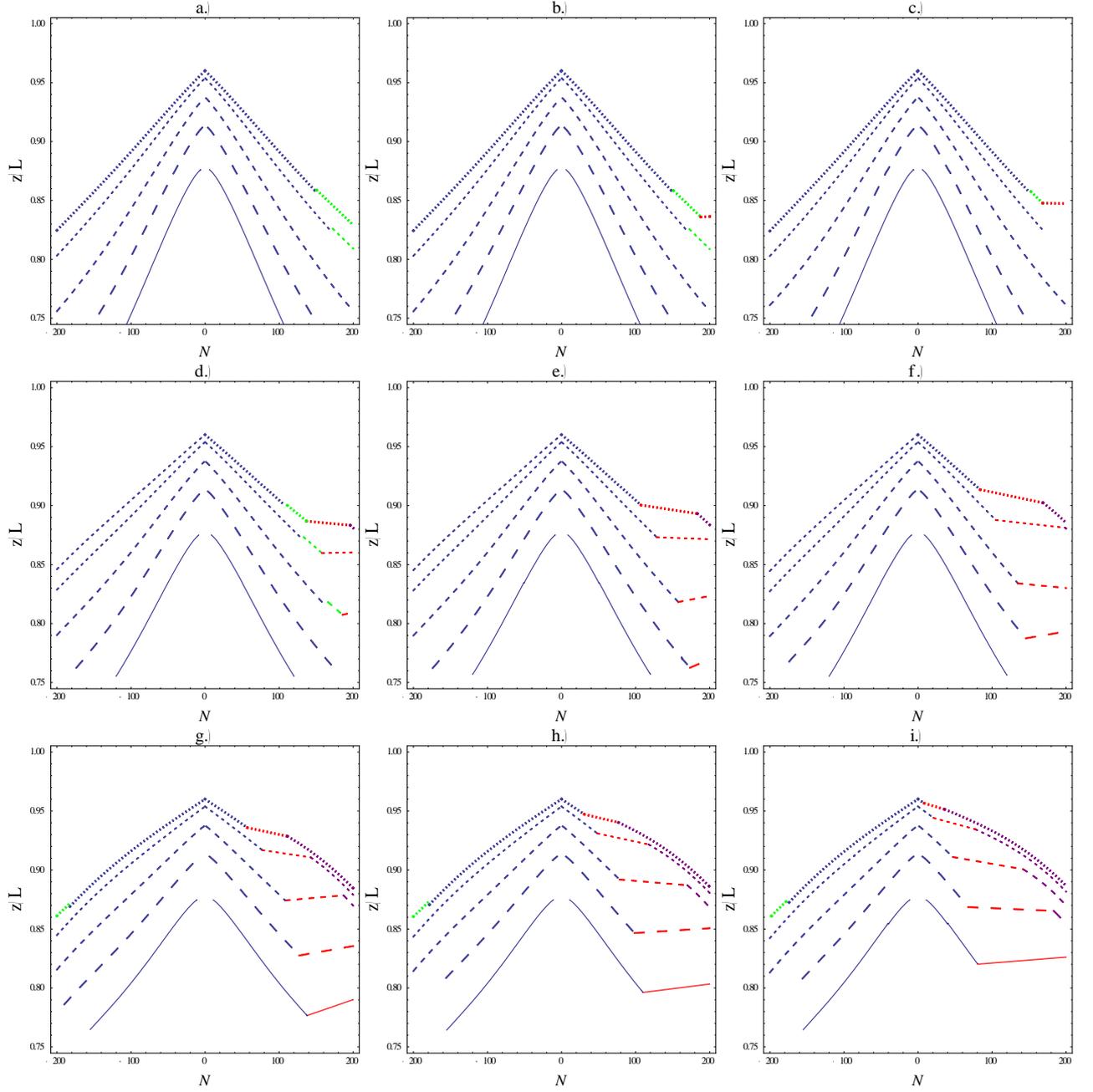

Fig. B.9 Extension curves as a function of $N$ for varying force and interaction parameter values. An attractive helix specific interaction term described by Eq. (2.53) and (2.55) is considered here. All plots are generated at a value $\kappa_D^{-1} = 15.5\,\text{Å}$. The parameter values for each panel are: a.) $\theta = 0.6$, $f_H = 0.5$; b.) $\theta = 0.6$, $f_H = 0.7$; c.) $\theta = 0.6$, $f_H = 0.9$; d.) $\theta = 0.7$, $f_H = 0.5$; e.) $\theta = 0.7$, $f_H = 0.7$; f.) $\theta = 0.7$, $f_H = 0.9$; g.) $\theta = 0.8$, $f_H = 0.5$, h.) $\theta = 0.8$, $f_H = 0.7$ and i.) $\theta = 0.8$, $f_H = 0.9$. The blue curves correspond to the $\lambda_h^* = \infty$ state, the green curves to the finite $\lambda_h^*$, $\Delta\bar{\Phi}=0$ state, and the purple curves to the finite $\lambda_h^*$, $\Delta\bar{\Phi}\approx\pi/2$ state. The red lines correspond to extension curves in the coexistence regions, where the two states at the end points coexist, and behave in a linear fashion. The solid, long dashed, medium dashed, short dashed and dotted lines refer to force values $F=3.5\,\text{pN}, 7\,\text{pN}, 14\,\text{pN}, 28\,\text{pN}$ and $40\,\text{pN}$, respectively.

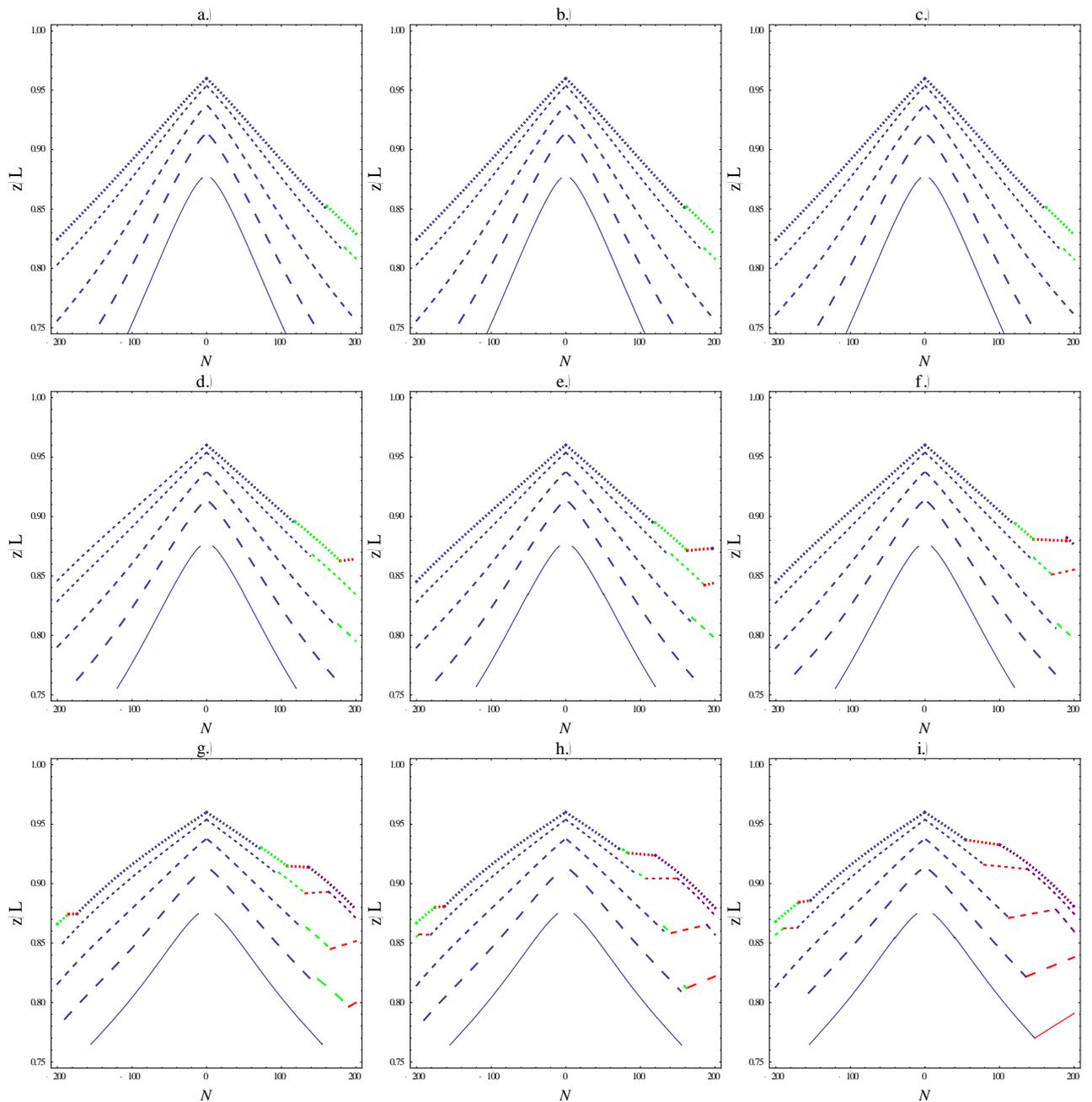

Fig. B.10 Extension curves as a function of $N$ for varying force and interaction parameter values. An attractive helix specific interaction term described by Eqs. (2.53) and (2.56) is considered here. This supposes that $\sin\eta(s)$ is limited by the helix geometry. All plots are generated using a Debye screening value $\kappa_D^{-1}=15.5\,\text{Å}$. The parameter values for each panel are: a.) $\theta=0.6$, $f_H=0.5$; b.) $\theta=0.6$, $f_H=0.7$; c.) $\theta=0.6$, $f_H=0.9$; d.) $\theta=0.7$, $f_H=0.5$; e.)

$\theta = 0.7$, $f_H = 0.7$ ; f.) $\theta = 0.7$, $f_H = 0.9$; g.) $\theta = 0.8$, $f_H = 0.5$, h.) $\theta = 0.8$, $f_H = 0.7$ and i.) $\theta = 0.8$, $f_H = 0.9$. The blue curves correspond to the $\lambda_h^* = \infty$ state, the green curves to the finite $\lambda_h^*$, $\Delta\bar{\Phi} = 0$ state, and the purple curves to the finite $\lambda_h^*$, $\Delta\bar{\Phi} \approx \pi/2$ state. The red lines correspond to extension in the coexistence regions, where the two states at the end points coexist, which behaves in a linear fashion. The solid, long dashed, medium dashed, short dashed and dotted lines refer to force values $F = 3.5\,\text{pN}, 7\,\text{pN}, 14\,\text{pN}, 28\,\text{pN}$ and $40\,\text{pN}$, respectively.

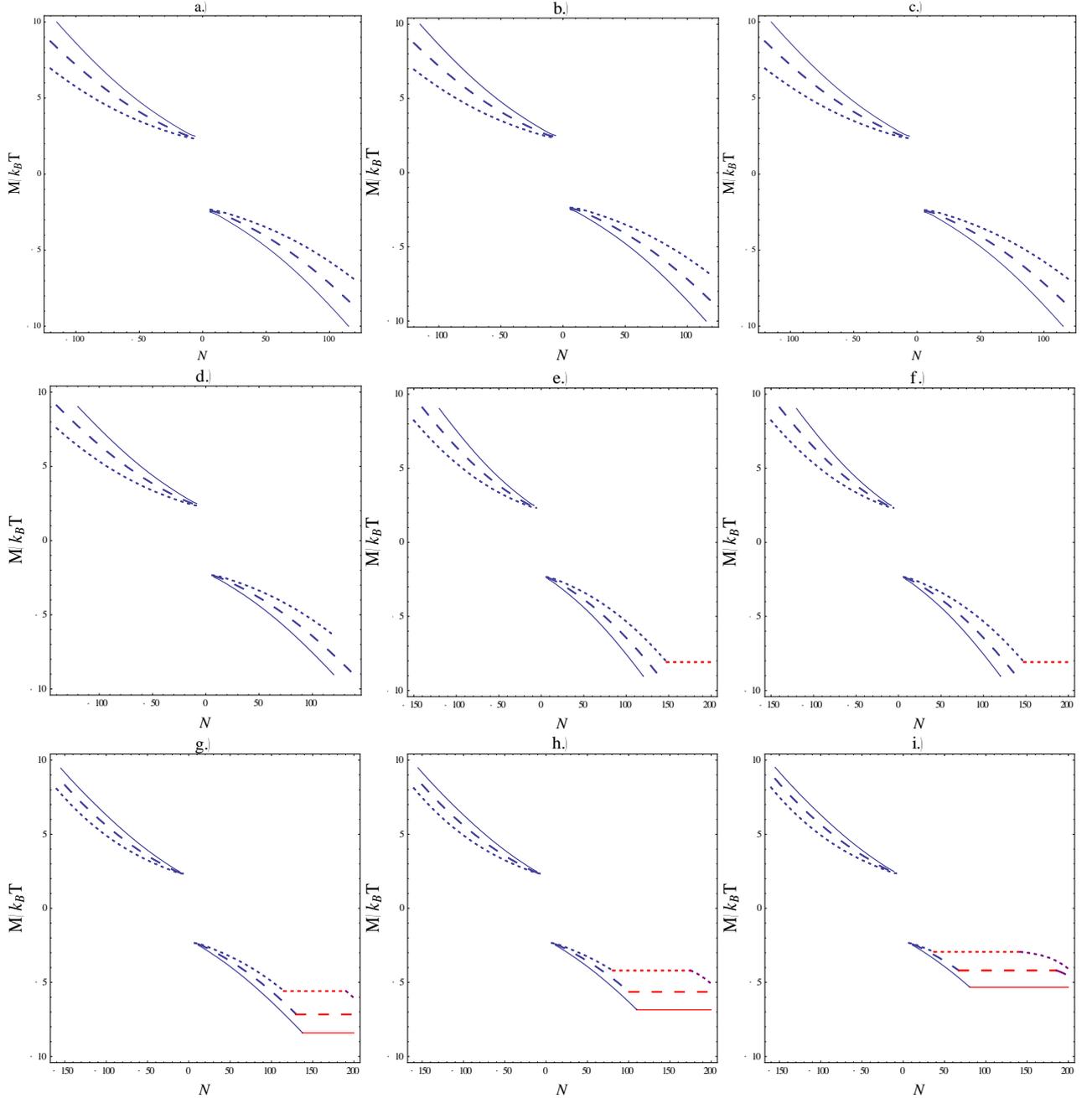

Fig. B.11. Moment curves at $F = 3.5\,\text{pN}$ as a function of $N$ for varying salt concentration and interaction parameter values. An attractive helix specific interaction term described by Eqs. (2.53) and (2.55) is considered here. The parameter values for each panel are: a.) $\theta = 0.6$, $f_H = 0.5$ ; b.) $\theta = 0.6$, $f_H = 0.7$; c.) $\theta = 0.6$, $f_H = 0.9$; d.) $\theta = 0.7$, $f_H = 0.5$; e.) $\theta = 0.7$, $f_H = 0.7$ ; f.) $\theta = 0.7$, $f_H = 0.9$; g.) $\theta = 0.8$, $f_H = 0.5$, h.) $\theta = 0.8$, $f_H = 0.7$

and i.) $\theta = 0.8$, $f_H = 0.9$. The blue curves correspond to the $\lambda_h^* = \infty$ state, the green curves to the finite $\lambda_h^*$, $\Delta\bar{\Phi} = 0$ state, and the purple curves to the $\lambda_h^*$, $\Delta\bar{\Phi} \approx \pi/2$ state. The red lines correspond to coexistence regions, where the two states at the end points coexist, and where the moment is constant. The solid, long dashed and medium dashed lines correspond to the Debye screening length values $\kappa_D^{-1} = 15.5\text{Å}, 11.7\text{Å}$ and $7.75\text{Å}$, respectively.

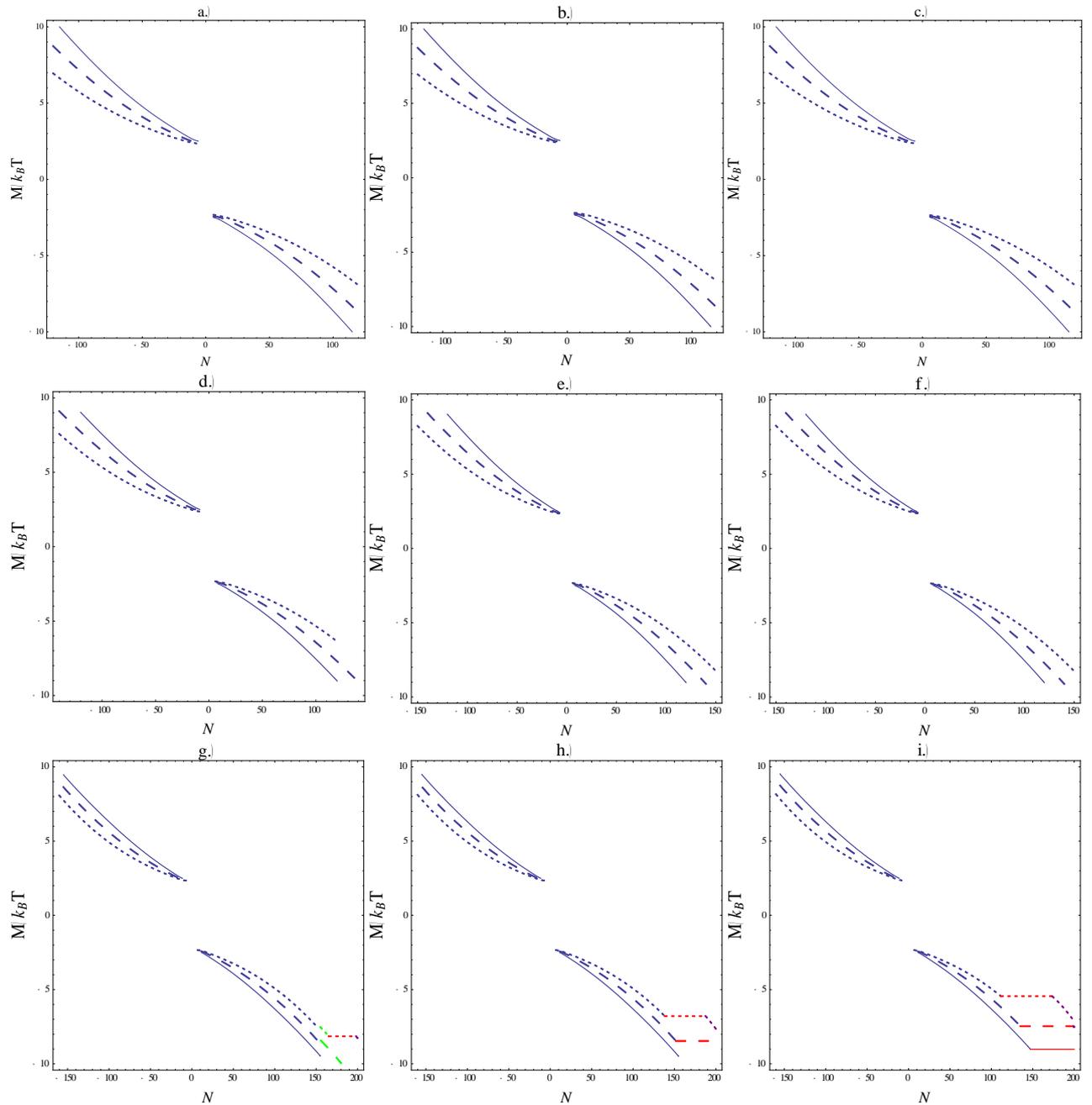

Fig. B.12. Moment curves at $F = 3.5\,\text{pN}$ as a function of $N$ for varying salt concentration and interaction parameter values. An attractive helix specific interaction term, described by Eqs. (2.53) and (2.56), is considered here. This includes an additional $\sin^2\eta(s)$ term in the interaction potential due to helix geometry. The parameter values for each panel are: a.) $\theta = 0.6$, $f_H = 0.5$; b.) $\theta = 0.6$, $f_H = 0.7$; c.) $\theta = 0.6$, $f_H = 0.9$; d.) $\theta = 0.7$, $f_H = 0.5$; e.)

$\theta = 0.7$, $f_H = 0.7$ ; f.) $\theta = 0.7$, $f_H = 0.9$; g.) $\theta = 0.8$, $f_H = 0.5$, h.) $\theta = 0.8$, $f_H = 0.7$ and i.) $\theta = 0.8$, $f_H = 0.9$. The blue curves correspond to the $\lambda_h^* = \infty$ state, the green curves to the finite $\lambda_h^*$, $\Delta\bar{\Phi} = 0$ state, and the purple curves to the $\lambda_h^*$, $\Delta\bar{\Phi} \approx \pi/2$ state. The red lines correspond to coexistence regions, where the two states at the end points coexist, and where the moment is constant. The solid, long dashed and medium dashed lines correspond to the Debye screening length values $\kappa_D^{-1} = 15.5\text{Å}, 11.7\text{Å}$ and $7.75\text{Å}$, respectively.

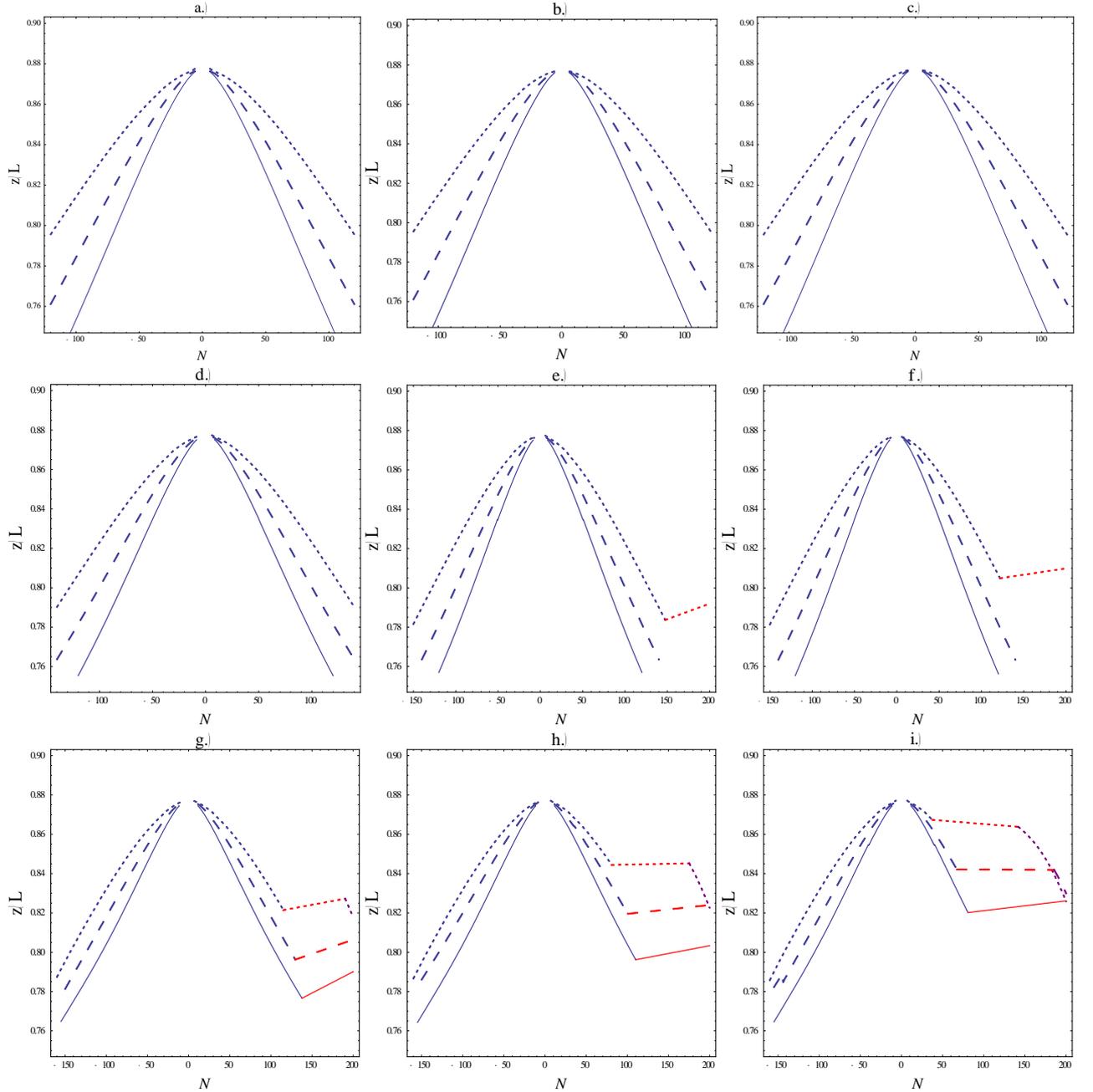

Fig. B.13 Extension curves at $F = 3.5\,\text{pN}$ as a function of $N$ for varying salt concentration and interaction parameter values. An attractive helix specific interaction term that consists Eqs. (2.53) and (2.55) is considered here. The parameter values for each panel are: a.) $\theta = 0.6$, $f_H = 0.5$ ; b.) $\theta = 0.6$, $f_H = 0.7$; c.) $\theta = 0.6$, $f_H = 0.9$; d.) $\theta = 0.7$, $f_H = 0.5$; e.) $\theta = 0.7$, $f_H = 0.7$ ; f.) $\theta = 0.7$, $f_H = 0.9$; g.) $\theta = 0.8$, $f_H = 0.5$, h.) $\theta = 0.8$, $f_H = 0.7$

and i.) $\theta = 0.8$, $f_H = 0.9$. The blue curves correspond to the $\lambda_h^* = \infty$ state, the green curves to the finite $\lambda_h^*$, $\Delta\bar{\Phi} = 0$ state, and the purple curves to the $\lambda_h^*$, $\Delta\bar{\Phi} \approx \pi/2$ state. The red lines correspond to the extension in the coexistence regions, where the two states at the end points coexist, which behaves in a linear fashion. The solid, long dashed and medium dashed lines correspond to the Debye screening length values $\kappa_D^{-1} = 15.5\text{Å}, 11.7\text{Å}$ and $7.75\text{Å}$, respectively.

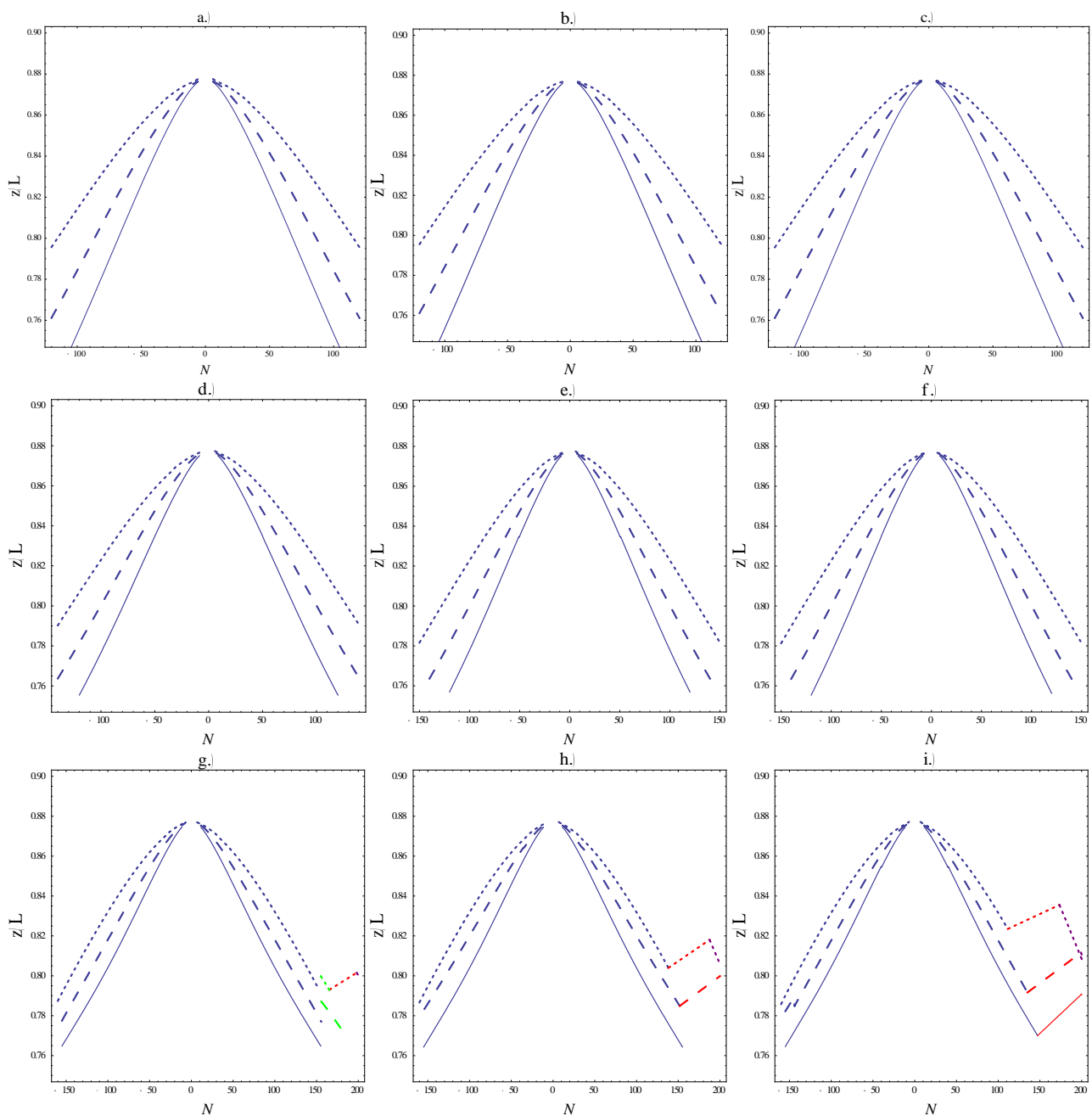

Fig. B.14 Extension curves at $F = 3.5\,\text{pN}$ as a function of $N$ for varying salt concentration and interaction parameter values. An attractive helix specific interaction term consisting of Eqs. (2.53) and (2.56) is considered here. This includes an additional $\sin^2\eta(s)$ term in the interaction energy due to the helix geometry. The parameter values for each panel are: a.) $\theta = 0.6$, $f_H = 0.5$; b.) $\theta = 0.6$, $f_H = 0.7$; c.) $\theta = 0.6$, $f_H = 0.9$; d.) $\theta = 0.7$, $f_H = 0.5$; e.)

$\theta = 0.7$, $f_H = 0.7$ ; f.) $\theta = 0.7$, $f_H = 0.9$; g.) $\theta = 0.8$, $f_H = 0.5$, h.) $\theta = 0.8$, $f_H = 0.7$ and i.) $\theta = 0.8$, $f_H = 0.9$. The blue curves correspond to the $\lambda_h^* = \infty$ state, the green curves to the finite $\lambda_h^*$, $\Delta\bar{\Phi} = 0$ state, and the purple curves to the $\lambda_h^*$, $\Delta\bar{\Phi} \approx \pi/2$ state. The red lines correspond to the extension in the coexistence regions, where the two states at the end points coexist, which behaves in a linear fashion. The solid, long dashed and medium dashed lines correspond to the Debye screening length values $\kappa_D^{-1} = 15.5\text{Å}, 11.7\text{Å}$ and $7.75\text{Å}$, respectively.

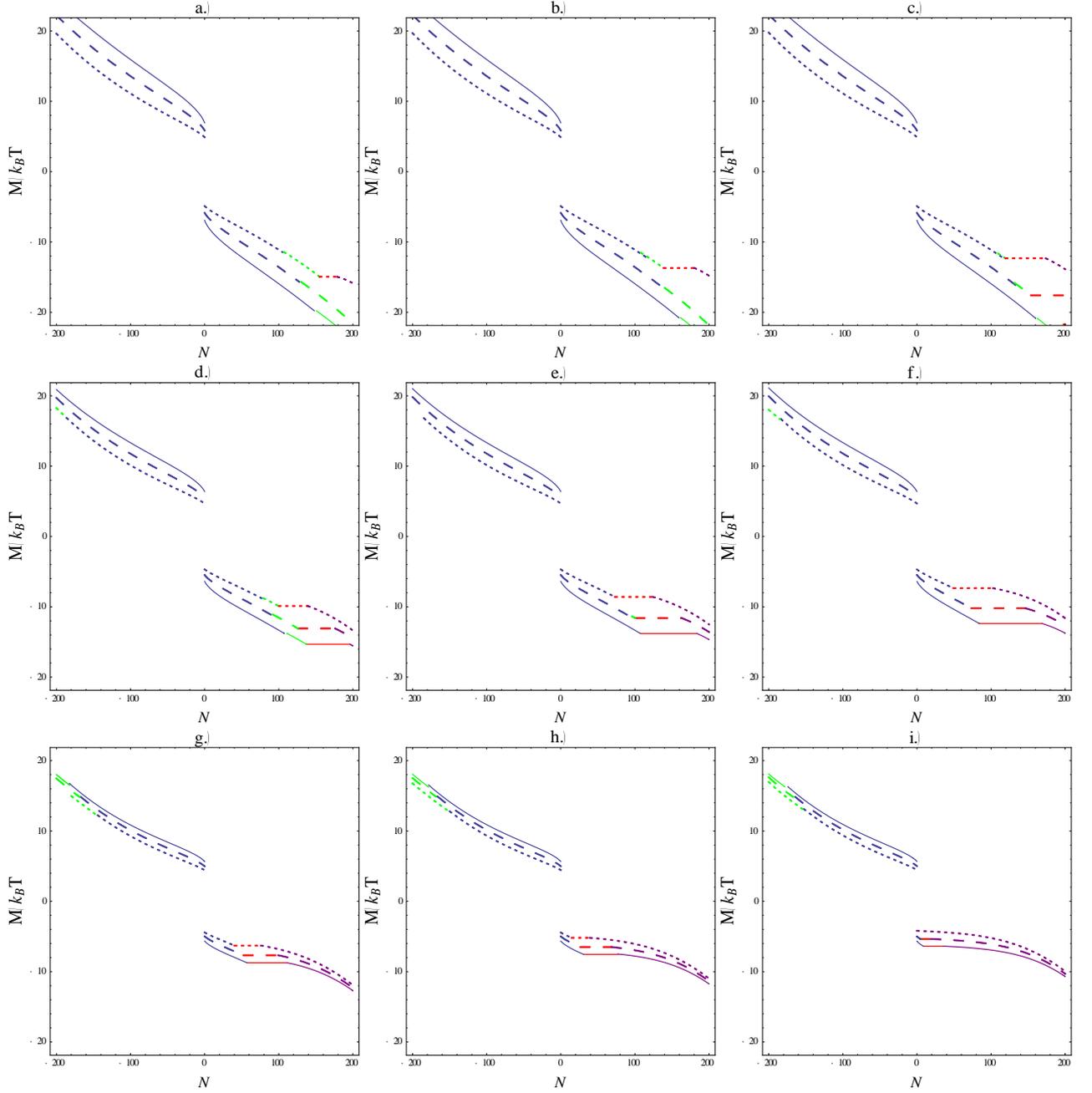

Fig. B.15. Moment curves at $F = 40\,\text{pN}$ as a function of $N$ for varying salt concentration and interaction parameter values. An attractive helix specific interaction term consisting of Eqs. (2.53) and (2.55) is considered here. The parameter values for each panel are: a.) $\theta = 0.6$, $f_H = 0.5$; b.) $\theta = 0.6$, $f_H = 0.7$; c.) $\theta = 0.6$, $f_H = 0.9$; d.) $\theta = 0.7$, $f_H = 0.5$; e.) $\theta = 0.7$, $f_H = 0.7$ ; f.) $\theta = 0.7$, $f_H = 0.9$; g.) $\theta = 0.8$, $f_H = 0.5$, h.) $\theta = 0.8$, $f_H = 0.7$ and i.) $\theta = 0.8$, $f_H = 0.9$ The blue curves correspond to the $\lambda_h^* = \infty$ state, the green curves to the finite $\lambda_h^*$,

$\Delta\bar{\Phi} = 0$ state, and the purple curves to the finite $\lambda_h^*$, $\Delta\bar{\Phi} \approx \pi/2$ state. The red lines correspond to coexistence regions, where the two states at the end points coexist, and where the moment is constant. The solid, long dashed and medium dashed lines correspond to the Debye screening length values $\kappa_D^{-1} = 15.5\text{Å}, 11.7\text{Å}$ and $7.75\text{Å}$, respectively.

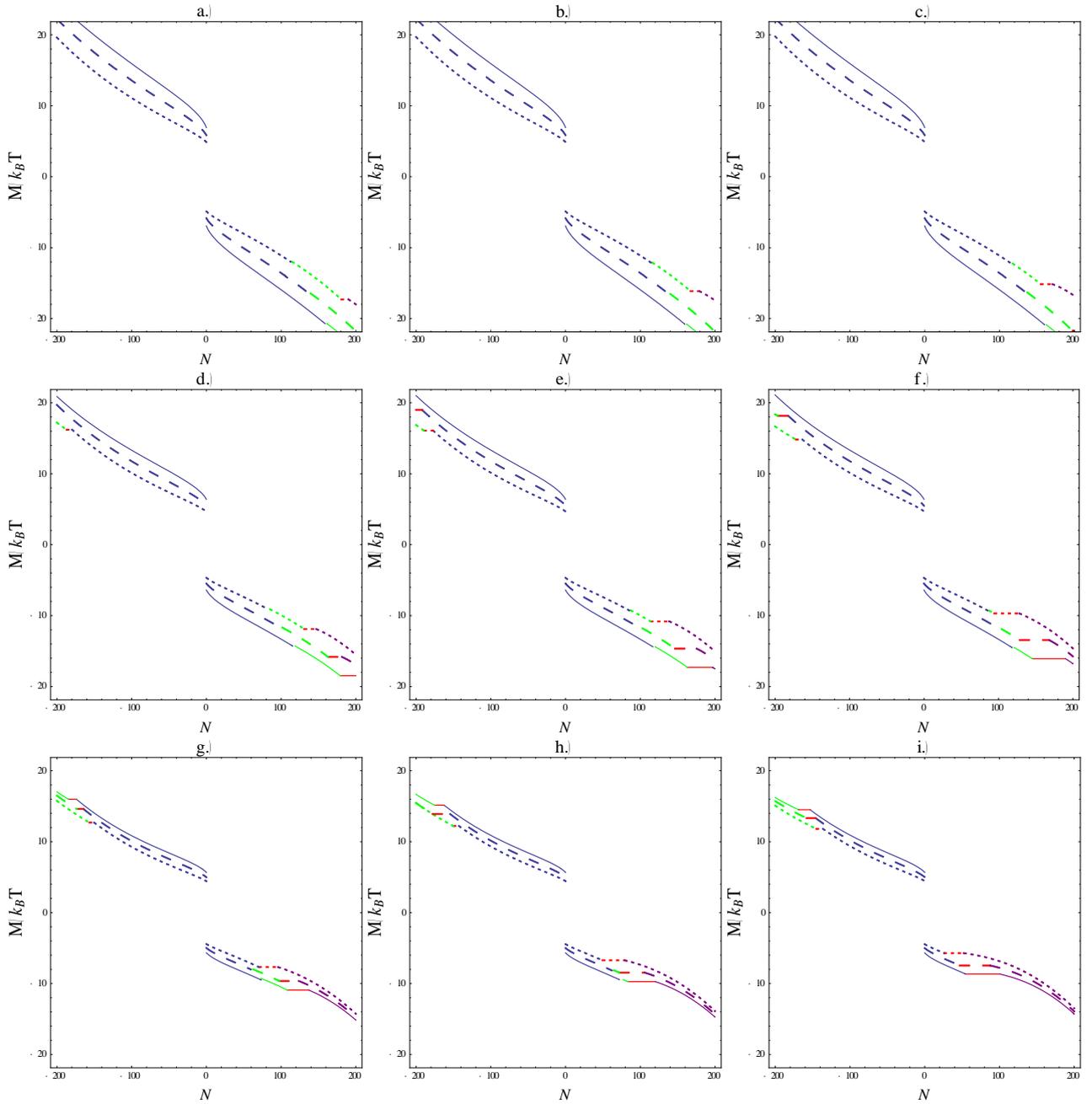

Fig. B.16 Moment curves at $F = 40\,\text{pN}$ as a function of $N$ for varying salt concentration and interaction parameter values. An attractive helix specific interaction term is considered here, described by Eqs. (2.53) and (2.56). This includes an additional $\sin^2 \eta(s)$ term in the interaction energy due to the helix geometry. The parameter values for each panel are: a.) $\theta = 0.6$, $f_H = 0.5$ ; b.) $\theta = 0.6$, $f_H = 0.7$; c.) $\theta = 0.6$, $f_H = 0.9$; d.) $\theta = 0.7$, $f_H = 0.5$; e.) $\theta = 0.7$, $f_H = 0.7$ ; f.) $\theta = 0.7$, $f_H = 0.9$; g.) $\theta = 0.8$, $f_H = 0.5$, h.) $\theta = 0.8$, $f_H = 0.7$ and i.) $\theta = 0.8$, $f_H = 0.9$. The blue curves correspond to the $\lambda_h^* = \infty$ state, the green curves to the finite $\lambda_h^*$, $\Delta\bar{\Phi} = 0$ state, and the purple curves to the finite $\lambda_h^*$, $\Delta\bar{\Phi} \approx \pi/2$ state. The red lines correspond to coexistence regions, where the two states at the end points coexist, and where the moment is constant. The solid, long dashed and medium dashed lines correspond to the Debye screening length values $\kappa_D^{-1} = 15.5\text{Å}, 11.7\text{Å}$ and $7.75\text{Å}$, respectively.

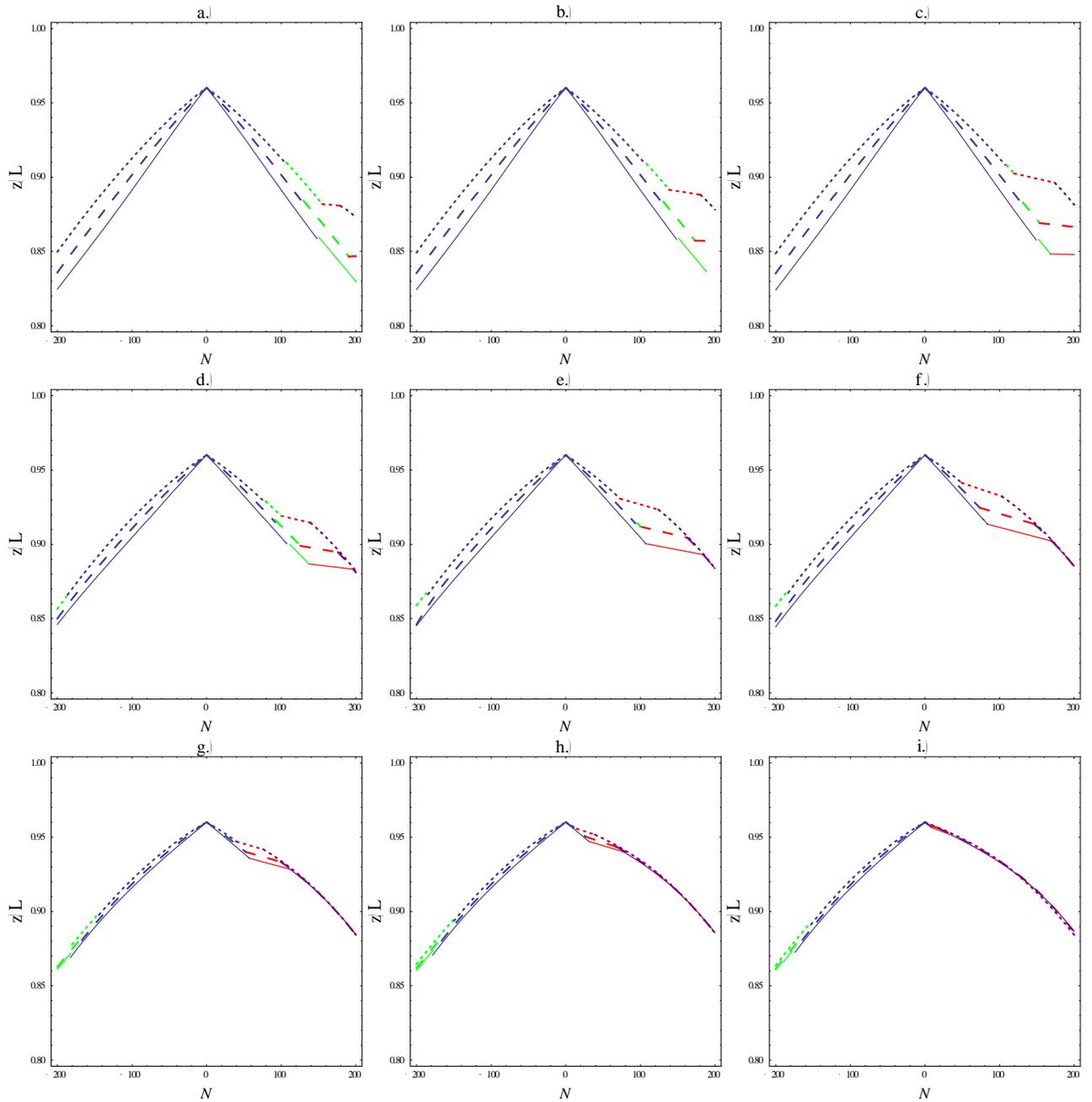

Fig. B.17. Extension curves at $F = 40\,\text{pN}$ as a function of $N$ for varying salt concentration and interaction parameter values. An attractive helix specific interaction term consisting of Eqs. (2.53) and (2.55) is considered here. The parameter values for each panel are: a.) $\theta = 0.6$, $f_H = 0.5$ ; b.) $\theta = 0.6$, $f_H = 0.7$; c.) $\theta = 0.6$, $f_H = 0.9$; d.)

$\theta = 0.7$, $f_H = 0.5$; e.) $\theta = 0.7$, $f_H = 0.7$ ; f.) $\theta = 0.7$, $f_H = 0.9$; g.) $\theta = 0.8$, $f_H = 0.5$, h.) $\theta = 0.8$, $f_H = 0.7$ and i.) $\theta = 0.8$, $f_H = 0.9$. The blue curves correspond to the $\lambda_h^* = \infty$ state, the green curves to the finite $\lambda_h^*$, $\Delta\bar{\Phi} = 0$ state, and the purple curves to the $\lambda_h^*$, $\Delta\bar{\Phi} \approx \pi/2$ state. The red lines correspond to the extension in the coexistence regions, where the two states at the end points coexist, which behaves in a linear fashion. The solid, long dashed and medium dashed lines correspond to the Debye screening length values $\kappa_D^{-1} = 15.5\,\text{Å}, 11.7\,\text{Å}$ and $7.75\,\text{Å}$, respectively.

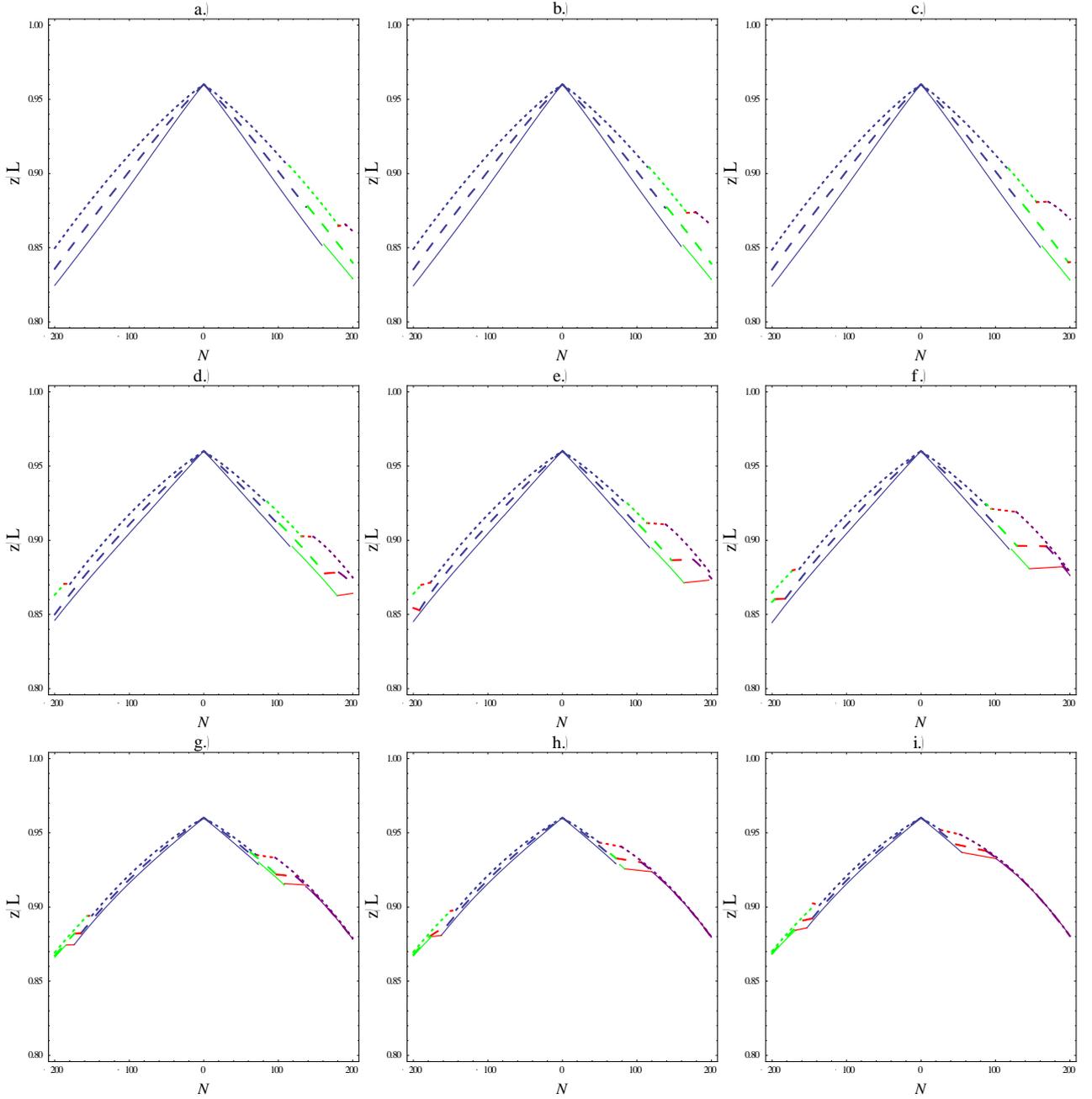

Fig. B.18 Extension curves at $F = 40\,\text{pN}$ as a function of $N$ for varying salt concentration and interaction parameter values. An attractive helix specific interaction term consisting of Eqs. (2.53) and (2.56) is considered here. This includes an additional $\sin^2 \eta(s)$ term in the interaction energy due to the helix geometry. The parameter values for each panel are: a.) $\theta = 0.6$, $f_H = 0.5$; b.) $\theta = 0.6$, $f_H = 0.7$; c.) $\theta = 0.6$, $f_H = 0.9$; d.) $\theta = 0.7$, $f_H = 0.5$; e.)

$\theta = 0.7$, $f_H = 0.7$ ; f.) $\theta = 0.7$, $f_H = 0.9$; g.) $\theta = 0.8$, $f_H = 0.5$, h.) $\theta = 0.8$, $f_H = 0.7$ and i.) $\theta = 0.8$, $f_H = 0.9$. The blue curves correspond to the $\lambda_h^* = \infty$ state, the green curves to the finite $\lambda_h^*$, $\Delta\bar{\Phi} = 0$ state, and the purple curves to the finite $\lambda_h^*$, $\Delta\bar{\Phi} \approx \pi/2$ state. The red lines correspond to the extension in the coexistence regions, where the two states at the end points coexist, which behaves in a linear fashion. The solid, long dashed and medium dashed lines correspond to the Debye screening length values $\kappa_D^{-1} = 15.5\text{Å}, 11.7\text{Å}$ and $7.75\text{Å}$, respectively.